\documentclass[aps,rmp,reprint,amsmath,amssymb,graphicx,longbibliography]{revtex4-2}

\usepackage{bm}
\usepackage{array}
\usepackage{color}
\usepackage{dcolumn}
\usepackage{bm}
\usepackage{epsf}
\usepackage{graphicx}
\usepackage{amsmath} 
\usepackage{amssymb}
\usepackage{dsfont} 
\usepackage{tikz} 
\usepackage{hyperref}
\usepackage{bbm}
\usepackage{underscore}

\newcommand{\beq}{\begin{equation}}
\newcommand{\eeq}{\end{equation}}
\newcommand{\bea}{\begin{eqnarray}}
\newcommand{\eea}{\end{eqnarray}}

\begin{document}
\topmargin-0.8in
\textheight 9.5in 

\title{Neutrinos from dense environments : Flavor mechanisms, theoretical approaches, observations, and new directions}

\author{M. Cristina Volpe}
\affiliation{CNRS, Universit\'e Paris Cit\'e, Astroparticule et Cosmologie, F-75013 Paris, France}
\email{volpe@apc.in2p3.fr}

\begin{abstract}
Neutrino masses and mixings produce vacuum oscillations, an established quantum mechanical phenomenon. 
In matter, the Mikheev-Smirnov-Wolfenstein effect, due to neutrino interactions with the background particles,
triggers resonant flavor modification.  In dense environments, such as core-collapse supernovae or compact mergers,
sizable neutrino-neutrino interactions, shock waves and turbulence impact the neutrino flavor content under a variety of phenomena. Theoretical approaches of neutrino propagation range from the mean-field approximation to the full quantum kinetic equations. Interesting connections have been uncovered between weakly interacting dense neutrino gases and other many-body systems and domains, from condensed matter and nuclear physics to quantum computing. Besides the intrinsic theoretical interest, establishing how neutrinos change flavor contributes to answering the longstanding open questions of how massive stars explode and of the $r$-process sites. It is also important for future observations of core-collapse supernova neutrinos and of the diffuse supernova neutrino background that should be discovered in the foreseeable future. 

\end{abstract}

\date{\today}
\maketitle
\tableofcontents

\section{General and historical background}

\subsection{The birth of neutrino astronomy}

In his famous letter to Lise Meitner, and to {\it "Dear Radioactive Ladies and Gentlemen"}, \textcite{Pauli:1930, Pauli:1930b} hypothesized the existence of a new fermion, the {\it neutron}. He wanted to explain the observed continuous beta spectrum in the $\beta$ decay of atomic nuclei and {\it "to save the laws of energy conservation and the statistics"}. This particle had to be as light as the electron with a mass not heavier than 0.01 that of the proton. Renamed {\it neutrino} ("small neutral particle" in Italian), it remained elusive until \textcite{Cowan:1992xc} detected electron antineutrinos via inverse $\beta$ decay from nearby reactors, the most powerful man-made neutrino sources in terrestrial experiments. 

The same year \textcite{Lee:1956qn} examined the question of parity conservation in weak interactions, inspired by the so-called $\theta$-$\tau$ meson puzzle. They suggested, as a possible experimental test of the parity nonconservation hypothesis, the measurement of a pseudoscalar observable, namely the angular distribution of electrons emitted in polarized $^{60}$Co decay. In a few months \textcite{Wu:1957my} successfully performed the experiment, demonstrating weak interaction differentiated the right from the left. The following year
\textcite{Goldhaber:1958nb} measured neutrinos from electron capture in $^{152}$Eu and found them to be left-handed.  
In the \textcite{Glashow:1961tr}, \textcite{Weinberg:1967tq}, \textcite{Salam:1957st} (GWS) model, there are  three neutrino flavors, $\nu_e$, $\nu_{\mu}$ and $\nu_{\tau}$
and neutrinos are massless.

In his seminal work \textcite{Bethe:1939bt} suggested that carbon and nitrogen act as catalysts in a chain reaction and are mainly responsible for hydrogen burning 
into helium in luminous main sequence stars (later known as the CNO cycle). Afterward, solar models predicted 
sizable $\nu_e$ fluxes from energy generation due mostly to hydrogen burning into helium in the proton-proton ({\it pp}) reaction chain \cite{Bahcall:1964gx}. 
\textcite{Davis:1968cp} first detected solar neutrinos with his pioneering radiochemical experiment in the Homestake mine, using neutrino capture on $^{37}$Cl \cite{Davis:1964hf}.
In a few months the measurement revealed fewer neutrinos than expected according to the predictions of \cite{Bahcall:1968hc}: the {\it solar neutrino problem} was born. 
Based on these observations it was deduced that only a small portion of the solar radiated energy was coming from the CNO cycle \cite{Davis:1968cp,Bahcall:1968hc}. 

For more than three decades, 
radiochemical, water Cherenkov and scintillator experiments showed that, depending on neutrino energy, one-third to one-half of the predicted solar neutrino fluxes were actually reaching Earth (see for example \cite{Raffelt:1996wa,Giunti:2007ry,Robertson:2012ib}). Both the standard solar model and neutrino properties were questioned. 
Helioseismology brought an important clue in support of the Standard Solar Model (see for example \cite{TurckChieze:1993dw}). In particular, the solar sound speed, measured at a few percent level, agreed with predictions. 

Among the debated solutions of the solar neutrino problem was the possibility that neutrinos could oscillate, as earlier pointed out by \textcite{Pontecorvo:1957cp,Pontecorvo:1957qd} who first suggested that $\nu$ could transform into $\bar{\nu}$. Later on, \textcite{Gribov:1968kq} considered the possibility of $\nu_e$ oscillations into $\nu_{\mu}$ in analogy with oscillations of neutral $K_0$-$\bar{K}_0$ mesons.  

\textcite{Wolfenstein:1977ue} pointed out that in matter neutrinos can change flavor due to coherent forward scattering and a flavor-dependent refractive index. 
In a subsequent work, \textcite{Wolfenstein:1979ni} explained that matter at high density in collapsing stars can inhibit vacuum oscillations. 
Later on, \textcite{Mikheev:1986wj} realized that flavor conversion in matter could be resonantly amplified: an adiabatic evolution at the resonance location could solve the solar neutrino problem (see also \cite{Mikheev:1986wj,Bouchez:1986kb,Bethe:1986ej,Haxton:1986dm,Parke:1986jy}). This phenomenon came to be known as the Mikheev-Smirnov-Wolfenstein (MSW) effect. 

\begin{figure}
\begin{center}
\includegraphics[width=.45\textwidth]{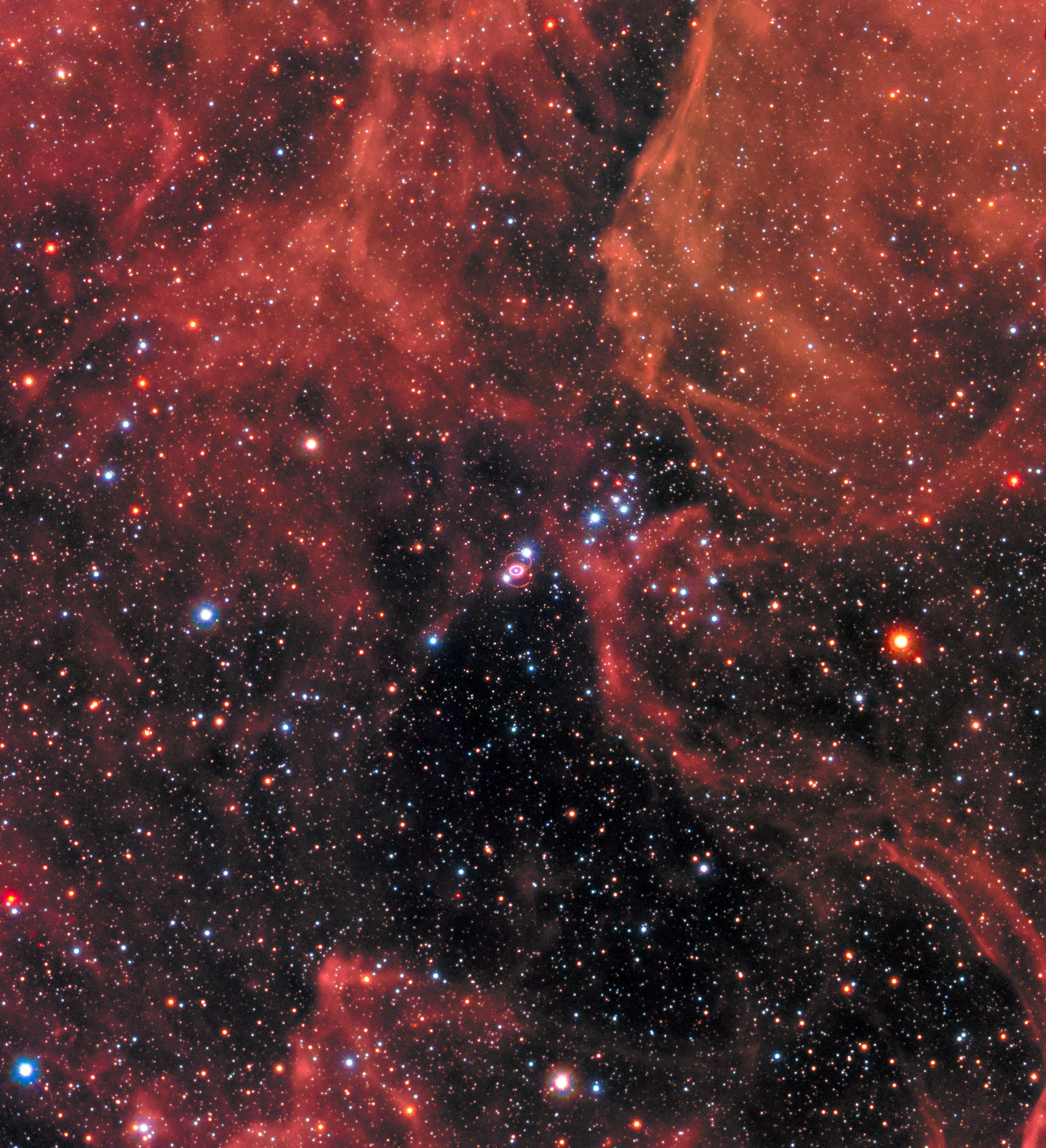}
\caption{Hubble Space Telescope image of SN1987A, in the Large Magellanic Cloud, a neighboring galaxy of the Milky way, 30 yr after its explosion. The central part of the image -- SN1987A -- shows three rings that correspond to earlier events. The central inner ring is material that has been ejected by the star about 20 000 yr before SN1987A. The material ejected by the supernova explosion has reached the inner ring and makes it glow. From \cite{HST:SN1987A}.}
\label{fig:SN1987A}
\end{center}
\end{figure}

In 1987, the explosion of the blue supergiant Sk-69$^{\circ}$202 brought evidence that core-collapse supernovae\footnote{Supernovae are massive stars that, at the end of life, undergo either thermonuclear explosions -- SNe type Ia -- or gravitational core-collapse -- SNe types II and Ib/c.  Stars having 8-11 ${\rm ~M}_{\odot}$ develop degenerate O-Ne-Mg cores that eventually undergo gravitational collapse producing supernovae (see e.g. \cite{Nomoto:1984,Ritossa:1999,Janka:2012wk}. More massive stars develop an iron core before collapse. SN type II show hydrogen in their spectra; whereas SNe type Ia do not. Type Ia contain Si in their spectrum, contrary to types Ib and Ic. Type Ic is also poor in He, whereas type Ib is rich. SNe types II and Ib/c become powerful neutrino sources when they undergo gravitational core collapse and explode.} emit $\nu$ at the end of their life (Figure \ref{fig:SN1987A}). SN1987A was in the Large Magellanic Cloud (LMC), a satellite galaxy of the Milky Way. Kamiokande-II (KII) \cite{Kamiokande-II:1987idp}, Irvine-Michigan-Brookhaven (IMB) \cite{Bionta:1987qt}  detectors and the Baksan Scintillator Telescope (BST)  \cite{Alekseev:1988gp} recorded a 10 seconds burst of about 24 events, with a few tens of MeV energy. The Mont Blanc Liquid Scintillator Detector (LSD) \cite{Aglietta:1987it} detected five events, 5 hours before the others, so the connection of LSD events to SN1987A remains controversial.

The neutrino events from SN1987A confirmed that neutrinos take away most of the gravitational energy, as \textcite{Colgate:1966ax} conjectured, and agreed overall with the predicted neutrino fluxes and spectra. Moreover, the Bayesian analysis of 
the SN1987A time signal by \textcite{Loredo:2001rx} corroborated a two-component (accretion plus cooling) model at $2 \sigma$ to  $3 \sigma$, which was confirmed by the subsequent analysis by \textcite{Pagliaroli:2008ur}. This supported the {\it delayed neutrino-heating mechanism} of \textcite{Bethe:1984ux}, thus rejecting the favored {\it prompt bounce-shock} model by \textcite{Colgate:1966ax}. On the particle physics side, the approximately two dozens events brought a large amount of constraints on unknown neutrino properties (such as the neutrino magnetic moment, charge radius or decay), on nonstandard interactions and particles such as axions (see for example \cite{Raffelt:1996wa,ParticleDataGroup:2020ssz,Mohapatra:1998rq,Payez:2014xsa,Chang:2018rso,Ivanez-Ballesteros:2023lqa,Mahoney:2017jqk,Fiorillo:2022cdq}). 

The observation of neutrinos from the Sun and from SN1987A pioneered neutrino astronomy\footnote{R. Davis (Homestake) and M. Koshiba (Kamiokande Collaboration) were the recipients of the 2002 Nobel Prize with R. Giacconi (X-ray astronomy).}. The detection of PeV neutrinos in the IceCube detector at the South Pole \cite{Aartsen:2014gkd}  opened
a new observational window. One of the events detected thus far is consistent with blazar TXS 0506+056 \cite{IceCube:2018cha}. Furthermore,
$79^{+22}_{-20}$ TeV neutrino events have been associated with the active galaxy NGC1068 and a supermassive black hole, with a statistical significance of 4.2 $\sigma$ \cite{IceCube:2022der}. With these observations, neutrino astronomy now runs from MeV to PeV neutrino energies. 

\subsection{The oscillation discovery}
Primary cosmic rays interacting with the Earth's atmosphere produce twice as many $\nu_{\mu}$ as $\nu_{e}$ from $\pi$ and $\mu$ decay.
Underground experiments searching for proton instability, which was expected in some unified theories, reported
a reduced $\nu_{\mu}/\nu_e$ ratio in the atmospheric background, with respect to Monte Carlo simulations. This was known as the {\it atmospheric anomaly}
(see for example \citealp{Giunti:2007ry}).

In 1998 the Super-Kamiokande (SK) Collaboration \cite{Fukuda:1998mi} discovered\footnote{T. Kajita (SK Collaboration) and A.B. McDonald (SNO Collaboration) were recipients of the Nobel Prize in 2015.} that atmospheric $\nu_{\mu}$ traversing Earth (upgoing) were less than expected, whereas upgoing $\nu_{e}$ stayed unaffected. The zenith angle dependence of the $\mu$-like and $e$-like events  gave unambiguous proof that $\nu_{\mu}$ oscillated into $\nu_{\tau}$.

The Sudbury Neutrino Observatory (SNO) \cite{Ahmad:2001an} and the Kamioka Liquid Scintillator Antineutrino Detector (KamLAND) \cite{Eguchi:2002dm} experiments brought two further milestones in the clarification of the solar neutrino problem. The first experiment \cite{SNO:2001kpb}, which used heavy water, found $^8$B solar neutrinos to be in agreement with the standard solar model predictions. The different sensitivities of $\nu_e$, $\nu_{\mu}$ and $\nu_{\tau}$ to elastic scattering, combined with neutral- and charged-current  $\nu$ interactions on deuterium allowed the solar $\nu_{\mu}, \nu_{\tau}$ fluxes to be identified at $5.3~\sigma$ \cite{Ahmad:2002jz}. Moreover KamLAND \cite{Eguchi:2002dm} measured $\bar{\nu}_e$ disappearance at an average distance of 200 km from Japanese reactors and unambiguously identified the MSW solution called large-mixing-angle (LMA).\footnote{At that time, discussed were the MSW solutions called small mixing angle with $\Delta m^2_{21} = (4$-$10) \times 10^{-6} $ eV$^2$ and $\sin^2 2 \theta_{21} = (2$-$10) \times 10^{-3} $, large mixing angle (LMA) with $\Delta m^2_{21} = (1$-$10) \times 10^{-5} $ eV$^2$ and $\sin^2 2 \theta_{21} =$ 0.7-0.95, "low $\delta m^2$" (LOW) with $\Delta m^2_{21} \sim (0.5$-$2) \times 10^{-7} $ eV$^2$ and $\sin^2 2 \theta_{21} = 0.9$-1  and the vacuum oscillation (VO) solution also referred to as "just so" with $\Delta m^2_{21} \sim 10^{-10}$-$10^{-11} $ eV$^2$ and $\sin^2 2 \theta_{21} = 0.8$-1 (see \cite{Dighe:1999bi, Giunti:2007ry}, and references therein).} .  

These observations established that only half of low energy (less than 2 MeV) solar $\nu_e$ reach Earth because of averaged vacuum oscillations, whereas high energy $^{8}$B neutrinos are reduced to one-third due to the MSW effect.  The solar neutrino problem was finally solved. 
 
 The solution of the solar neutrino problem had required more than three decades of searches, concerning in particular the origin of the energy dependence of the solar neutrino deficit, the 
 day-night effect and seasonal variations, and was also the result of global data analysis and fits. These investigations excluded other explanations due to nonstandard physics, investigated along the years, such as neutrino decay, nonstandard interactions and the neutrino magnetic moment (see for the review by \textcite{Robertson:2012ib}).

Furthermore the Borexino experiment measured for the first time the solar $^7$Be \cite{Borexino:2008dzn}, the $pep$ \cite{Borexino:2011ufb} and the $pp$ neutrinos from the keystone reaction of the $pp$ reaction chain.  The corresponding fluxes are consistent with vacuum-averaged oscillations. In particular, the measurement of the $^7$Be flux confirmed that the $\nu_e $ survival probability increases in the vacuum-dominated region. Besides neutrinos from the CNO cycle were first detected \cite{BOREXINO:2020aww}, confirming that it contributes to solar fusion at 1 $\%$ level, favoring Standard Solar models with high metallicity.

Vacuum oscillations imply that neutrinos are elementary particles with nonzero masses and mixings. Hence, the flavor and mass bases are related by the Pontecorvo-Maki-Nakagawa-Sakata (PMNS) unitary matrix, which is analogous to the Cabibbo-Kobayashi-Maskawa matrix in the quark sector (although with large mixing angles).   

Since 1998, atmospheric, solar,  reactor and accelerator experiments have determined most of the neutrino oscillation parameters in the theoretical framework with three active neutrinos. The mixing angles,  $\theta_{23} \approx 45^{\circ} $, $\theta_{12} \approx 35^{\circ}$ and $\theta_{13} \approx 8.2^{\circ} $, as well as the mass-squared differences $\Delta m^2_{32} = m^2_{3} - m^2_{2} = 2.4 \times 10^{-3} $ eV$^2$ (atmospheric) and $\Delta m^2_{21} = m^2_{2} - m^2_{1} =  7.5 \times 10^{-5} $ eV$^2$ (solar) \cite{ParticleDataGroup:2020ssz} are known with good accuracy (i.e. at a few percent precision). The PMNS matrix also depends on three phases, one Dirac and two Majorana phases. The Dirac phase is currently being measured. It can produce a difference between neutrino and antineutrino oscillations, thus breaking the CP symmetry (C for charge conjugation and P for parity). The Majorana phases remain unknown. 
The discovery of neutrino vacuum oscillations was a breakthrough: it opened a window beyond the standard model and had an impact in astrophysics and cosmology. 

\subsection{Unknown neutrino properties}
Key neutrino properties remain unknown and will be the object of an intense experimental program. 
Sixty years after \textcite{Christenson:1964fg} discovered that weak interaction breaks the CP symmetry in $K_0$ decay, there are indications that neutrinos do not oscillate in the same way as antineutrinos. If confirmed by future experiments, this will point to the presence of CP violation in the lepton sector and to CP breaking values of the Dirac phase (see \cite{Capozzi:2021fjo} for an analysis).

The ordering of the neutrino mass eigenstates needs to be established, since the atmospheric mass-squared difference sign has not yet been measured. The neutrino mass ordering (or hierarchy) might be normal  ($ \Delta m^2_{32} > 0$), or inverted ($ \Delta m^2_{32} < 0$). On the other hand, the sign of the solar mass-squared difference is determined by the presence of the MSW resonance in the Sun. Currently data show a preference  (at 2.5 $\sigma$) for normal ordering, i.e. the third mass eigenstate is likely more massive than the others \cite{Capozzi:2021fjo}.

Vacuum oscillations are sensitive to mass-squared differences but do not give information on the scale of the neutrino masses.
The absolute neutrino mass scale is not identified yet. The KATRIN experiment obtained sub-eV upper limits ($m < 0.8$ eV at 90 $\%$ confidence level) on the effective $\nu_e$ mass with tritium $\beta$-decay \cite{KATRIN:2021uub}. Complementary information comes from cosmological observations which give (model dependent) information on the sum of the neutrino masses (at the sub-eV level) \cite{ParticleDataGroup:2020ssz}.

The ensemble of results from oscillation experiments cannot be fully interpreted in the theoretical framework with three active neutrinos. It presents some {\it anomalies}. 
The {\it reactor antineutrino anomaly} refers to a discrepancy, at 5-6 $\%$ level, between the predicted and measured ${\bar{\nu}}_e$ fluxes from nuclear power plants, due to a reevaluation of the fluxes \cite{Mention:2011rk}. The {\it gallium anomaly} refers to a $\nu_e$ deficit observed in the solar GALLEX and SAGE experiments when the fluxes from a radioactive source were measured \cite{Giunti:2010zu}. The last debated anomalies come from neutrino accelerator experiments, namely the liquid scintillator neutrino detector experiment that measured vacuum oscillations at $\Delta m^2 \approx 1$ eV$^2$ and the MiniBooNE experiment that found an excess of $\nu_e$ events at low energy.    

Recent evaluations of the reactor neutrino fluxes and a campaign of nuclear measurements have lowered the statistical significance of the reactor anomaly \cite{Giunti:2021kab,Zhang:2023zif}, whereas the gallium one, confirmed by the counting BEST experiment \cite{Barinov:2021asz}, gives sterile mixing parameters in tension with the reactor anomaly. Moreover, the first results from the MicroBooNE experiment  \cite{Arguelles:2021meu} disfavor some explanations and part of the parameter space identified by the MiniBooNE low energy excess. Clearly further work is needed to elucidate the origin of such anomalies.

Among the debated solutions of the anomalies is the existence of a fourth nonstandard sterile neutrino.
Sterile neutrinos do not interact with matter (they do not couple to the standard model gauge bosons) and can manifest themselves 
in neutrino vacuum oscillations because of their coupling to active neutrinos through a PMNS matrix with $N \ge 4$. The existence of sterile
neutrinos is actively debated (see \cite{Acero:2022wqg}). 

The origin of the neutrino masses remains an open issue. 
Seesaw mechanisms constitute a possibility of explaining the smallness of neutrino masses and are investigated in numerous theories beyond the Standard model (see for example the reviews by \cite{Altarelli:2010gt,King:2015aea}). In the simplest (type-I) seesaw models, neutrinos acquire a small mass because of the existence of very heavy right-handed neutrinos.

Moreover, as pointed out long ago by \textcite{Majorana:1937vz}, neutrinos could well be their own antiparticles. Searches for a rare nuclear process called 2$\beta$(0$\nu$) decay
that implies total lepton-number violation \cite{Giunti:2007ry,Agostini:2022zub} appear to be the most feasible path to uncover the neutrino nature and give access to the Majorana CP violating phases. As for neutrino electromagnetic properties, such as the neutrino magnetic moment, only bounds exist  (see \cite{Giunti:2015gga}). 

In conclusion, the key open issues in neutrino physics include the neutrino absolute mass and mass ordering, the origin of neutrino masses, the existence of CP violation in the lepton sector and of sterile neutrinos, the neutrino Dirac versus Majorana nature and neutrino electromagnetic properties. 

Neutrino properties are intertwined with neutrino flavor evolution in dense sources and influence observations. Therefore, as we later discuss, neutrinos from such environments offer ways to learn about some of the unknowns on the one hand, and constitute a unique probe in astrophysics and cosmology on the other hand. 

\subsection{Future supernova neutrino observations}
To date SN1987A remains the only core-collapse supernova observed through its neutrinos.
Supernovae\footnote{Note that the detection of the emitted neutrinos could help in elucidating the precise mechanism for the thermonuclear explosion of SNe Ia \cite{Wright:2016gar}.} type II and Ib/c are an interesting and rich laboratory for particle physics and astrophysics, requiring both multipurpose and dedicated neutrino observatories that can run over long time periods.  

A network of neutrino detectors around the world, based on different technologies, is awaiting for the next (extra)galactic supernova explosion. 
Among the detectors included in the network are SK, the Helium and Lead Observatory (HALO, 76 tons of lead), KamLAND (1 kton, liquid scintillator), IceCube (cubic-km, Cherenkov detector) and in the future the Deep Underground Neutrino Experiment (DUNE, 40 kton, liquid argon), the Jiangmen Underground Neutrino Observatory (JUNO, 20 kton, liquid scintillator), KM3NeT/ARCA \cite{KM3Net:2016zxf}, the  Hyper-Kamiokande (Hyper-K, water Cherenkov, 248 ktons) and dark matter detectors. 
The Supernova Early Warning System (SNEWS) \cite{Scholberg:1999tm,SNEWS:2020tbu} should alert astronomers if a supernova explosion takes place.

In the Milky Way the spatial probability distribution of objects that are likely to become supernovae has its maximum at the Galaxy Center at 8 kpc and its mean at 10 kpc. The latter is adopted mostly to making predictions. 

Neutrino observatories will measure the time signal and the spectra of $\nu_e$, $\bar{\nu}_e$, $\nu_x$ and $\bar{\nu}_x$ ($x = \mu, \tau$) with charged-current $\nu_e$ scattering on nuclei, inverse $\beta$ decay, elastic scattering on electrons and protons \cite{Beacom:2002hs}, and coherent neutrino-nucleus scattering \cite{COHERENT:2017ipa}.  
If a supernova explodes in our galaxy (10 kpc), detectors will  
measure\footnote{For the event rates see also https://github.com/SNOwGLoBES/snowglobes \cite{Scholberg:2012id} and SNEWPY \cite{SNEWS:2021ewj}.} about 40 (540) events in HALO (HALO-2, 1 kton) \cite{Vaananen:2011bf}, hundreds in KamLAND, up to $3 \times 10^3$ in DUNE \cite{DUNE:2020zfm}, up to about $ 8 \times 10^3$ in JUNO \cite{JUNO:2015zny}, almost 10$^{4}$ in Super-K \cite{Beacom:1998ya}, $10^5$ in Hyper-Kamiokande \cite{Hyper-Kamiokande:2018ofw} and $10^{6}$ in IceCube\footnote{The rates correspond to a luminosity of $3 \times 10^{53}$ ergs (or close to it) and an average energy between 12 and 18 MeV (depending on the neutrino species) with 100 $\%$ or more realistic efficiencies.}. From a supernova in Andromeda galaxy (M31, 773 kpc) that has a low supernova rate, 12 events are expected in Hyper-K. 

Dark matter detectors will also measure supernova neutrinos through coherent neutrino-nucleus scattering, sensitive to all neutrino flavors. 
Xenon nT/LZ (7 tons) and DARWIN (40 tons) will measure 120 and 700 events, and have a discovery potential for supernova neutrinos up to the Milky Way edge and
the small Magellanic Cloud (SMC), respectively (a 27 ${\rm M}_{\odot}$ progenitor) \cite{Lang:2016zhv}. For the same progenitor, a liquid argon dark matter detector such 
as Darkside-20k (50 tons) will detect 336 events (supernova at 10 kpc) and be able to detect supernova neutrinos up to the LMC, whereas Argo (360 tons) will detect 2592 events
and be sensitive to a supernova explosion up to the SMC \cite{DarkSide20k:2020ymr}. 

Supernovae are rare in our galaxy. Typical quoted number for the core-collapse supernova rate in our Galaxy is 1 to 3 per century. \textcite{Rozwadowska:2020nab},
obtained a mean time of occurrence of $61 ^{+ 24}_{- 14}$ based on neutrino and electromagnetic observations of collapse events in the Milky Way and the Local Group.

Supernovae are frequent in the Universe. 
With a 1 Mton detector, about one supernova per year is expected within 10 Mpc, due to nearby galaxies with higher rates than the Milky Way. Within 4 Mpc, less than one neutrino event per year would be detected \cite{Ando:2005ka}. 

Neutrinos from past supernova explosions form a relic, or diffuse, supernova neutrino background (DSNB) (see the reviews by \textcite{Ando:2004hc,Beacom:2010kk,Lunardini:2010ab,Mathews:2019klh,Suliga:2022ica,Ando:2023fcc}). Its flux, integrated over cosmological redshift, depends on the redshifted supernova neutrino fluxes, the evolving core-collapse supernova rate and the fraction of failed supernovae that turn into a black hole without an electromagnetic counterpart \cite{Lunardini:2009ya,Keehn:2010pn}. At present we only have upper limits.

SK \cite{Malek:2002ns} set the first upper limit of 1.2 $\bar{\nu}_e$ cm$^{-2}$s$^{-1}$ ($E_{\nu} > 19.3~$MeV, 90 $\%$ C.L.) on the supernova relic flux. 
The bound was improved with SK-IV data \cite{Zhang:2013tua} using $\bar{\nu}_e$ detection via inverse $\beta$ decay and neutron tagging on protons.
The DSNB search combining SK-I to SK-IV data yields 2.7 $\bar{\nu}_e$ cm$^{-2}$s$^{-1}$ ($E_{\nu} > 17.3~$MeV, 90 $\%$ C.L.). 
The KamLAND experiment obtained the upper value of 139 $\bar{\nu}_e$ cm$^{-2}$s$^{-1}$ (90 $\%$ C.L.) encompassing 8.3 to 31.8 MeV \cite{KamLAND:2011bnd}. 
The Borexino Collaboration extracted a model-dependent limit of 112.3 $\bar{\nu}_e$ cm$^{-2}$s$^{-1}$ (90 $\%$ C.L.)  in the interval between 7.8 and 16.8 MeV \cite{Borexino:2019wln}. 

As for $\nu_e$, the ensemble of SNO data provided the upper limit of 19 $\nu_e$ cm$^{-2}$s$^{-1}$ encompassing 22.9 to 36.9 MeV (90 $\%$ C.L.) \cite{SNO:2006dke}. 
The loosest limits are $\phi_{\nu_x, \bar{\nu}_x} <$ 1.3-1.8 10$^3 $ cm$^{-2}$s$^{-1}$ ($E_{\nu} > 19$ MeV for $x = \mu, \tau$ flavors)  \cite{Lunardini:2008xd}.
With neutrino-nucleus coherent scattering in dark matter detectors, one could improve this bound to 10 $\nu_x$ or 10 $\bar{\nu}_x$ \cite{Suliga:2021hek}. 

\textcite{Beacom:2003nk} suggested to add gadolinium (Gd) to water Cherenkov detectors\footnote{The idea was named GADZOOKS ! for Gadolinium Antineutrino Detector Zealously Outperforming Old Kamiokande, Super!}. Neutron capture by Gd improves inverse $\beta$ decay tagging through the 8 MeV photons following the capture\footnote{An efficiency of 90 $\%$ is expected with a 0.1 $\%$ Gd concentration.}. The SK-Gd experiment is currently accumulating data. 

DSNB predictions are close to current bounds (see \cite{Ando:2004hc,Fukugita:2002qw,Lunardini:2005jf,Yuksel:2007mn,Galais:2009wi,Chakraboty:2010sz,Vissani:2011kx,Priya:2017bmm,Horiuchi:2017qja,Moller:2018kpn,Kresse:2020nto,Ivanez-Ballesteros:2022szu,Tabrizi:2020vmo,DeGouvea:2020ang,Ekanger:2022neg,Ashida:2022nnv,Ashida:2023heb}).
The analysis from the combined SK-I to IV data showed an excess at 1.5 $\sigma$ over a background prediction \cite{Super-Kamiokande:2021jaq}. The related sensitivity analysis is on par with four of the most optimistic predictions \cite{Ando:2004hc,Horiuchi:2020jnc,Kresse:2020nto,Galais:2009wi} and a factor of about 2-5 larger than the most conservative ones. 

The first result from the SK-Gd experiment (SK VI-SK VII phases) combined with SK I-SK IV data now shows an excess at 2.3 $\sigma$ in the DSNB signal window \cite{Harada:2024}.
With the SK-Gd, the upcoming JUNO and the more distant future Hyper-K and DUNE experiments, the DNSB should be discovered in the future. Indeed, its discovery could be imminent. 

\subsection{The $r$ process and GW170817}
Besides direct observations of neutrinos from supernovae, the study of indirect effects produced by neutrinos in dense environments
has also stimulated intense investigations of neutrinos and of neutrino flavor evolution in dense media. 
Neutrinos in dense environments are tightly connected to two unresolved issues in astrophysics:  the death of massive stars and the origin of $r$-process elements ($r$-process stands for rapid neutron capture process). 

Currently two- and three-dimensional simulations include realistic neutrino transport, convection, turbulence and hydrodynamical instabilities such as the standing accretion-shock instability (SASI) \cite{Blondin:2002sm} and the lepton-number emission self-sustained asymmetry (LESA) \cite{Tamborra:2014aua} (see \textcite{Kotake:2005zn,Janka:2012wk,Janka:2017vcp,Mezzacappa:2020oyq,Radice:2017kmj,Burrows:2019zce,Takiwaki:2021dve,Foglizzo:2015dma}). The 
delayed neutrino-heating mechanism is believed to trigger most core-collapse supernova explosions. The most energetic supernovae might require a magnetohydrodynamical mechanism (see \textcite{Janka:2012wk}). The future observation of neutrinos from a galactic or extragalactic supernova could confirm or refute the current paradigm and elucidate a six-decade quest. 

The $r$ process is a nucleosynthesis process that takes place in a neutron-rich environment and during short timescales (i.e. seconds). In this process nuclei capture neutrons faster than they undergo $\beta$ decay toward the stability line. The $r$-process produces thousands of exotic nuclei far from the neutron drip line and about half of the heavy elements in our Galaxy\footnote{The other half is produced in the $s$ process ($s$ stands for slow), where nuclei undergo $\beta$-decay towards the stability line faster than they capture neutrons. A small part of the heavy elements comes from the so-called $p$ process ($p$ stands for proton).}.  A weak $r$ process produces elements in the first peak around mass number A=80-90 and the second peak around A=130-138. A strong $r$-process reaches the third peak at A=195-208.

\textcite{Burbidge:1957vc} and \textcite{Cameron:1958vx} first linked the $r$-process to core-collapse supernovae which have long been thought the main $r$-process site\footnote{With a rule of thumb, if each supernova produces $10^{-4} M_{\odot}$ elements and there are three such events per century, in $10^{10}$ yr there are 3 $10^{4} M_{\odot}$ $r$-process elements ejected in the Milky Way.} (see \textcite{Qian:2013fsa}, \textcite{Kajino:2019abv}). While simulations show that entropies are typically too low, the most energetic supernovae seem to provide the right conditions to attain a successful nucleosynthesis (see \textcite{Cote:2018qku,Cowan:2019pkx} for a comprehensive review). 

Another candidate site for the $r$ process is binary neutron star mergers (BNS) as first suggested by
\textcite{Lattimer:1974slx,Lattimer:1976kbf} (see for example \cite{Kajino:2019abv,Cote:2018qku,Goriely:2015fqa} for reviews). As supernovae, BNS mergers are powerful sources of MeV
 neutrinos. Indeed they emit 10$^{51}$ to 10$^{53}$ erg in $\nu_e$, ${\nu}_{\mu}$, ${\nu}_{\tau}$ and their antiparticles, with tens of MeV. Contrary to supernovae, binary neutron star mergers are more neutron-rich and produce an excess of $\bar{\nu}_e$ over $\nu_e$ (see \textcite{Cusinato:2021zin}). As for ${\nu}_{\mu}$ and ${\nu}_{\tau}$, their fluxes are predicted to be small compared to those of core-collapse supernovae and have large theoretical uncertainties\footnote{See Table VII of \textcite{Frensel:2016fge}).}. If supernovae are more frequent than BNS mergers, simulations show that BNS mergers offer more suitable astrophysical conditions for a strong $r$ process. Moreover, studies have shown that some $r$-process elements are synthetized in accretion disks around black holes \cite{Surman:2005kf} and black hole-neutron star mergers \cite{Surman:2008qf,Caballero:2011dw}. 

\begin{figure}
\begin{center}
\includegraphics[width=.45\textwidth]{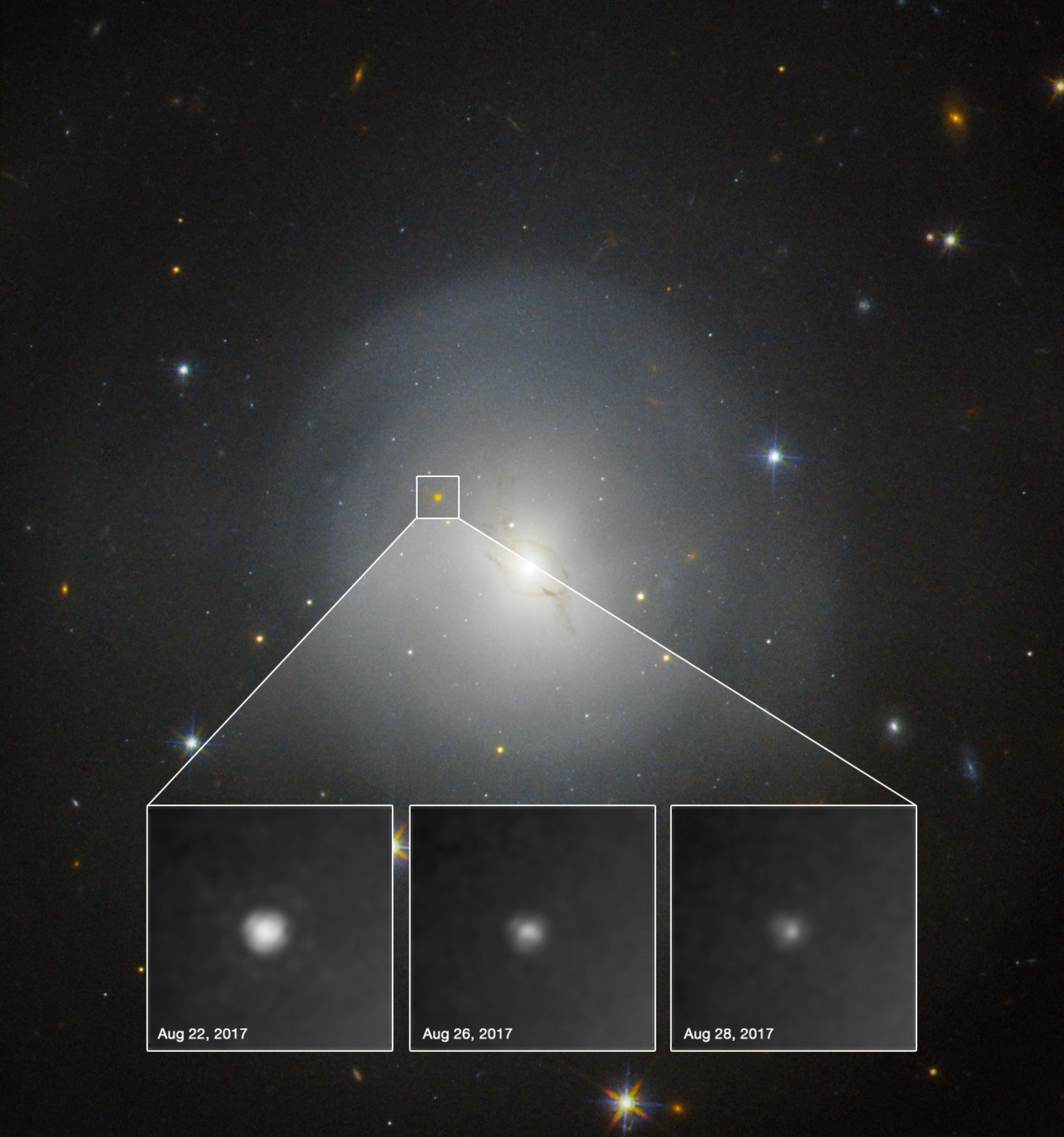}
\caption{Hubble Space Telescope image of kilonova gradually fading, in the lenticular galaxy NGC 4993 (40 Mpc). The event GW170817 was seen concomitantly in gravitational waves, in electromagnetic emission and with a short gamma-ray burst. It represented the first observation from merging binary neutron stars. Adapted from \cite{NASA}.}
\label{fig:kilonova}
\end{center}
\end{figure}

The first detection of gravitational waves from the fusion of two black-holes  by the Virgo-LIGO Collaboration  \cite{Abbott:2016blz} has opened
the era of gravitational wave astronomy. To date, GW170817 \cite{TheLIGOScientific:2017qsa,GBM:2017lvd} is a unique multimessenger event in which gravitational waves from a binary neutron star merger were first detected, also concomitantly with a short gamma-ray burst and a kilonova. The electromagnetic signal of a kilonova \cite{Metzger:2010sy}, or macronova \cite{Kulkarni:2005jw} is between those of novae and supernovae. Since the afterglow emission of the kilonova associated with GW170817 extends over some days, it appears that radioactivity injects some energy, thereby powering the kilonova. The optical or IR spectra in the IR emission peak are compatible with elements heavier than iron to be responsible for absorption and reemission of the radiation \cite{Cowan:2019pkx} (Fig. \ref{fig:kilonova}). In particular,
the ejecta opacities indicate the presence of actinides and lanthanides (see the studies by \cite{Cote:2018qku,Tanaka:2017lxb,Cowan:2019pkx}). 
This represents the first evidence for $r$-process elements in binary neutron star mergers.

Before GW170817 dynamical ejecta were thought to contribute mainly to a strong $r$ process (see the discussion of \textcite{Martin:2015hxa}). But the kilonova observation has changed this paradigm. Indeed the comparison of the electromagnetic emission with most models shows two components: the early and fast premerger contribution from dynamical ejecta and the later 
postmerger one due to viscosity and neutrino-driven winds. The former is associated with red emission, while the latter is associated with the blue one \cite{Radice:2018pdn}. The role of neutrinos on the premerger and postmerger ejecta and of flavor evolution appears to be crucial and is currently object of debate; see \textcite{Nedora:2020hxc}.

Indeed, numerous $r$-process studies have included not only neutrinos but also neutrino flavor evolution. Many find that matter becomes proton rich and tends to harm the $r$ process in core-collapse supernovae\footnote{Note that there are other nucleosynthesis processes where neutrinos influence element abundances, including neutrino nucleosynthesis \cite{Langanke:2019ggn} and the $\nu p$ process \cite{Frohlich:2005ys}.}. 
One should keep in mind though the complexity of studying flavor evolution in dense environments in a consistent manner that tracks the evolution both of the neutrino flavor and of the nucleosynthetic abundances. This is true both for core-collapse supernovae and for BNS mergers. Depending on the site and the assumptions made, one can find situations in which flavor modification favors or harms the $r$ process. In general, what emerges from investigations is that neutrino flavor evolution impacts the nucleosynthetic abundances when one includes standard or nonstandard $\nu$ properties and interactions.  
In conclusion, the quest for the identification of the $r$-process site(s) and the supernova mechanism as well as the need for predictions regarding future observations has motivated
an in-depth investigation of flavor evolution in dense environments for many years, as we now discuss. 

\subsection{Theoretical developments}
In an astrophysical environment such as binary neutron star merger remnants or core-collapse supernovae, the environments are dense not only in matter but also in neutrinos. Indeed, in such sites the neutrino density becomes comparable to that of electrons or nucleons. For example, during the explosion of a core-collapse supernova, about $10^{58} \nu$ with an average energy of 10 MeV are emitted, so the neutral-current $\nu\nu$ interaction becomes sizable.   
 
Understanding flavor evolution is an interesting theoretical problem that prompts many interesting questions. 
In dense environments, do new phenomena emerge? What are the conditions to trigger them and what is their impact?
Do novel flavor mechanisms introduce extra heating and help supernova explosions? How do neutrinos behave in the presence of shock waves and turbulence? Are there flavor mechanisms that favor the $r$ process? Is the commonly employed mean-field approximation sufficient? 
Do weakly interacting neutrino gases behave like known many-body systems? What is the interplay with unknown neutrino properties? What is the role of strong gravitational fields? 
Are many-body correlations important? The list encompasses many others.  Over 30 years theoretical studies have paved the way to answers to these questions. 

To begin, investigations have shown that a variety of novel conversion phenomena can occur due to matter, shock waves, turbulence and $\nu\nu$ interactions.
There is the established MSW effect that takes place in both astrophysical and cosmological environments. 
In particular, \textcite{Dighe:1999bi} pointed out that, because of the large densities and of radiative corrections \cite{Botella:1986wy}, three MSW resonances
occur in core-collapse supernovae. 
While the MSW effect is certainly a reference in studies of flavor modification,
the novel phenomena uncovered go far beyond it.
 
\textcite{Schirato:2002tg} reported that shock waves could modify the time signal of supernova neutrinos. 
\textcite{Tomas:2004gr} demonstrated the presence of front and reverse shocks in supernova simulations. 
\textcite{Fogli:2003dw} and then \textcite{Dasgupta:2005wn} found that front and reverse shocks produce multiple MSW resonances and {\it phase effects}. 
Several authors studied their impact; see \cite{Takahashi:2002yj,Fogli:2004ff,Choubey:2007ga,Kneller:2007kg,Gava:2009pj}. 

Concerning noisy media, \textcite{Loreti:1994ry} first studied the influence of matter fluctuations in relation with solar neutrinos. 
\textcite{Balantekin:1996pp,Nunokawa:1996qu,Loreti:1995ae} showed that  
fluctuations of matter density profiles could induce neutrino depolarization. 
The impact of turbulence on neutrinos was then explored in the context of supernovae  with
similar conclusions reached \cite{Fogli:2006xy,Friedland:2006ta,Kneller:2010sc,Lund:2013uta,Abbar:2020ror}, as well as opposite ones \cite{Borriello:2013tha}. 

Besides shock waves and turbulence, neutrino-neutrino interactions have attracted a strong interest. In the early 1990s, \textcite{Pantaleone:1992eq} pointed out that such interactions become sizable when the neutrino number densities are sufficiently large while introducing an off-diagonal refractive index. As a consequence, neutrino propagation becomes a nonlinear many-body problem. \textcite{Samuel:1993uw} showed that such interactions could trigger new flavor effects. 
At first, studies for the early Universe implemented $\nu\nu$ interactions \cite{Kostelecky:1993yt,Kostelecky:1994dt,Dolgov:2002ab,Pastor:2001iu,Abazajian:2002qx,Mangano:2005cc}; see also \cite{Gava:2010kz}. Moreover, \textcite{Rudzsky:1990}, \textcite{Sigl:1992fn} and  \textcite{McKellar:1992ja} derived
neutrino quantum kinetic equations including neutrino interactions with matter and neutrinos. 

\textcite{Duan:2006an} attracted attention to $\nu\nu$ interactions in core-collapse supernovae.
\textcite{Balantekin:2004ug} also showed that such interactions could produce significant effects on the $r$ process; see also the early work of \cite{Qian:1994wh}). These works stimulated intense activity on $\nu \nu$ interactions; see the reviews by \textcite{Duan:2009cd}, \textcite{Duan:2010bg}, \textcite{Mirizzi:2015eza}, \textcite{Volpe:2015rla}. 
The first numerical simulations, which were based on the stationary {\it bulb} model, studied in great detail the mechanisms under which neutrinos first synchronized, then underwent bipolar oscillations and finally spectral splits\footnote{These are often called slow modes, as we discuss in section II.F.}. In the literature, flavor modes due to $\nu\nu$ interactions are often referred to as {\it collective neutrino oscillations}\footnote{Note however, that, in the presence of a neutrino background, the effects of $\nu\nu$ interactions are not necessarily collective or oscillatory.}.

Moreover, \textcite{Malkus:2012ts} showed that in black hole accretion disks the interplay between $\nu \nu$ and matter interactions produced a new mechanism later called {\it neutrino-matter resonance}. This was studied in the context of merging compact objects (black hole-neutron star and neutron star-neutron star) \cite{Malkus:2014iqa,Malkus:2015mda,Wu:2015fga,Vaananen:2015hfa,Frensel:2016fge,Vlasenko:2018irq} and of core-collapse supernovae, with nonstandard interactions \cite{Stapleford:2016jgz}.

Before the creation of the {\it bulb} model, \textcite{Sawyer:2005jk} showed that neutrino-neutrino interactions could trigger significant flavor conversion on short scales; see also \textcite{Sawyer:2008zs}. It was only ten years later that \cite{Sawyer:2015dsa}, when he considered different neutrinospheres for $\nu_e$ and $\bar{\nu}_e$ and found flavor modes with a few nanosecond scale, triggered excitement. 
Indeed, for a long period, theorists had searched for mechanisms that could take place behind the shock wave and impact the explosion dynamics of core-collapse supernovae. 
These modes were called fast, in contrast to the ones found in the bulb model. 

Subsequent studies showed that the occurrence of fast modes is when nontrivial angular distributions of $\nu_e$ and $\bar{\nu}_e$ produce a crossing (a change of sign) of the angle distribution of the electron lepton number (ELN) of the neutrino flux, as first pointed out by \textcite{Izaguirre:2016gsx}.
The conditions to have fast modes and their impact is actively investigated; see
\cite{Chakraborty:2016lct,Dasgupta:2016dbv,Abbar:2018beu,Abbar:2018shq,Abbar:2019zoq,Glas:2019ijo,Wu:2017qpc,Wu:2017drk,George:2020veu,Nagakura:2021hyb,Just:2022flt}, and the review by \textcite{Tamborra:2020cul}. \textcite{Morinaga:2021vmc} demonstrated that the presence of ELN crossings is a necessary and sufficient condition for fast modes in an inhomogeneous medium. Moreover \textcite{Fiorillo:2023hlk} proved that ELN crossings are not only necessary but also sufficient in homogeneous media. 
 
While most of the developments focused on the novel flavor mechanisms and their impact, numerous studies concentrated on the neutrino evolution equations themselves. Indeed, the majority of the literature employs the mean-field equations derived by \cite{Samuel:1993uw,Sawyer:2005jk,Balantekin:2006tg,Volpe:2013uxl,Serreau:2014cfa}.
However, in dense stellar regions, or in the early Universe, where collisions matter, neutrino quantum kinetic equations are necessary. Such equations were obtained using different approaches;  see \cite{Rudzsky:1990,Sigl:1992fn,McKellar:1992ja,Vlasenko:2013fja,Blaschke:2016xxt,Froustey:2020mcq}.
In principle, the theoretical framework should consistently evolve from the collision-dominated to the mean-field regime.

Keep in mind that, even in models with reduced degrees of freedom and approximations, the description of neutrino propagation requires the solution of a large number of stiff and coupled 
nonlinear equations (in the presence of $\nu\nu$ interactions). Instead of focusing on the solution to the full nonlinear problem, one can gain some information on the occurrence of flavor instabilities from linearized equations of motion, as first pointed out by \textcite{Sawyer:2008zs}.  

\textcite{Banerjee:2011fj} provided a linearization of the equations of motion that give eigenvalue equations. If solutions are complex they point to unstable normal modes whose amplitude can grow exponentially. \textcite{Vaananen:2013qja} provided an alternative derivation inspired by the random-phase approximation that is used, for example, in the study of collective vibrations in atomic nuclei or metallic clusters. \textcite{Izaguirre:2016gsx} cast the linearized equations in a dispersion-relation approach that is commonly used in the study of fast modes.

New numerical methods based on deep learning techniques have recently been employed; see \cite{Rrapaj:2020lmx,Armstrong:2021bxq,Armstrong:2022tsj,Abbar:2023zkm}). They go beyond theoretical approaches using forward integration.

But are the commonly employed mean-field equations enough when neutrinos start free streaming?
This aspect has been actively debated. Extended mean-field equations were suggested first. In particular,
\textcite{Balantekin:2006tg} discussed corrections to the mean-field approximation using the coherent-state path-integral approach.
\textcite{Volpe:2013uxl} pointed out that the most general mean-field equations include both pairing correlations, analogous to those of Bardeen-Cooper-Schrieffer \cite{Bardeen:1957mv} and wrong-helicity contributions due to the absolute neutrino mass. For the latter, in an early work \textcite{Rudzsky:1990} derived quantum kinetic equations for Wigner distribution functions in both flavor and spin space. The wrong-helicity contributions were also revived by \textcite{Vlasenko:2013fja}, and called {\it spin coherence}, in their derivation of the neutrino quantum kinetic equations based on the closed-time-path formalism.
\textcite{Serreau:2014cfa} referred to them as {\it helicity coherence} and provided the most general mean-field equations for anisotropic and inhomogeneous media.

The commonly used mean-field approximation involves a coherent sum of neutrino forward-scattering amplitudes. \textcite{Cherry:2012zw} showed that a small fraction of backward neutrinos (a {\it neutrino halo}) can influence the neutrino flavor content. Backscattered neutrinos can be consistently accounted for only with solutions to the full quantum kinetic equations. The result by \textcite{Cherry:2012zw} questioned the commonly used description of neutrino flavor evolution as a boundary-value problem. Halo effects were further studied in O-Ne-Mg \cite{Cherry:2013mv} and iron core-collapse supernovae; see \cite{Sarikas:2012vb}.

Moreover, \textcite{Pehlivan:2011hp} showed that the use of algebraic methods and of the Bethe ansatz opens
the way to exact solutions of the quantum many-body problem of neutrino propagation in dense media (without the matter term and collisions). Further investigation 
this has uncovered the importance of many-body correlations \cite{Pehlivan:2014zua,Birol:2018qhx} and brought interesting connections to quantum information theory \cite{Cervia:2019res,Patwardhan:2021rej,Lacroix:2022krq,Roggero:2021fyo,Roggero:2021asb,Roggero:2022hpy} and quantum devices \cite{Hall:2021rbv,Amitrano:2022yyn}.

As for neutrino quantum kinetic equations, which consistently include collisions, their full solution is achievable if the medium is homogeneous and isotropic, as in the early Universe. Such a solution was recently obtained by \cite{Froustey:2020mcq,Bennett:2020zkv}. 
In contrast, this becomes a formidable numerical task in dense stellar environments. Indeed, the neutrino quantum kinetic equations depend on time and the neutrino position and momentum. Not only is the phase space seven dimensional, the $\nu\nu$ interactions are sizable, and we face a nonlinear many-body problem.

There are currently serious efforts to explore the role of collisions and their interplay with flavor mechanisms; see \cite{Capozzi:2018clo,Richers:2019grc,Hansen:2022xza,Xiong:2022vsy,Ehring:2023abs}).
Moreover a new mechanism called the collisional instability,  which can occur on the microsecond scale, was pointed out by \cite{Johns:2021qby}. This new instability takes place 
if  the collision rates for neutrinos and antineutrinos are unequal in deep supernova regions where neutrinos have not fully decoupled from the medium. Collisional instabilities are currently being investigated; see \cite{Johns:2022yqy,Xiong:2022vsy,Xiong:2022zqz}.
 
Furthermore, theoretical studies brought interesting connections between a weakly interacting neutrino gas in a dense environment and other many-body systems.
\textcite{Pehlivan:2011hp} showed that the neutrino Hamiltonian can be related to the (reduced) Bardeen-Cooper-Schrieffer Hamiltonian of Cooper pairs in superconductivity. 
\textcite{Volpe:2013uxl} applied the Born-Bogoliubov-Green-Kirkwood-Yvon (BBGKY) hierarchy to an interacting neutrino gas in a medium, thereby introducing contributions from the
neutrino mass and pairing correlators, and established a formal connection with atomic nuclei and condensed matter. \textcite{Vaananen:2013qja} linearized the
extended mean-field equations and introduced a description in terms of quasiparticles. Moreover, 
with the SU(2) [SU(3)] formalism for $2 \nu$ ($3\nu$) flavors of Bloch vectors, \textcite{Fiorillo:2023mze} analyzed differences and similarities of both slow and fast modes, 
and the connection with the BCS theory for superconductivity based on the work of \textcite{Anderson:1958zza} and \textcite{Yuzbashyan:2018gbu} on the BCS Hamiltonians. 
Finally, \textcite{Mirizzi:2015fva} pointed out a connection between the flavor evolution of an interacting neutrino gas and the transition from 
laminar to turbulent regimes in nonlinear fluids.

Flavor evolution in dense objects is also interesting because it is tightly linked to neutrino properties and nonstandard physics or particles, for example 
sterile neutrinos \cite{McLaughlin:1999pd,Fetter:2002xx,Tamborra:2011is,Wu:2013gxa,Xiong:2019nvw}, nonstandard interactions \cite{Wolfenstein:1977ue,EstebanPretel:2007yu,Blennow:2008er,Stapleford:2016jgz,Chatelain:2017yxx,Das:2017iuj}, neutrino mass ordering \cite{Engel:2002hg,Dighe:2003be,Barger:2005it,Gava:2009pj,Serpico:2011ir} and CP violation \cite{Akhmedov:2002zj,Balantekin:2007es,Gava:2008rp,Pehlivan:2014zua,Popov:2021icg}.
  
Numerous reviews on core-collapse supernova neutrinos are available. They focus on oscillations in media \cite{Kuo:1989qe}, the diffuse supernova $\nu$ background \cite{Ando:2004hc,Beacom:2010kk,Lunardini:2010ab,Mathews:2019klh,Suliga:2022ica,Ando:2023fcc}, presupernova neutrinos \cite{Kato:2020hlc}, $\nu\nu$ interactions \cite{Duan:2010bg,Tamborra:2020cul}, $\nu\nu$ interactions and turbulence \cite{Duan:2009cd}, supernova $\nu$ detection \cite{Scholberg:2012id}, observations \cite{Horiuchi:2018ofe}, production, oscillations and detection \cite{Mirizzi:2015eza} and the neutrino evolution equations \cite{Volpe:2015rla}.
 
The goal of this review is to highlight the richness and complexity of neutrino flavor evolution in dense media while summarizing the status and the challenges that lie ahead. The review encompasses in particular two aspects of this interesting problem, namely flavor mechanisms and the theoretical frameworks including the connections to other domains, and discusses aspects of supernova neutrino observations. For more than 15 years this has been a fast-developing field where new approaches, interesting connections and  ideas continue to be proposed which makes the writing of this review challenging.

The structure of the review is as follows. Section II focuses on flavor mechanisms in media. Section III presents the theoretical approaches and developments in the description of neutrino propagation in dense media, as well as the connections to other domains. Section IV addresses past and future observations of supernova neutrinos. Finally Section V presents conclusions and perspectives. 

\section{Neutrino flavor mechanisms in dense environments} 
Neutrino flavor mechanisms are quantum mechanical phenomena\footnote{From now on we use $\hbar = c = 1$.}. 
Neutrino oscillations in vacuum, first suggested by \textcite{Pontecorvo:1957cp}, arise because the flavor (or interaction) and mass (or propagation) bases 
do not coincide. This produces an interference phenomenon among the mass eigenstates when neutrinos propagate. Vacuum oscillations are analogous to Rabi oscillations in atomic physics and $K_0-\bar{K}_0$ oscillations in meson systems.

The flavor and mass bases are related by the PMNS matrix $U$, that is
\beq \label{eq:VO}
\vert \nu_{\alpha} \rangle = U^{*}_{\alpha i } ~ \vert \nu_{i} \rangle \, ,   
\eeq
with $i = 1, 2, 3, ..., N$  and $\alpha = e, \mu, \tau, \dots $ the mass and flavor indices\footnote{A sum on repeated indices is intended.}, respectively, with $N$ an arbitrary number of neutrino families. 
The matrix is unitary ($U^{-1} = U^{\dagger}$). For antineutrinos the same relation holds, with $U_{\alpha i } $ instead of $U^{*}_{\alpha i }$.

For $N$ neutrino families, the PMNS matrix depends on $N(N-1)/2$ angles and $N(N+1)/2$ phases.
However, only $(N-1)(N-2)/2$ phases are left if neutrinos are Dirac particles, since some phases can be reabsorbed by a redefinition of both the charged lepton 
and the neutrino fields in the GWS Lagrangian. In contrast, $N-1$ supplementary phases remain if neutrinos are Majorana particles, since some of the phases cannot be reabsorbed by a redefinition of the neutrino fields. 

More explicitly, the PMNS matrix for 3$\nu$ flavors can be parametrized as \cite{ParticleDataGroup:2020ssz}
\begin{equation} \label{eq:pmns}
U_{\rm PMNS} = U V = U {\rm diag} (1, e^{i \eta_1}, e^{i \eta_2}) \ ,
\end{equation}
with $\eta_1$and $\eta_2$ the Majorana CP violating phases\footnote{Note that $\eta_1$and $\eta_2$ do not influence vacuum oscillations.} and, 
\begin{equation} \label{eq:PP}
U =  \left(
\begin{tabular}{ccc}
$c_{12} c_{13}$  & $s_{12} c_{13} $   &$ s_{13} e^{-i \delta}$ \\
$ - s_{12} c_{23} - c_{12} s_{13} s_{23}  e^{i \delta} $ & $ c_{12} c_{23} - s_{12}  s_{13} s_{23} e^{i \delta} $  &$ c_{13}  s_{23} $ \\
$s_{12} s_{23} - c_{12}  s_{13} c_{23} e^{i \delta}$  &$  - c_{12} s_{23} - s_{12} s_{13} c_{23} e^{i \delta}  $ &  $ c_{13}  c_{23}$ \\
\end{tabular}
\right) \, ,
\end{equation}
where $c_{ij} = \cos \theta_{ij}$ and $s_{ij} = \sin \theta_{ij}$ with $i, j = 1, 3$; $\delta$ is the Dirac CP violating phase.

The massive neutrino states are eigenstates of the free Hamiltonian ${\cal H}_{\rm vac} = {\rm diag}(E_i)$, 
with eigenenergies 
\beq \label{eq:nueneergy}
E_i = \sqrt{{\bf p}^2_i + m_i^2} \, ,   
\eeq
momentum ${\bf p}_i$ and mass $m_i$. 
Assuming the equal momentum approximation and that neutrinos are ultra-relativistic,
the neutrino energy can be written as
\beq\label{eq:relap}
E_i \approx E + {m_i^2 \over {2 E}} \ , 
\eeq
where $E = \vert  {\bf p} \vert $. 

In the flavor basis, the free Hamiltonian reads  
\beq \label{eq:Hvac}
{\cal H}^f_{\rm vac} = U {\cal H}_{\rm vac} U^{\dagger} .
\eeq
This Hamiltonian is responsible for the well established phenomenon of neutrino oscillations in vacuum.

We recall the neutrino equations of motion used mostly to determine how neutrino flavor changes in media.
Based on such equations several flavor conversion mechanisms emerge, as we later see. 

\subsection{Mean-field equations}\label{sec:MF}
Here we consider the notion that neutrinos evolve in a medium and coherently scatter on the particles composing it. Incoherent scattering and general relativistic effects
(for example spacetime curvature due to the presence of strong gravitational fields) are neglected. Moreover neutrinos are described using plane waves.  
Under these assumptions, a neutrino flavor state evolves according to the Schr\"odinger-like equation\footnote{We assume that $t=x$ (light-ray approximation).} \cite{Halprin:1986pn}
\beq\label{e:Schro}
i {d  \over dt} \vert \nu_{\alpha} (t) \rangle = {\cal H}  \vert \nu_{\alpha} (t) \rangle \, ,
\eeq
with the initial condition $\vert  \nu_{\alpha} (0) \rangle = \vert  \nu_{\alpha} \rangle$.
When neutrinos traverse dense matter, the neutrino Hamiltonian receives different contributions, namely
\beq\label{e:H}
{\cal H} = {\cal H}_{\rm vac}^f + {\cal H}_{\rm mat} + {\cal H_{\nu \nu}} + {\cal H_{\rm NSI}} \, ,
\eeq
where $ {\cal H}_{\rm mat}$ comes from neutrino interactions with matter and ${\cal H_{\nu \nu}} $ comes from $\nu\nu$ interactions. The last term ${\cal H}_{\rm NSI}$ is present if nonstandard interactions exist between neutrinos and neutrinos, or neutrinos and matter.

We now take a closer look at the different contributions to the neutrino Hamiltonian \eqref{e:H}. When neutrinos traverse
a dense astrophysical medium, they interact with the background electrons, protons, neutrons and neutrinos. The mean field is the simplest and most widely used approximation to implement such interactions. 

\begin{figure}
\begin{center}
\includegraphics[width=0.5\textwidth]{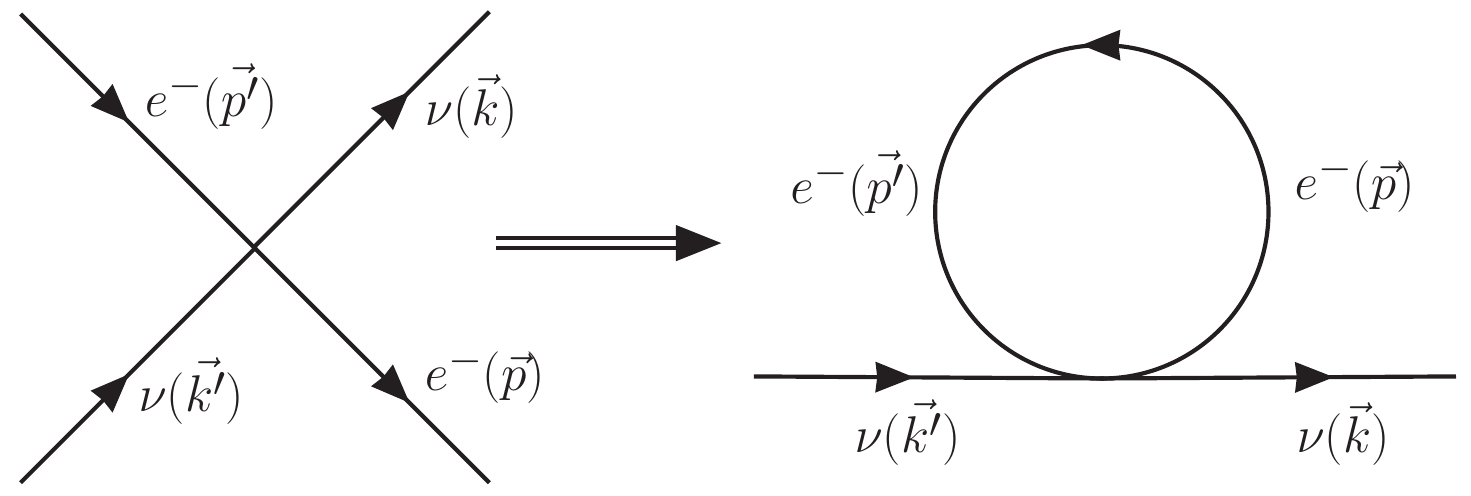} 
\includegraphics[width=0.5\textwidth]{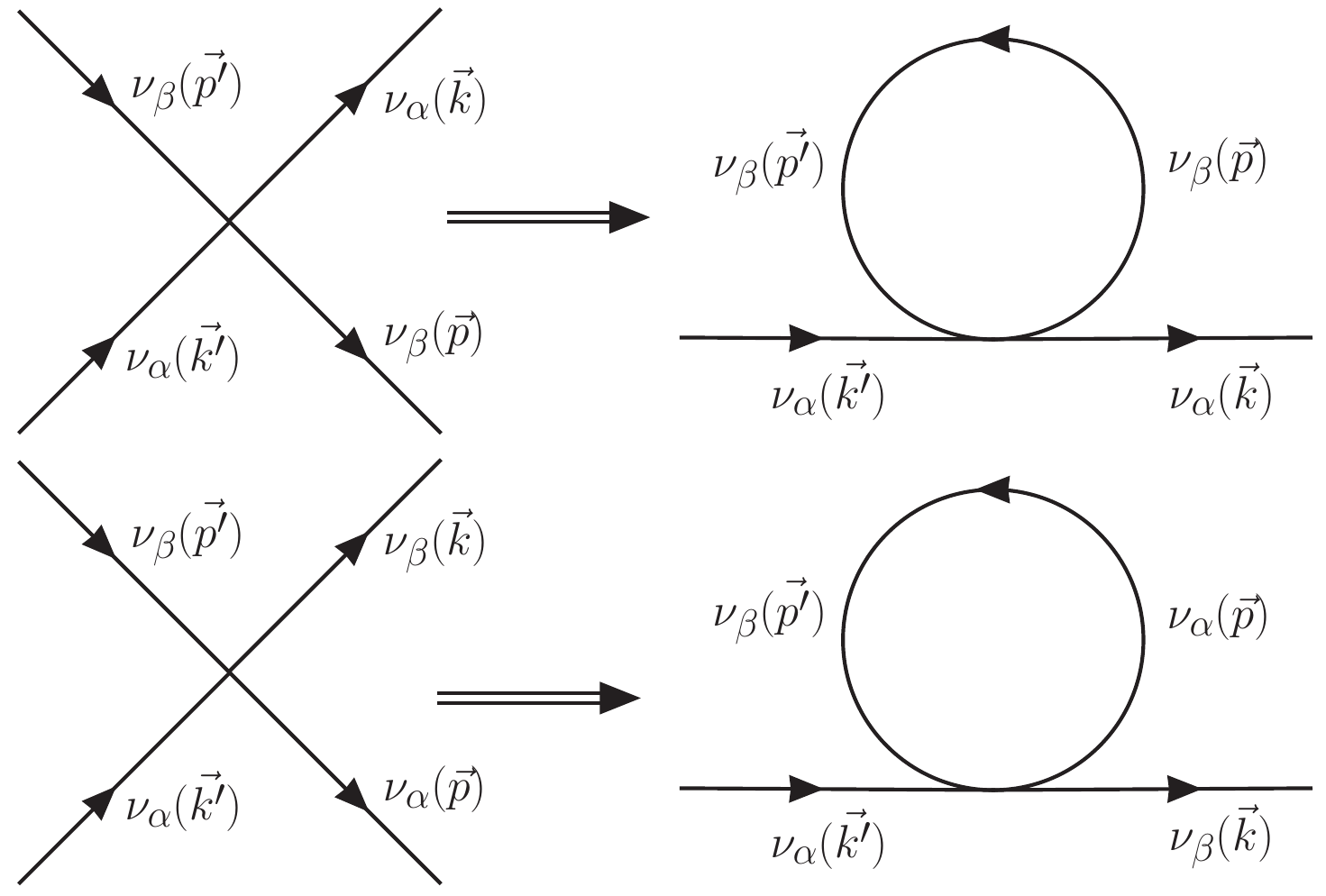}
 \caption{Mean-field approximation. Left side: low energy diagrams for $\nu$-electron (top diagram) and $\nu$-$\nu$ scattering (middle and bottom diagrams)
 that contribute to the neutrino evolution equations in dense media. Right side: tadpole diagrams corresponding to the left diagrams in the mean-field approximation with ${\vec p} = {\vec p}'$.
 For the $\nu\nu$ interaction, the mean-field contributions have a diagonal (middle diagram) and an off-diagonal term (bottom diagram); see the text. From \textcite{Volpe:2013uxl}.}
\label{Fig:MF}
 \end{center}
\end{figure}

Many have derived mean-field evolution equations including neutrino interactions with matter and neutrinos  
\cite{Samuel:1993uw,Sawyer:2005jk,Balantekin:2006tg,Volpe:2013uxl,Serreau:2014cfa}. 
To put it in words, the mean-field approximation consists in adding the amplitudes associated with neutrino scattering on a background particle, weighted by the quantum expectation value of the particle number operator over the background. Integrating such a quantity, over the degrees of freedom of the background particle, generates a potential that acts on the neutrino propagating through the medium; see Fig. \ref{Fig:MF}. Since only forward scattering is included, one commonly says that the mean-field approximation corresponds to {\it coherent forward scattering}.

We first consider neutrino-matter interactions whose contribution can be derived from the charged- and neutral-current interactions terms of the GWS model.
Associated with charged-current $\nu$-$e$ scattering is the well-known mean-field Hamiltonian\footnote{Note that Eqs.\eqref{eq:Hmsw} and \eqref{eq:HmswNC} 
are often referred to as potentials, and are denoted as $V_{\rm CC}$ and $V_{\rm NC}$ respectively.}
\beq\label{eq:Hmsw}
H_{\rm CC}^{\nu e} = \sqrt{2} G_{\rm F} (n_e - n_{\bar{e}}) \ , 
\eeq
where $G_{\rm F} $ is the Fermi coupling constant and $n_e$ ($n_{\bar{e}}$) is the electron (positron) number density. 
For antineutrinos, the rhs has a negative sign; see Section III.A.
Neutral-current interactions of $\nu_e$, $\nu_{\mu}$ and $\nu_{\tau}$ on neutrons give the following mean-field
contribution
\beq\label{eq:HmswNC}
H_{\rm NC}^{\nu n} = - \sqrt{2} G_F n_n / 2 \ , 
\eeq
which is equal for all neutrino flavors, with $n_n$ the neutron number density. The neutral-current contributions on electrons and protons cancel each other in a neutral medium.  

Putting these contributions together, one gets the Hamiltonian in flavor space that accounts for neutrino scattering with electrons, protons and neutrons in the background 
\beq\label{eq:Hmswfull}
{\cal H}_{\rm mat} =  \left(
\begin{tabular}{ccc}
$ \sqrt{2} G_{\rm F} (n_e - {n_n \over 2} )  $  &   0 & 0 \\
0   &   $- {G_F \over {\sqrt{2}}} n_n $ &  0 \\
0 &  0 & $-  {G_F \over {\sqrt{2}}} n_n $ \\
\end{tabular}
\right) \ ,   
\eeq
(positron number densities are usually small and thus are neglected here).
Since the neutral-current contributions on neutrons are proportional to the identity matrix, they do not influence neutrino oscillations and are usually not explicitly shown. 
Note that recent work in core-collapse supernovae (see \cite{Bollig:2017lki}) showed that there is a significant density of muons giving a supplementary contribution to the matter Hamiltonian. 

Equation \eqref{eq:Hmsw} holds for a homogenous, isotropic and unpolarized medium. This is a good approximation for the Sun. 
If the assumptions are relaxed, more mean-field terms appear due for example to polarization, as discussed by \textcite{Nunokawa:1997dp}.
In dense media like supernovae or compact object remnants interesting features arise due to anisotropy, as we later see. 

For $2\nu$ flavors,  the evolution equation for the neutrino amplitudes is
\beq\label{eq:2nu}
i {d \over{dt}}
\left(
\begin{array}{c}
\vert \nu_{\alpha} \rangle \\
\vert \nu_{\beta} \rangle
\end{array}
\right)
= {\cal H} \left(
\begin{array}{c}
\vert \nu_{\alpha} \rangle\\
\vert \nu_{\beta} \rangle
\end{array} 
\right) \, .   
\eeq

The neutrino Hamiltonian in Eq.\eqref{eq:2nu} including the vacuum and matter terms is
\beq\label{eq:Hvpm}
{\cal  H} = {\cal H}_c +  \Big(
\begin{tabular}{cc}
 - ${\Delta m^2 \over{4E}} \cos 2 \theta + \sqrt{2} G_{\rm F}  n_e $  &   ${\Delta m^2 \over{4E}} \sin 2 \theta$  \\
 ${\Delta m^2 \over{4E}} \sin 2 \theta$  & ${\Delta m^2 \over{4E}} \cos 2 \theta$ \\
\end{tabular}
\Big) \ , 
\eeq
where $\theta$ is the neutrino mixing angle and $\Delta m^2 = m_2^2 - m_1^2$ is the difference of the mass squared of the
mass eigenstates 2 and 1. The first term in Eq.\eqref{eq:Hvpm} is a term common to all flavors, which has been subtracted from the Hamiltonian in Eq.\eqref{eq:2nu}. It reads
\beq\label{eq:Hc}
{\cal H}_c = \left[ E + {(m_1^2 + m_2^2) \over 4E} \right] {\bf 1} \, ,
\eeq
where ${\bf 1} $ is the identity matrix. 
The second term in Eq.\eqref{eq:Hvpm} depends on the neutrino momentum through the vacuum oscillation frequency, i.e. $\omega = \Delta m^2 /2E$.

Remember that to investigate flavor evolution instead of neutrino flavor states one often evolves neutrino amplitudes, effective spins (Appendix A) or density matrices (see for example \cite{Giunti:2007ry}).  \textcite{Harris:1980zi} who first discussed the density matrix (and polarization vector) approach in relation to neutrinos while describing the coherence properties of a two state system undergoing random fluctuations in a medium. For $3 \nu$ flavors the neutrino density matrix reads 
\beq\label{eq:rhonu}
\varrho_{\bf p} =  \left(
\begin{tabular}{ccc}
$ \langle a^{\dagger}_{e} a_{e}  \rangle $  &  $ \langle a^{\dagger}_{\mu}  a_{e} \rangle $ & $ \langle a^{\dagger}_{\tau }  a_{e}  \rangle $ \\
$ \langle a^{\dagger}_{e}  a_{\mu}  \rangle $   &  $ \langle a^{\dagger}_{\mu} a_{\mu} \rangle $ &  $ \langle a^{\dagger}_{\tau} a_{\mu }  \rangle $ \\
$ \langle a^{\dagger}_{e} a_{\tau }  \rangle $   &  $ \langle a^{\dagger}_{\mu } a_{\tau}  \rangle $ &  $ \langle a^{\dagger}_{\tau }  a_{\tau }  \rangle $  \\
\end{tabular}
\right) \, ,   
\eeq
where the quantum expectation values\footnote{The operators $a^{\dagger}$ and $a$ are the particle creation and annihilation operators that satisfy the equal-time anticommutation rules $\{a({\bf p}, h), a^{\dagger}({\bf p}', h')\} =( 2 \pi)^3 \delta^3({\bf p} - {\bf p}') 2 E_{p} \delta_{h,h'} $ and  $\{a({\bf p}, h), a({\bf p}', h')\} = \{a^{\dagger}({\bf p}, h), a^{\dagger}({\bf p}', h')\} = 0$ ($h$ and  $h'$ are helicities). Similar rules hold for the antiparticle creation and annihilation operators, $b^{\dagger}$ and $b$.} are over the background that the neutrinos are traversing. Here $\varrho_{\bf p} = \varrho({\bf p}, t)$ and on the rhs we omit the explicit dependence on time and on neutrino quantum numbers that characterize the neutrino states, like momentum (or helicity), not to overburden the text. An expression similar to Eq.\eqref{eq:rhonu} holds for antineutrinos,  but with $\bar{\varrho}_{ij} = \langle b^{\dagger}_i b_j  \rangle$\footnote{Note that with this convention $\varrho$ and $\bar{\varrho}$ have the same equations formally; inversely to $\bar{\varrho}_{ij} = \langle b^{\dagger}_j b_i  \rangle$, which introduces complex conjugates of $\bar{\varrho}$ to the equations of motion; see  \cite{Sigl:1992fn,Volpe:2013uxl}.} instead of $\varrho_{ij} = \langle a^{\dagger}_j a_i  \rangle$. The diagonal entries of Eq.\eqref{eq:rhonu} are the quantum expectation values of the occupation number operator. Note also that the general form of Eq.\eqref{eq:rhonu}
involves bilinear products of the type $a^{\dagger}_i({\bf p}) a_j({\bf p}')$ or $b^{\dagger}_j({\bf p}) b_i({\bf p}')$; see Eq.\eqref{eq:rho1} and Sec. III.

Instead of evolving the neutrino states as in Eq.\eqref{e:Schro}, one can solve the Liouville Von-Neumann equation\footnote{From now on, a dot indicates a derivative with respect to time.}  for the neutrino or the antineutrino density matrix\footnote{Note that, in the equation of motion for antineutrinos the vacuum contribution to the Hamiltonian has a minus sign.}, i.e.
\beq\label{eq:LVN}
i {\dot \varrho_{\bf p}}  = [{\cal H}_{\bf p}, \varrho_{\bf p}] ~~~~~~~
i {\dot{\bar \varrho}_{\bf p}}  = [\bar{{\cal H}}_{\bf p}, \bar{\varrho}_{\bf p}] . 
\eeq

Besides the contributions from the neutrino mixings and the $\nu$-matter interactions, dense media have the peculiarity that neutral-current $\nu \nu$ interactions are sizable. In the mean-field approximation, the $\nu\nu$ Hamiltonian reads
\beq\label{eq:nunu}
{\cal H_{\nu \nu, {\bf p} }} = \sqrt{2} G_{\rm F}  \int_{p'} (1 - \hat{\bf p} \cdot \hat{\bf p}') \left[ \varrho_{{\bf p}'} - \bar{\varrho}_{{\bf p}'}  \right] \, ,   
\eeq
with
\begin{equation}\label{notations}
 \int_{p'} =  \int d {\bf p}' / (2 \pi)^3 \ ,  
 \eeq
In the angular term on the rhs of Eq.\eqref{eq:nunu}, $\hat{\bf p} = {\bf p}/ \vert {\bf p} \vert $ (one proceeds similarly with $\hat{\bf p}'$).
The term $ \hat{ \bf p} \cdot \hat{ \bf p}'$ originates from the $V$-$A$ structure of the weak interactions and contributes in anisotropic dense media,
where it plays a significant role. 

More generally, the mean-field equations of motion that describe neutrino propagation in dense environments (and similarly antineutrinos) read
\beq\label{eq:MF}
i (\partial_t + \mathbf{v} \cdot \mathbf{\nabla}_{\bf x} + \mathbf{F} \cdot \mathbf{\nabla}_{\bf p}) \varrho_{\bf x, p} = [{\cal H}_{\bf x, p}, \varrho_{\bf x, p}] \ , 
\eeq
with $ \mathbf{v} = {\bf p}/E$ the neutrino velocity. In the Liouville operator on the lhs of Eq.\eqref{eq:MF}, 
the second term is an advective term that contributes in the presence of spatial inhomogeneities. The third term in Eq.\eqref{eq:MF} depends on 
 a possible external force $\mathbf{F}$, such as the gravitational one, that acting on the neutrinos can change its momentum or energy (because of trajectory bending for example). 
Since the Liouville operator depends on time, position and momentum, describing neutrino evolution in dense media is a seven-dimensional problem, and therefore is extremely challenging numerically\footnote{Note that \textcite{Fiorillo:2024fnl} discussed that Eq.\eqref{eq:MF} does not conserve the  $\nu\nu$ refractive energy in the presence of inhomogeneities. They discussed the fact that supplementary gradient terms enhance energy conservation.}.

The solution of the neutrino mean-field equations reveals flavor mechanisms
that arise from the interplay among the vacuum, the matter, and the $\nu\nu$ contributions, as we now describe. 

\subsection{The MSW effect}
The MSW effect is a reference phenomenon for flavor evolution studies. Several of the uncovered mechanisms are either MSW-like or multiple MSW phenomena. 
To clarify this link, we review some of the basics.

The MSW effect arises when the following three conditions hold, namely the MSW resonance condition is fulfilled, the resonance width is large and evolution through it is adiabatic. It is equivalent to the two-level problem in quantum mechanics \cite{Cohen}.
 
We now introduce a new basis made of the eigenvectors for which the Hamiltonian describing neutrino propagation in an environment
is diagonal at each instant.  This basis is called the matter basis and the corresponding eigenvalues are called the matter eigenvalues.  
In this section the neutrino Hamiltonian comprises only 
the vacuum and the matter contributions. More generally a "matter" basis can be introduced regardless of which terms are included in the Hamiltonian. 

The flavor basis is related to the matter basis through the relation
\beq \label{eq:matterbasis}
\vert \nu_{\alpha} \rangle = \tilde{U}_{\alpha i}^* ~ \vert \tilde{ \nu}_{i} \rangle \, ,   
\eeq
with $i = 1, 2, 3, ..., N$ and $\alpha = = e, \mu, \tau, \dots $. In the unitary matrix $\tilde{U}$, effective mixing parameters in matter replace the vacuum ones.

From Eqs. \eqref{e:Schro} and \eqref{eq:matterbasis}, one gets the following equation of motion for the matter basis
\beq\label{eq:eommb}
i {d  \over dt} \vert \tilde{\nu} (t) \rangle = {\cal \tilde{H}}~ \vert \tilde{\nu}(t) \rangle =  \Big[{\cal {\tilde K}}+ i \tilde{U}^{\dagger} {d \tilde{U} \over {dt}} \Big] \vert \tilde{\nu}(t) \rangle \, ,
\eeq
where ${\cal {\tilde K}} = {\rm diag}(\tilde {k}_1, \tilde {k}_2, \dots \tilde {k}_N)$ depends on the matter eigenvalues $\tilde {k}_i$ ($i=1,N$) and the matter Hamiltonian
$\cal \tilde{H}$ now includes the derivatives of the effective mixing parameters in matter. These depend on the specific environment neutrinos are traversing.

We now consider the explicit expressions for 2$\nu$ flavors for which Eq. \eqref{eq:matterbasis} reads
 \beq\label{eq:matterbasis2}
\left(
\begin{array}{c}
\vert  \nu_e \rangle \\
\vert  \nu_x \rangle
\end{array}
\right)
=\Big(
\begin{tabular}{cc}
1  & 0  \\
0  & $ e^{i \tilde{\beta}} $ \\
\end{tabular}
\Big)
\Big(
\begin{tabular}{cc}
$\cos \tilde{\theta} $  & $\sin\tilde{ \theta} $  \\
$- \sin \tilde{\theta} $ & $ \cos \tilde{\theta} $ \\
\end{tabular}
\Big)
\left(
\begin{array}{c}
\vert  \tilde{\nu}_1 \rangle \\
\vert \tilde{\nu}_2 \rangle
\end{array} 
\right) \, ,   
\eeq
with $\tilde{\theta}$  and $\tilde{\beta}$ the effective mixing angle and phase respectively. 
Neglecting the phase, the evolution equation of the matter basis \eqref{eq:eommb} reads
\beq\label{eq:eqmatterbasis}
i {d \over{dt}}
\left(
\begin{array}{c}
\vert \tilde{\nu}_1 \rangle \\
\vert \tilde{\nu}_2 \rangle
\end{array}
\right)
=\Big(
\begin{tabular}{cc}
$\tilde {k}_1 $ &  $ i \dot{\tilde{\theta}} $  \\
   -$ i \dot{\tilde{\theta}} $   & $ \tilde {k}_2  $ \\
\end{tabular}
\Big)
\left(
\begin{array}{c}
\vert \tilde{\nu}_1\rangle\\
\vert \tilde{\nu}_2 \rangle
\end{array} 
\right) \, ,   
\eeq
where the difference between $\tilde {k}_1$ and  $\tilde {k}_2$ is given by
\beq\label{eq:eqmattervalues}
\tilde {k}_2 - \tilde {k}_1 = \sqrt{(\Delta m^2 \cos2\theta - A)^2 + (\Delta m^2 \sin 2 \theta)^2 }\, . 
\eeq       
In matter neutrinos acquire an effective mass.
Figure \ref{fig:MSW} shows how $\tilde {k}_1$ and $\tilde {k}_2$ evolve as a function of the electron number density in an environment.

The effective mixing angle diagonalizing the $2 \times 2$ matrix given in Eq.\eqref{eq:Hvpm} satisfies
\beq\label{eq:thetaeff}
\sin^2 2 \tilde{\theta} = {{(\Delta m^2 \sin 2 \theta)^2} \over {(\Delta m^2 \cos 2 \theta - A)^2 + (\Delta m^2 \sin 2 \theta)^2}} \, ,   
\eeq
with $A = 2 E H_{\rm CC}^{\nu e}$ Eq.\eqref{eq:Hmsw}. 

\begin{figure}
\begin{center}
\includegraphics[scale=0.5]{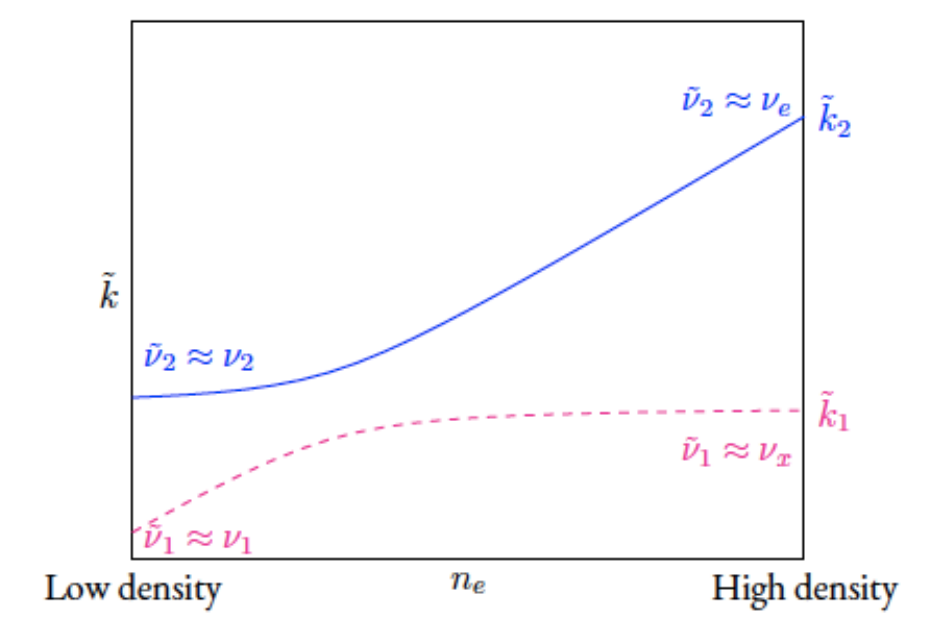}
\caption{The Mikheev-Smirnov-Wolfenstein effect: matter eigenvalues of the heavier (solid) and lighter (dashed) matter eigenstates, as a function of the electron number density $n_e$. The matter eigenvalues $\tilde{k}_2$ and $\tilde{k}_1$ evolve from the inner dense regions of a star (right) to the star surface (left), where $\tilde{\nu}_2$ and $\tilde{\nu}_1$ become the neutrino mass eigenstates in vacuum $\nu_2$ and $\nu_1$ respectively. The MSW resonance is at the location where  the matter eigenvalues are the closest and the matter mixing angle maximal. The MSW effect is a two-level problem in quantum mechanics. (Figure from \cite{Chatelain:2016th}).}
\label{fig:MSW}
\end{center}
\end{figure}

\noindent
One can see that, when\footnote{Note that the $n_{\bar{e}}$ contribution is neglected here.}
\beq\label{e:MSW}
\sqrt{2} G_{\rm F}  n_e = {\Delta m^2 \over {2E}} \cos2\theta \, ,
\eeq
holds the matter mixing angle is maximal, i.e. $\tilde{\theta} = \pi/4$, and the distance between the matter eigenvalues is minimal; see Fig. \ref{fig:MSW}. 

Equation \eqref{e:MSW} corresponds to the difference between the diagonal elements in the Hamiltonian \eqref{eq:Hvpm} being equal to zero (or to a minimal distance of the matter eigenvalues in the matter basis). It is the  MSW {\it resonance condition}. 
In the usually adopted convention, the fulfillment of \eqref{e:MSW} implies that, since  $\theta < \pi/4$ and $\cos2\theta > 0 $, $\Delta m^2 > 0$ to make
the rhs positive. Therefore the occurrence of the MSW effect gives the sign of the mass-squared difference. 

When the resonance condition is satisfied, and its width large, the fate of neutrinos 
depends on the adiabaticity of the neutrino evolution through the resonance. As in the two-level problem in quantum mechanics, adiabaticity through the resonance
can be quantified via the adiabaticity condition (see \cite{Giunti:2007ry})
\beq\label{eq:MSWadiabaticity}
\gamma^{-1} = { 2 \vert  \dot{\tilde{\theta}}   \vert \over {\tilde{k}_2 - \tilde{k}_1 }}  \ll 1 \, , 
\eeq
and
\beq\label{eq:adiacros}
\gamma_{\rm res} = {{\Delta m^2} \over {2 E}} {{\sin^2 2 \theta} \over {\cos2 \theta}}  {1 \over {\vert {1 \over {n_e}} {{dn_e}  \over {dr }} \vert_{r = r_{\rm res} } \vert } }  \, , 
\end{equation}
where $\gamma$ is the adiabaticity parameter and $\gamma_{\rm res}$ its value at the resonance.
Eq.\eqref{eq:MSWadiabaticity} corresponds to adiabatic evolution; whereas  
if $\gamma^{-1} \gg 1$ evolution is fully nonadiabatic.  
Indeed, it is the derivative of the matter mixing angle that governs the mixing between the matter eigenstates Eq.\eqref{eq:eqmatterbasis}, but how significant its impact is
on the neutrino evolution depends on the ratio of the off-diagonal terms of the Hamiltonian over the difference of the diagonal terms. 
If the former are much smaller than the latter, the adiabaticity condition \eqref{eq:MSWadiabaticity} holds and each mass eigenstate acquires a phase only during the evolution from the inner regions to the surface of the star. 

We now again consider the case of solar neutrinos. 
In the dense inner regions of the Sun, the matter mixing angle $\tilde{\theta}$ is close to $\pi/2$: the matter and flavor eigenstates practically coincide. 
If a neutrino, initially produced as $\nu_e$ coincides with $ \tilde{\nu}_2$ evolves through the MSW resonance adiabatically
it emerges as $\nu_2$ on the surface of the star (Fig. \ref{fig:MSW}). A significant fraction of the $\nu_e$ flux is then detected as $\nu_{x}$ on Earth.
Note that, if the vacuum mixing angle had been small (as historically believed), the MSW effect would have produced a noteworthy matter-enhanced conversion of $\nu_e$ into the other flavors. 

The survival probability of solar $\nu_e$, averaged over the production region, reads \cite{Parke:1986jy}
\beq\label{eq:Parke}
\langle P(\nu_e \rightarrow \nu_e) \rangle = {1 \over 2} + ( {1 \over 2} - P_{h}) \cos 2 \tilde{\theta}^{(i)} \cos 2 \theta \ , 
\eeq
where $ \tilde{\theta}^{(i)}$ is the mixing angle produced at high density. 
The quantity $P_{h}$ is the probability of $\tilde{\nu}_2$ to $\tilde{\nu}_1$ transition and is thus related to the mixings of the matter eigenstates at the MSW resonance. 
In the adiabatic case, the mixing is suppressed, $P_h = 0$. 
For the large mixing angle $\theta = 35^{\circ}$,  equation \eqref{eq:Parke} gives $ \langle P(\nu_e \rightarrow \nu_e) \rangle = \sin^2 \theta = 0.33$,  
as observed for solar $^8$B neutrinos. 

The observation of solar neutrinos in the SNO experiment, through charged-current and neutral-current interactions of neutrinos on deuterium and neutrino-electron elastic scattering, allowed the total solar neutrino flux to be measured and showed an unambiguous conversion of $\nu_e$ to the other active $\nu_{\mu}$ and $\nu_{\tau}$ flavors \cite{Ahmad:2002jz}.
Moreover KamLAND measurement of reactor $\bar{\nu}_e$ disappearance showed oscillations with 
$\Delta m^2 = 6.9 \times 10^{-5}$ eV$^2$ and $\sin 2 \theta = 1.0$ (best fit) thus identifying the MSW solution to the solar neutrino problem,
with the large mixing angle \cite{Eguchi:2002dm}.   

Thus, the MSW effect is responsible for a reduction of the $^8$B solar neutrino flux to 1/3 the value predicted by the standard solar model. 
Since the MSW resonance condition is satisfied for the high energy solar neutrinos, 
this gives us the sign of the corresponding mass-squared difference $\Delta m_{12}^2 > 0$.

We now return to the concept of adiabaticity more generally. The adiabaticity condition, defined as the ratio of twice the modulus of the off-diagonal term of the neutrino Hamiltonian over the difference of the diagonal terms, can be generalized in presence of $\nu\nu$ interactions as done by \textcite{Galais:2011jh}.
In this case the condition also involves the derivatives of the phase that arise because of the complex contribution, to the
neutrino Hamiltonian in the presence of to the neutrino-neutrino interaction \eqref{eq:nunu}.

When the matter density changes smoothly, neutrinos evolve through the resonance "adjusting" to the density variation:
evolution is adiabatic. This contrasts with what happens when steep variations of the density profiles are present, as  in presence of shock waves.

\begin{figure}
\begin{center}
\includegraphics[scale=0.4,width=.4\textwidth]{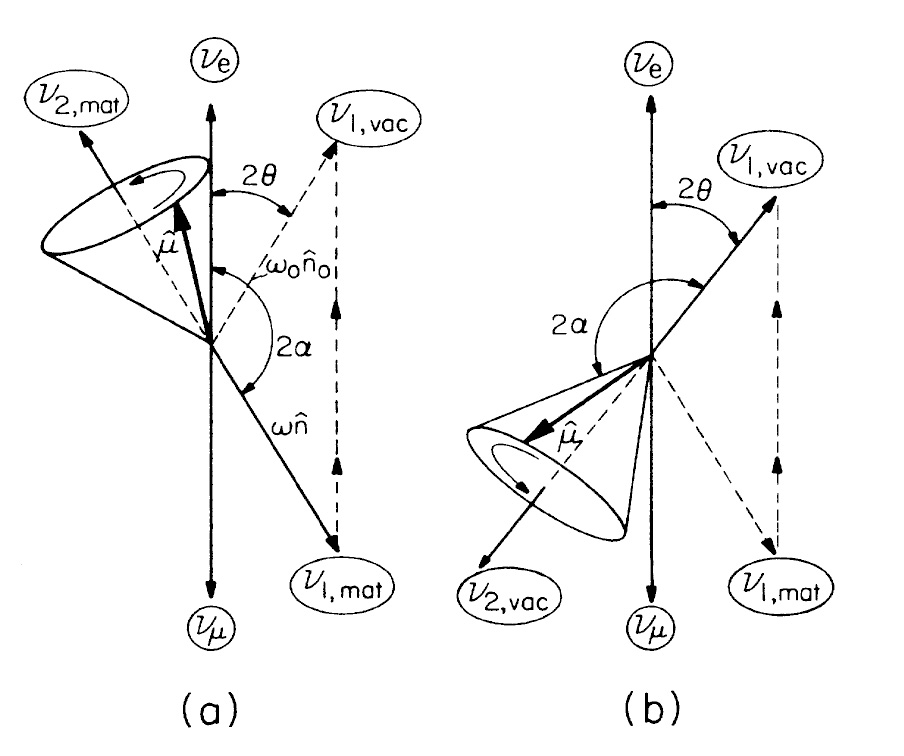}
\caption{Spin formalism:  picture of the neutrino evolution in flavor space for $2 \nu$ flavors. The third component of the neutrino spin vector corresponds to either $\nu_e$ (upward) or
$\nu_{\mu}$ (downward). When electron neutrinos are produced in the Sun, the vector ${\bf \mu}$ (${\bf P}$ in the text) precess around 
the matter vector ${\bf n}$ ((a), ${\bf B_{\rm mat}}$ in the text) and, if evolution is adiabatic, follows ${\bf n}$ until it becomes the vacuum vector ${\bf n}_0$ ((b), ${\bf B_{\rm vac}}$ in the text). The latter is tilted by $2 \theta$ with respect to the $z$ axis. From \textcite{Kim:1987bv}.}
\label{fig:spins}
\end{center}
\end{figure}

With the spin formalism, we look at flavor phenomena with different eyes, since we follow neutrinos through the evolution of an effective spin ${\rm {\bf P}}$\footnote{These are also called neutrino isospins or polarization vectors.}, which is subject to effective magnetic fields; see Appendix A and \cite{Cohen}. In this context, vacuum oscillation is a precession of neutrino spins in flavor space, around the vacuum effective magnetic field ${\rm {\bf B}}_{\rm vac}$ tilted by $2 \theta$ \cite{Kim:1987ss,Kim:1987bv}; see Fig. \ref{fig:spins}. One can show that, from the evolution of the third component $P_z$, one recovers the vacuum oscillation formula. 

As for the MSW effect, it takes place in matter when ${\rm {\bf P}}$ goes through the $x$-$y$ plane, since the MSW resonance condition \eqref{e:MSW} corresponds to $P_z = 0$. Adiabatic evolution occurs when the precession frequency of ${\rm {\bf P}}$ around  ${\rm {\bf B}} = {\rm {\bf B}}_{\rm vac} + {\rm {\bf B}}_{\rm mat}$ is fast compared to the rate at which ${\rm {\bf B}}$ changes, so neutrino spins follow the magnetic field during propagation. On the contrary, if evolution is nonadiabatic, ${\rm {\bf P}}$ "lags behind". 
The theoretical description in terms of neutrino isospins has largely been exploited to study neutrino flavor evolution in dense environments and in particular when the $\nu\nu$ interactions are sizable; see the review by \textcite{Duan:2010bg}. 
 
\subsection{The MSW effect in dense media}
Neutrinos face more than one resonance if the density is large as in astrophysical environments or the early Universe. 
Although here we mostly refer to supernova neutrinos as an example, the aspects that we cover can be transposed to other
dense environments, such as binary neutron star mergers.   
In our discussion, we assume that, unless otherwise stated, the $\nu_{\mu}, \nu_{\tau}$ and $\bar{\nu}_{\mu}, \bar{\nu}_{\tau}$ fluxes (referred to as $\nu_x$ and $\bar{\nu}_x$ respectively) are equal, as in nearly all of the available literature.
 
\textcite{Dighe:1999bi} pointed out that there are three MSW resonances in supernovae: 
the high ($H$-), the low ($L$-) and $V_{\mu\tau}$. 
As the MSW resonance condition shows, the resonance location depends on the neutrino energy
and mixing parameters. From Eq. \eqref{e:MSW} one finds that the $H$- and $L$-resonances, associated with $(\theta_{31}, \Delta m^2_{31})$ and $(\theta_{21}, \Delta m^2_{21})$ respectively, take place at 
\beq\label{e:HL-res}
\rho_{\rm res} \approx 1.4 \times 10^6~ {\rm g\over{cm^{3}}} \Big({{\rm \Delta m^2} \over 1 ~{\rm eV}^2}\Big) \Big({15~ {\rm MeV} \over E }\Big) \Big({0.5 \over Y_e }\Big) \cos 2 \theta \, .
 \eeq
 where the electron fraction $Y_e$ is given by
\beq\label{eq:Ye}
Y_e = {{n_e - n_{\bar e}} \over{n_B}} \ , 
\eeq
That is, it is the net number of electrons (the difference between the electron and positron number densities) per baryon $n_B = n_p + n_n$ ($n_p$ is
the proton number density). Therefore, for a 40 MeV neutrino, one finds that $ \rho_{res} \approx 1.2 \times 10^{3}$ g/cm$^{3}$ for the H--resonance and $ \rho_{res} \approx 13 $ g/cm$^{3}$ for the L-resonance ($Y_e=0.5$).    
\begin{figure}
\begin{center}
\includegraphics[scale=0.15]{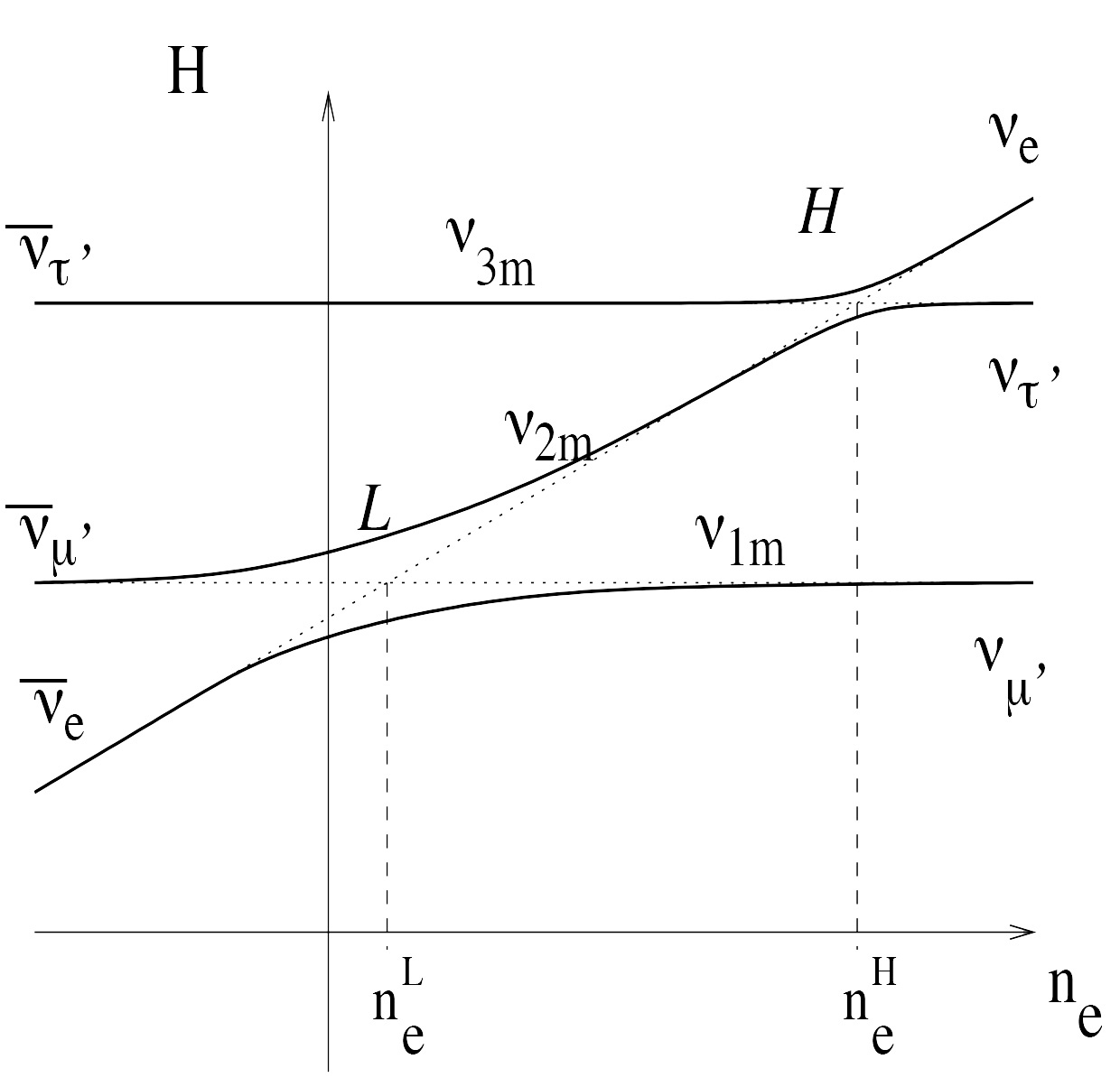}
\caption{MSW effect in core-collapse supernovae: effective neutrino masses as a function of the electron number density in the absence (dashed) or in the presence (solid lines) of neutrino mixings. The high ($H$-) and low ($L$-) MSW resonances are indicated, whereas the $V_{\mu \tau}$ resonance (at higher density) is not shown. Note that the antineutrino case is represented in the part of the plot with negative electron number density. The diagram corresponds to normal mass ordering. Adapted from \cite{Dighe:1999bi}.}
\label{fig:MSWSN}
\end{center}
\end{figure}
Moreover, radiative corrections that differentiate $\nu_{\mu}$ from $\nu_{\tau}$ \cite{Botella:1986wy}, introduce the potential $V_{\mu\tau} = 10^{-5}V_{\rm CC}$ where $V_{\rm CC}$ 
is as given in Eq.\eqref{eq:Hmsw}. The quantity $V_{\mu\tau}$
becomes relevant if the medium is very dense (i.e. for densities larger than $10^{7}$-$10^{8}$ g/cm$^{3}$). 
The $V_{\mu\tau} $ resonance is associated with the atmospheric mixing parameters $(\theta_{32}, \Delta m^2_{32})$.
Figure \ref{fig:MSWSN} shows the eigenvalues of the neutrino Hamiltonian given by Eq.\eqref{eq:Hmswfull} and including the $V_{\mu\tau}$ potential in the ($\nu_e, \nu_{\mu}', \nu_{\tau}'$) basis which is a rotated ($\nu_e, \nu_{\mu}, \nu_{\tau}$) basis where the ($ \nu_{\mu}, \nu_{\tau}$) submatrix of the neutrino Hamiltonian is diagonal. 

As is well known, the supernova neutrino time signal comprises three characteristic phases of the supernova dynamics and explosion called the neutronization burst, the accretion phase and the cooling phase of the newly born protoneutron star; see Fig.\ref{fig:3DSN}, in section IV.B and the reviews by \textcite{Mirizzi:2015eza} and \textcite{Vitagliano:2019yzm}). The neutrino spectra depend on these phases, the supernova progenitor and on the direction of observation, in particular because of the SASI and LESA instabilities.  For our purposes, the explicit time dependence of the neutrino signals is not necessary.

A characteristic feature of flavor phenomena is that they induce spectral modifications. We now consider the neutrino spectra at the neutrinosphere, $\phi^0_{\bar{\nu}_e}$ and $\phi^0_{\bar{\nu}_x}$ (thought of as a function of time, or as fluences, i.e. time-integrated fluxes) and assume $\nu$ evolution through $H$- and $L$-resonances only in terms of probabilities\footnote{This simple approach assumes that the evolution at each resonance is factorizable and  neglects the role of phases both from the neutrino amplitudes and from the neutrino mixing matrix $U$.}. For example, if $\bar{\nu}_e$, produced in the inner stellar regions, traverse the three resonances, their spectra become
\beq\label{e:MSWswap}
\phi_{\bar{\nu}_e} = \bar{p} \phi^0_{\bar{\nu}_e} + (1 - \bar{p}) \phi^0_{\bar{\nu}_x} \, , 
\eeq
where $\bar{p}$ is the spectral swapping probability. In particular, $\bar{p} = 0.68$ and $\bar{p} = 0$ for normal and inverted mass ordering respectively\footnote{Note that this includes both the effect due to the MSW mechanism and the projection on the flavor basis due to the PMNS matrix; see \cite{Dighe:1999bi}.}.   

As supernova simulations show, the neutrino spectra at the neutrinospheres, $\phi^0_{\bar{\nu}_e}$ and $\phi^0_{\bar{\nu}_x}$, are well described by pinched Fermi-Dirac distributions \cite{Dighe:1999bi} or by power laws \cite{Keil:2002in}.  Because of their microscopic interactions, the neutrino average energies often satisfy the (approximate) hierarchy $ \langle E_{\nu_e} \rangle  < \langle E_{\bar{\nu}_e} \rangle < \langle E_{\nu_x} \rangle$, with typical energies $E_{\nu_e} \in [8, 14]$ MeV, $E_{\bar{\nu}_e} \in [14, 18]$ MeV and $E_{\nu_x}  \in [16, 20] $ MeV. 
In fact, $\nu_x$ undergo neutral-current interactions and decouple from deeper hotter regions. Unlike $\nu_x$, $\nu_e$ and $\bar{\nu}_e$ interact via charged- and neutral-current interactions and decouple from colder outer shells. 

In Eq.\eqref{e:MSWswap}, one sees that, in a case of inverted mass ordering, due to the MSW effect, $\bar{\nu}_e$ can acquire the hotter spectra of the nonelectron flavor neutrinos (if the latter have a higher average energy at the neutrinosphere). A similar mechanism occurs in the $\nu_e$ spectra
 if the mass ordering is normal. 
If a supernova blows off, such spectral modifications will impact charged-current events associated with inverse-$\beta$ decay or neutrino-nucleus interactions, in a scintillator, Cherenkov, lead, or liquid argon supernova neutrino detector. On the contrary, neutral-current events are "flavor blind" and therefore not sensitive to spectral swapping.

Keep in mind that if, under some conditions, the supernova fluxes of the different neutrino flavors become practically degenerate, then 
spectral distortions due to flavor mechanisms, according to \eqref{e:MSWswap}, impact observations neither directly nor indirectly. For this reason, the possibility of flavor equilibration is often discussed in the literature and theorists have been actively looking for this simplifying possibility. 

We now add a few more remarks. First, in core-collapse supernovae, for typical matter profiles (in the absence of shock waves), the evolution through the $L$-resonance is adiabatic. Second, 
as typically assumed in the literature, if $\phi^0_{\bar{\nu}_{\tau}} = \phi^0_{\bar{\nu}_{\mu}}$ at tree level, the $V_{\mu\tau}$ resonance, which mixes $\nu_{\mu}$ and $\nu_{\tau}$, does not produce spectral modifications or therefore observable effects. Note however that, if one includes muons and six-species neutrino transport in supernova simulations  \cite{Bollig:2017lki}, $\nu_{\mu}$ and $\nu_{\tau}$ fluxes can differ at tree level. 
The study of the impact of the six-species neutrino transport on neutrino flavor evolution is ongoing. Third, the only unknown parameter that impacts the standard MSW effect is the neutrino mass ordering, since the sign of $\Delta m^2_{23} \approx \Delta m^2_{13}$ has not yet been determined. Therefore, the detection of the neutrino signal from a future galactic supernova could inform us about this key property\footnote{Similarly, numerous studies investigated ways to identify $\theta_{13}$ with a supernova neutrino signal, until it was measured by the Daya Bay \cite{DayaBay:2012fng}, RENO \cite{RENO:2012mkc} and Double Chooz experiments \cite{DoubleChooz:2011ymz}.}, as we later  see.
We now discuss multiple MSW resonances and MSW-like mechanisms that arise in dense astrophysical environments such as an exploding supernova.  

\subsection{Shock wave effects}
Shock-related structures in supernova neutrino observations could inform us on shock reheating and propagation, a unique observation of the explosion mechanism on its becoming. 
The availability of large-scale observatories and a close supernova would offer the possibility to observe such structures and other deviations from the expected exponential cooling of the newly formed neutron star. 
Even if there are variations among models, some features appear to be sufficiently generic to deserve investigation. 

\begin{figure}
\begin{center}
\includegraphics[scale=0.4]{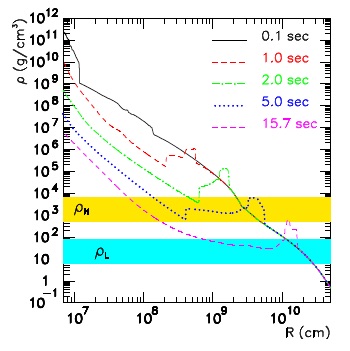}
\caption{Shock waves: matter density profiles,  at different postbounce times, as a function of distance in an exploding core-collapse supernova model. The front and reverse shocks are visible. The yellow and blue bands correspond to the densities and neutrino energies that fulfill the $H$- and $L$- MSW resonance conditions. From \cite{Tomas:2004gr}.}
\label{fig:FR-shocks}
\end{center}
\end{figure}

In an exploding supernova, shock waves constitute a major perturbation of the electron fraction $Y_e $ and of the presupernova matter density profiles.
The shock wave reaches the $H$-resonance region about 2 s after core bounce. \textcite{Tomas:2004gr} showed the presence of both a front and a reverse shock, due to the earlier slower ejecta meeting a hot supersonically expanding neutrino-driven wind; see Fig. \ref{fig:FR-shocks}. 
 
\textcite{Schirato:2002tg} first pointed out that shock waves could "shut off" flavor evolution when passing through an MSW resonance region. 
Because of the steepness of the density profile, $\nu_{\mu,\tau}\leftrightharpoons \nu_{\rm e}$ would be suppressed due to nonadiabatic evolution. Hence the $ \nu_{\rm e}$ and $\bar{\nu}_e$ fluxes would become colder thus producing dips in the supernova neutrino time signal. 

The passage of shock waves in MSW resonance regions engenders two effects: it makes the resonance temporarily nonadiabatic and induces multiple MSW resonances. The evolution through the different resonance locations can be treated as incoherent or coherent. In the former case, the MSW resonances are independent, while in the latter coherent evolution produces interference effects among the matter eigenstates called phase effects. 
Since the MSW resonance condition is necessary for shock wave effects, they occur either in the $\nu_e$ signal (for normal mass ordering), or in the $\bar{\nu}_e$ signal (for inverted mass ordering).

The change in the adiabaticity at the MSW resonance locations impacts the evolution in the $H$-resonance region\footnote{Note that the shock wave also influences the neutrino evolution through the $L$-resonance region. However its impact at low energies and at late times is negligible.} and modifies the neutrino average energies, thereby creating dips or bumps in the supernova time signal and the corresponding rates. These features were investigated in a series of works \cite{Takahashi:2002yj,Fogli:2003dw,Lunardini:2003eh,Fogli:2004ff,Kneller:2007kg}, see the review by \textcite{Duan:2009cd}). 

\textcite{Fogli:2003dw} pointed out that multiple resonances could produce phase effects that would average out for large values of $\theta_{13}$. \textcite{Dasgupta:2005wn} investigated them in detail. Phase effects require semiadiabatic and coherent evolution at the resonances\footnote{In a wave-packet description in flat spacetime, decoherence arises at distances larger than the coherence length. For a typical wave-packet width at production, i.e. $\sigma \approx 10^{-11}$-$10^{-12}$ cm and $E \in [5, 80]$ MeV (the average energy between two matter eigenstates) one gets $L_{coh} \approx 10^4 $ km. }.  They are difficult to see because, even when the coherence condition is met, the associated oscillations are smeared by the energy resolution of the detectors.

We now consider the presence of a dip in a supernova density profile as an example. A neutrino of energy $E$ encounters two resonances at locations $x_1$ and $x_2$.
If $\vert \nu_h \rangle$ and $\vert \nu_l \rangle$ are the heavier and lighter matter eigenstates respectively, at $x < x_1$, one has $\vert \nu_h \rangle (x \ll x_1) \approx \vert \nu_e \rangle$.
While evolution before the resonance is adiabatic, the resonance mixes the matter eigenstates just before the crossing $x < x_{1-}$ yielding new matter eigenstates 
\beq\label{e:mattereigen}
\left(
\begin{array}{c}
\vert \nu_h(x_{1+})   \rangle \\
\vert  \nu_l(x_{1+}) \rangle
\end{array}
\right)
=\Big(
\begin{tabular}{cc}
$\cos \chi_1 $  & $\sin\chi_1e^{i \phi}$  \\
$- \sin \chi_1e^{-i \phi} $ & $ \cos\chi_1$ \\
\end{tabular}
\Big)
\left(
\begin{array}{c}
\vert \nu_h(x_{1-})   \rangle \\
\vert  \nu_l(x_{1-}) \rangle
\end{array} 
\right) \, ,   
\eeq
where $P_i = \sin^2 \chi_1$ is the hopping probability for an isolated resonance. The matter eigenstates acquire a relative phase up to the second resonance 
at $x_2$. 
After the latter the $\nu_e$ survival probability is (far from $x_2$) 
\begin{align}\label{e:MSWmultiple}
P({\nu_e \rightarrow \nu_e}) & = \cos^2(\chi_1 - \chi_2) \\ \nonumber
& - \sin 2 \chi_1 \sin 2 \chi_2  \sin^2 \Big(\int_{x_1}^{x_2} {{\Delta \tilde{m}^2} \over {4E}}dx  \Big) \, . 
\end{align}
The last term in Eq.\eqref{e:MSWmultiple}, owing to the interference between the matter eigenstates, oscillates with the neutrino energy and the resonance locations. 
It produces fast oscillations (the phase effects) as a function of energy or, for a given energy, as a function of distance or time because the shock wave propagation slightly shifts such locations. 
In the absence of coherence, the interference term averages out and the two resonances at $x_1$ and $x_2$ are independent.
 
Two studies implemented shock wave effects and $\nu\nu$ interactions in the bulb model.
Using a consistent treatment that retains phase information, \textcite{Gava:2009pj} 
showed that, depending on the neutrino energy, dips or bumps may be present in the positron time signal, due to inverse-beta decay (i.e. $\bar{\nu}_e + p  \rightarrow e^+ + n$), 
of scintillators or Cherenkov detectors (inverted mass ordering). Similar features may be present, with normal mass ordering, in the electron time signal of an argon-based detector such as DUNE, due to charged-current $\nu$-$^{40}$Ar interactions.
 In contrast, an analysis of the time signal for the lead detector HALO-2 by \textcite{Ekinci:2021miy}
showed changes  at the level of a few percent for the one-neutron and two-neutron emission rates in neutrino-lead interactions, that is, too small to be seen.

Besides shock waves, turbulence can play a significant role in supernova explosions; see \textcite{Radice:2017kmj}. The influence of turbulence on neutrino flavor content has features in common with shock wave effects, as we now discuss.

\subsection{Turbulence effects}
Noisy media, such as those originating from helioseismic g modes or temperature fluctuations, influence neutrino flavor evolution, as pointed out in relation to the solar neutrino problem; see \textcite{Sawyer:1990tw,Nunokawa:1996qu,Balantekin:1996pp}.  In particular, \textcite{Loreti:1994ry} showed that randomly fluctuating matter density and magnetic fields tend to depolarize neutrinos, i.e. the survival probability averages to one-half. Neutrino propagation in stochastic media was also discussed by \textcite{Torrente-Lujan:1998ddy,Burgess:1996mz}. 

Solar neutrino and KamLAND data constrain matter density fluctuations in our Sun at a few percent level. This result holds for delta-correlated (white) noise, and correlation lengths of 10-100 km (see \textcite{Balantekin:2003qm,Guzzo:2003xk}). Hence, one can extract the solar oscillation parameters independently of the fluctuations \cite{Burgess:2003su}.

Simulations of exploding core-collapse supernovae show that nonradial turbulent flows associated with convection and SASI have explosion supportive effects \cite{Mezzacappa:2015kya,Janka:2012wk,Couch:2014kza,Janka:2017vcp,Radice:2017kmj,Foglizzo:2015dma}. Hydrodynamic instabilities generate large-scale anisotropies between the protoneutron star and the supernova envelope. Therefore, supernova neutrinos reaching Earth  "see" stochastic matter density profiles. 

Noisy media might influence the supernova neutrino flavor content significantly.
First investigations evolved fluctuation-averaged density matrices, or probabilities\footnote{This gives a generalization of Parke's formula Eq.\eqref{eq:Parke} with a damping factor \cite{Burgess:1996mz}.}, with delta-correlated fluctuations and static \cite{Loreti:1995ae} or dynamic density profiles with front and reverse shocks \cite{Fogli:2006xy}. \textcite{Friedland:2006ta} argued for Kolmogorov-correlated fluctuations. 

\textcite{Kneller:2010sc} evolved neutrino amplitudes and built a statistical ensemble of instantiations for the neutrino survival probabilities using one-dimensional simulations with Kolmogorov fluctuations added. Retaining the phase information, the approach revealed the presence of multiple MSW resonances from turbulence and a transition, when the fluctuation amplitude was increased, from phase effects due to shock waves to a fluctuations-dominated regime. 
\textcite{Lund:2013uta} investigated the interplay among neutrino-neutrino interactions, shock waves, and turbulence using  one-dimensional dynamical simulations for three progenitors. 
These studies showed that large amplitude fluctuations resulted in depolarization of the neutrino probabilities \cite{Loreti:1995ae,Fogli:2006xy,Friedland:2006ta,Kneller:2010sc}.

\textcite{Borriello:2013tha} came to different conclusions. They performed the first investigation exploiting fluctuations from high resolution two-dimensional supernova simulations down to scales smaller than typical matter oscillation lengths\footnote{Note that small-scale fluctuations (i.e. less than 10 km) have smaller scales than what can be numerically resolved.}. 
These fluctuations followed broken power laws (with exponents 5/3 and 3) that were in agreement with two-dimensional Kolgomorov-Kraichnan theory of turbulence. Their analysis showed small damping of the neutrino probabilities due to matter fluctuations and the absence of strong or full depolarization\footnote{Three-dimensional simulations should bring turbulence spectra with a Kolmogorov exponent of 5/3 at all scales. Indeed Kolmogorov scaling seems to be recovered in 3D simulations; for a discussion of this aspect, see \textcite{Radice:2017kmj}).}. 
Further work is needed to determine the impact of turbulence on flavor evolution and to assess whether or not matter fluctuations introduce a loss of memory effects.

\subsection{MSW-like mechanisms}
The MSW effect arises from the cancellation of the vacuum and matter contributions.
New resonance conditions emerge from the interplay of the different terms of the neutrino Hamiltonian \eqref{e:H} describing neutrino propagation in a dense medium.  
Thus, various types of MSW-like phenomena have been uncovered, in particular the matter-neutrino resonance, helicity coherence and the $I$-resonance which we now describe.

\begin{figure}
\begin{center}
\includegraphics[width=.5\textwidth]{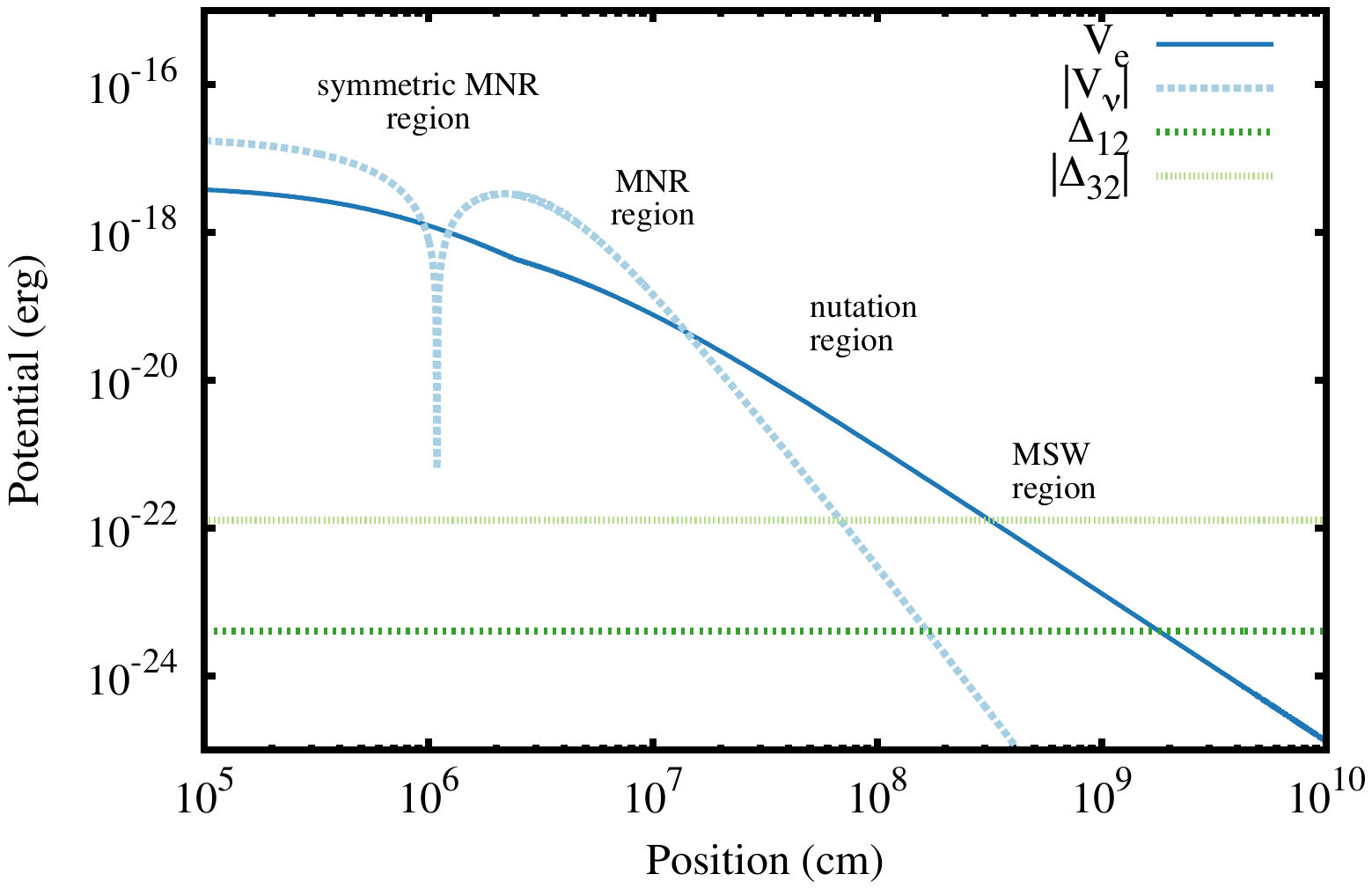}
\caption{Matter-neutrino resonances: potentials of the neutrino Hamiltonian, as a function of distance, in an accretion disk model of a black hole-neutron star merger. Regions are indicated where different neutrino flavor mechanisms take place. The $\nu\nu$ interaction and the matter terms ($V_e = H_{\rm CC}^{\nu e}$) cross at the locations of the symmetric and standard MNRs. Note that $\vert V_{\nu} \vert $ indicates the only nonzero term of the $\nu\nu$ Hamiltonian, in a disk geometry, at initial time. The nutation region is a region where oscillations occur due to a cancellation between the $\nu\nu$ interactions and the vacuum ($\Delta_{32}$) terms. The MSW region is also shown. 
Note that $\Delta_{12} = \Delta m^{2}_{12}$ and $\vert \Delta_{32} \vert = \vert \Delta m^2_{32} \vert $. Adapted from \cite{Malkus:2015mda}.}
\label{fig:MNR}
\end{center}
\end{figure}

\subsubsection{Matter-neutrino resonance}
Accretion disks around compact objects (binary neutron star merger remnants or black holes) \footnote{From collapsing stars or black hole-neutron star binaries} -- produce large amounts of neutrinos with luminosities and average energies similar to those of core-collapse supernovae. 
An important difference is that, in these environments, matter is neutron-rich which produces  an excess of the $\bar{\nu}_e$ flux over the $\nu_e$ one.
Computationally, even the simplest models require spherical symmetry breaking which is numerically more involved. 
Note that, in the context of core-collapse supernovae, spherical symmetry was assumed in numerous studies that yielded interesting results.

\begin{figure}
\begin{center}
\includegraphics[scale=0.45]{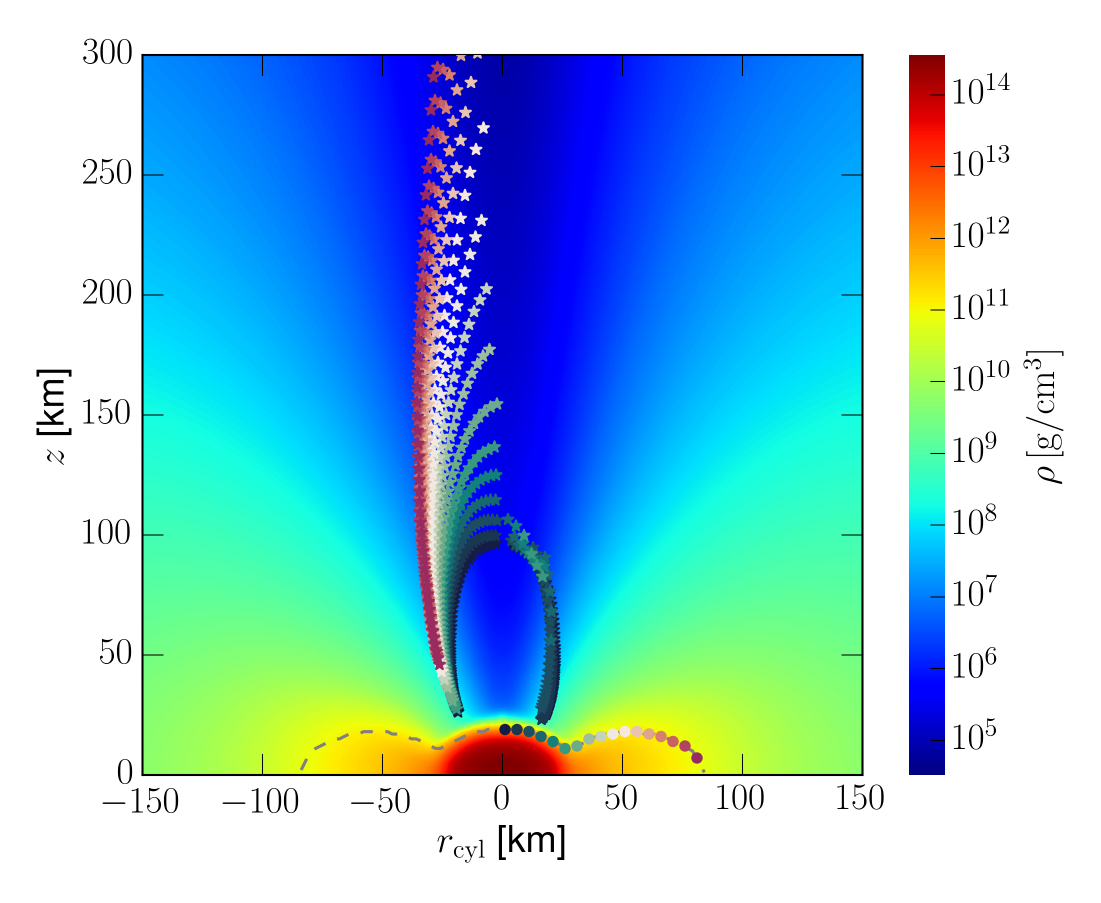}
\caption{Matter-neutrino resonances. The central object at the bottom is a binary neutron star merger remnant, with its accretion disk (dashed line). The stars indicate the locations where the MNR resonance condition is fulfilled along different trajectories above the remnant. From \cite{Frensel:2016fge}.}
\label{fig:MNRlocations}
\end{center}
\end{figure}

\begin{figure}
\begin{center}
\includegraphics[width=.5\textwidth]{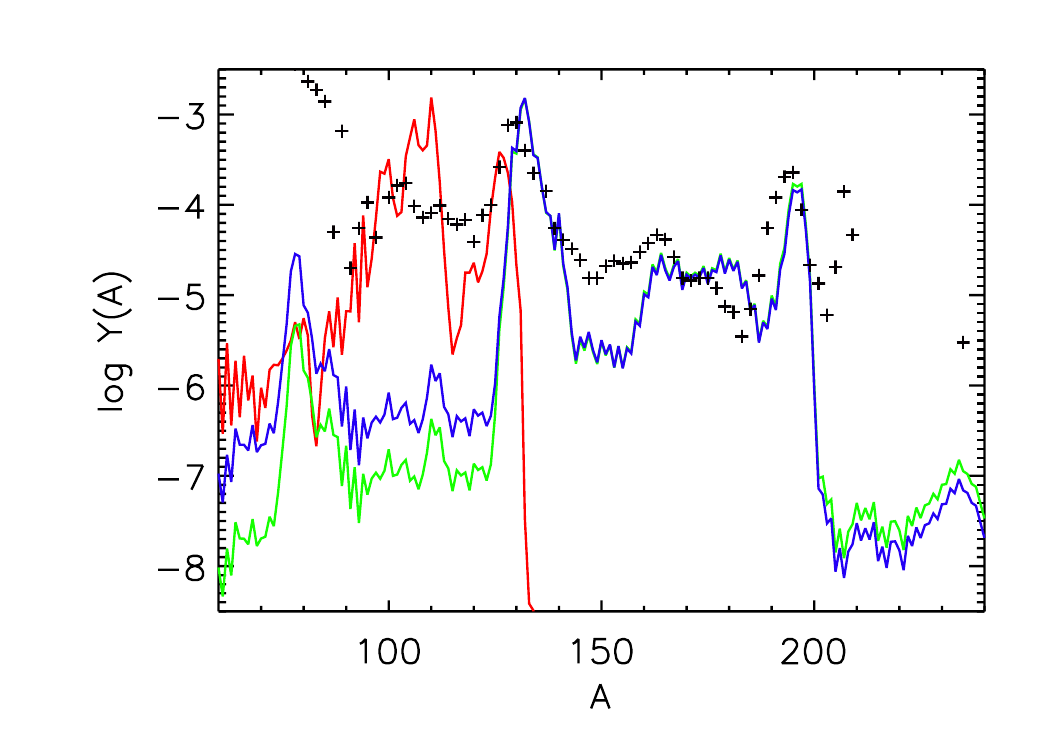}
\caption{Nucleosynthetic abundances. The crosses show the scaled solar abundances. Visible are the second peak (around A=130-138) and the third peak (around A=195-208) as well as the rare element plateau in between. The observations are compared to predictions for nucleosynthetic abundances in a black hole-accretion disk scenario. The line that ends at A $\approx$ 130 corresponds to results in the absence of neutrino oscillations (red line). The results with oscillations (blue line) give larger (smaller) abundances below A = 125 (A $>$ 200)
 than the results without the $\nu\nu$ interaction (green line). The results are for normal neutrino mass ordering.  From \cite{Malkus:2012ts}.}
\label{fig:BNSr}
\end{center}
\end{figure}

In a collapsar-type disk\footnote{A collapsar is a massive star that collapses to black holes forming a disk due to its large angular momentum. Collapsars can produce gamma-ray bursts; see \cite{Heger:2002by}.}, \textcite{Malkus:2012ts} found a novel conversion mechanism called the matter-neutrino resonance (MNR). The MNR arises in regions above the disk when the $\nu$-matter and the $\nu\nu$ interactions cancel each other; see Figs. \ref{fig:MNR} and \ref{fig:MNRlocations}. Indeed, the excess of the $\bar{\nu}_e$ flux over the $\nu_e$ one gives a different sign to the two contributions giving the possibility of a cancellation. Moreover, because of the geometry of the disks and the  $\bar{\nu}_e$s decoupling deeper than $\nu_e$, the sign of the only nonzero element of the $\nu\nu$ Hamiltonian at initial time (i.e. the $ee^{th}$) can flip at some location. If the flip in sign is not present the phenomenon is called standard MNR \cite{Malkus:2014iqa}; whereas if it is present, the process is called the symmetric MNR \cite{Malkus:2012ts,Malkus:2015mda}; see Fig. \ref{fig:MNR}. Adiabatic evolution through the MNRs produces efficient conversions of ${\nu}_e$ into ${\nu}_{\mu}$ and ${\nu}_{\tau}$ for the former and  $\nu_e$ and $\bar{\nu}_e$ for the latter. This can influence the electron fraction $Y_e$ and favor disk wind nucleosynthesis of $r$-process elements; see Fig. \ref{fig:BNSr}. 

 \textcite{Zhu:2016mwa} and \textcite{Frensel:2016fge} showed that patterns  of flavor evolution depend on the neutrino path\footnote{The first studies fixed the azimuthal angle  $\theta$ to $45^{\circ}$.}. Both studies were based on astrophysical inputs from the detailed two-dimensional simulations of a binary neutron star merger remnant by \textcite{Perego:2014fma}.  When considering different initial conditions and azimuthal angles, \textcite{Frensel:2016fge} found that the neutrino capture rates on neutrons and protons showed variations by tens of percent due to flavor mechanisms.

In these studies the $\nu\nu$ Hamiltonian \eqref{eq:nunu} is treated by taking the flavor history of one neutrino as representative of all trajectories\footnote{This is equivalent to the treatment of $\nu\nu$ interactions in the single-angle approximation in the supernova context (bulb model); see Sec. II.G.}. A consistent treatment of the neutrino-neutrino interaction term should also implement the neutrino evolution along different paths. \textcite{Vlasenko:2018irq} showed that, even when a more consistent treatment is used, the MNRs take place leading to significant neutrino conversion. 

What is the underlying mechanism of matter-neutrino resonances?
\textcite{Wu:2015fga}, with a schematic model, and \textcite{Chatelain:2016xva}, with detailed BNS simulations, showed that the MNRs are multiple MSW resonances. The crossing of the potentials shows the location where the MNR starts; see Fig. \ref{fig:MNR}. Furthermore if one looks at the full evolution of the Hamiltonian, the matter and $\nu\nu$ interaction terms cancel for tens of kilometers, concomitantly with the MNR process. Indeed, using a perturbative argument, \textcite{Chatelain:2016xva} showed that the $\nu\nu$ interactions adjust to the matter term over long distances: the MNR condition is fulfilled multiple times due to the nonlinearity of the equations and nonlinear feedback.

\subsubsection{Spin or helicity coherence}
The derivation of extended mean-field equations beyond the ones usually employed in flavor studies uncovered new terms (see Section II.A).
Thanks to these, new resonances become possible that can influence the neutrino content. 

Many years ago, \textcite{Rudzsky:1990} had pointed out wrong helicity contributions. 
\textcite{Volpe:2013uxl} derived mean-field equations including pairing correlators and wrong-helicity contributions, due to the neutrino mass. 
Afterward \textcite{Vlasenko:2013fja} obtained quantum kinetic equations for Majorana neutrinos using the closed-time-path formalism, and 
pointed to the wrong-helicity terms $\sim m/E$
naming them spin coherence. \textcite{Serreau:2014cfa} presented the most general mean-field equations and called such contributions helicity coherence. Present in anisotropic media, they couple neutrinos with antineutrinos but are suppressed, as expected, by the ratio $m/E$.  

 In a toy 2$\nu$ model \textcite{Vlasenko:2014bva} first studied whether helicity coherence modifies flavor. They found that it could trigger a significant $\nu$-$\bar{\nu}$ transformation through nonlinear feedback. Motivated by such findings, \textcite{Chatelain:2016xva} investigated these terms in binary neutron star mergers with inputs from detailed simulations. 
In contrast to the previous findings, their results showed that, while the resonance condition for helicity coherence (which was similar to the MNR one) was fulfilled, adiabatic evolution was absent for the ensemble of trajectories considered. 
Indeed, they were able to show that nonlinear feedback could not induce multiple matching of the resonance conditions\footnote{That is, unless peculiar matter density profiles are considered. Note that the argument holds for supernovae as well.}, unlike the MNR. The subsequent work of \textcite{Tian:2016hec} on the role of helicity coherence in core-collapse supernovae reached a similar conclusion.

\subsubsection{I-resonance}
Nonstandard interactions are present in theories beyond the standard model. Limits on nonstandard neutrino-neutrino interactions are rather loose \cite{Bilenky:1999dn}, whereas oscillations and scattering experiments provide tight constraints on non-standard neutrino-matter interactions (NSI); 
see the reviews by \textcite{Ohlsson:2012kf,Davidson:2003ha,Farzan:2017xzy,Biggio:2009nt}. After decades of attempts, \textcite{COHERENT:2017ipa} (COHERENT Collaboration) measured coherent neutrino-nucleus scattering, thereby giving new constraints \cite{Coloma:2019mbs,Giunti:2019xpr}.

NSIs are often evoked in the interpretation of neutrino oscillation experiments, as possible explanations of anomalies. If NSIs exist, mixing angles and mass-squared differences inferred by experiments are modified. In dense astrophysical environments, NSIs were studied by \cite{Fogli:2002xj,Esteban-Pretel:2007zkv,Esteban-Pretel:2009jqw,Stapleford:2016jgz,Chatelain:2017yxx}. Their results highlighted that NSIs significantly impact flavor evolution. Moreover the role of nonstandard neutrino-neutrino interactions in supernovae was studied by \textcite{Blennow:2008er}. In the context of primordial nucleosynthesis, NSIs give a subleading contribution to the effective number of degrees of freedom \cite{Mangano:2006ar}. 

\textcite{Esteban-Pretel:2007zkv,Esteban-Pretel:2009jqw} explored the combined effect of $\nu\nu$ interactions and $\nu$-matter NSI in core-collapse supernovae.  For NSI couplings $\vert  \epsilon \vert \ge 10^{-2}$, an MSW-like phenomenon called the $I$-resonance emerges\footnote{$I$ stands for "internal" since the phenomenon occurs close to the neutrinosphere, in the most deleptonized inner layers.}. It takes place  for $\nu$ and $\bar{\nu}$ simultaneously when the standard and nonstandard matter terms cancel each other. The $I$-resonance triggers efficient conversions of $\nu_{\rm e} \rightarrow \nu_{\mu,\tau} $ and $\bar{\nu}_{\rm e} \rightarrow \bar{\nu}_{\mu,\tau} $ independently of the neutrino energy.  

\textcite{Stapleford:2016jgz} performed an extensive investigation of NSI effects as a function of their couplings. They found that, even for NSI strengths well below bounds, NSIs produce symmetric and standard MNRs in core-collapse supernovae and impact the MSW $H$-resonance and $\nu\nu$ interaction effects (bulb model); see Sec. II.G.1. 

\begin{figure}
\begin{center}
\includegraphics[width=.5\textwidth]{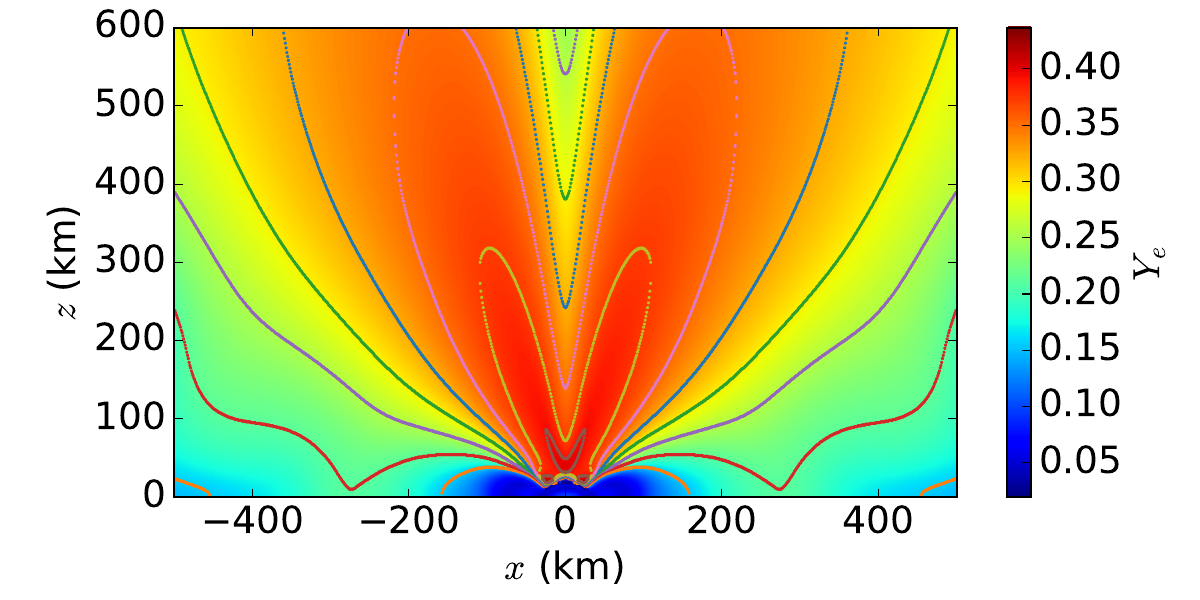}
\caption{$I$-resonances. The lines indicate the locations where the I-resonance condition is fulfilled above a binary neutron star merger remnant 
(bottom center).  In absence of flavor evolution values of the electron fraction $Y_e$ are color coded. From \cite{Chatelain:2017yxx}.}
\label{fig:I-locations}
\end{center}
\end{figure}

The first investigation of NSI effects in BNS mergers was performed by \textcite{Chatelain:2017yxx}. They showed that neutrino-neutrino interactions play a role in the $I$-resonance, in contrast to previous findings. Indeed, when the $\nu\nu$ interactions matter, the $I$-resonance becomes a synchronized MSW effect. The investigation of a large ensemble of trajectories, above a BNS remnant, uncovered the fact that, as in core-collapse supernovae, even small values of NSI parameters produce intricate patterns of mechanisms, including MNRs (Fig. \ref{fig:I-locations}), $I$- and synchronized $I$-resonances.

\subsection{Neutrino-neutrino interactions}
Dense environments have sizable $\nu\nu$ standard neutral-current interactions because neutrinos are emitted in large amounts. In a seminal work, \textcite{Pantaleone:1992eq} pointed out that $\nu\nu$ interactions introduce off-diagonal potentials\footnote{The $\nu\nu$ Hamiltonian has off-diagonal complex contributions, because of its dependence on the neutrino and antineutrino density matrices Eq.\eqref{eq:nunu}; see Fig. \ref{Fig:MF}.} and make $\nu$ evolution a nonlinear many-body problem.  

In the past 15 years theorists have worked to understand neutrino-neutrino refraction effects, novel flavor mechanisms, how they arise, and their impact. They have established connections with other many-body systems and figured out new approaches to deal with such interactions.  
Several reviews have provided a detailed picture of these developments \cite{Duan:2010bg,Duan:2009cd,Mirizzi:2015eza,Volpe:2015rla,Horiuchi:2018ofe,Tamborra:2020cul}. Here we highlight aspects that have emerged from the efforts to solve this complex problem.  

Flavor mechanisms due to $\nu\nu$ interactions are currently classified as slow or as fast. 
Slow modes occur at typical distances of $\mathcal{O}(10^2$-$10^3)$ km from the neutrinosphere, whereas fast modes have scales of $\mathcal{O}(1)$ m or much less and frequencies as large as $ \mu \sim \sqrt{2} G_{\rm F}  n_{\nu}$, $n_{\nu}$ the neutrino number density. Their rate can exceed the vacuum oscillation frequency by large factors, for example $\mu/\omega = 10^5$. 

\subsubsection{Slow modes}
Some 30 years ago \textcite{Samuel:1993uw} showed that $\nu\nu$ interactions stimulated new flavor effects. Studies in the cosmological context uncovered a stunning mechanism where neutrino spins "stuck together" precessing collectively around an effective magnetic field \cite{Kostelecky:1993yt,Kostelecky:1994dt,Pastor:2001iu}). Synchronized oscillations were included for example in investigations of cosmological neutrino-antineutrino asymmetries \cite{Dolgov:2002ab,Abazajian:2002qx} and also those with CP violation \cite{Gava:2010kz}. 

\begin{figure}
\begin{center}
\includegraphics[width=.45\textwidth]{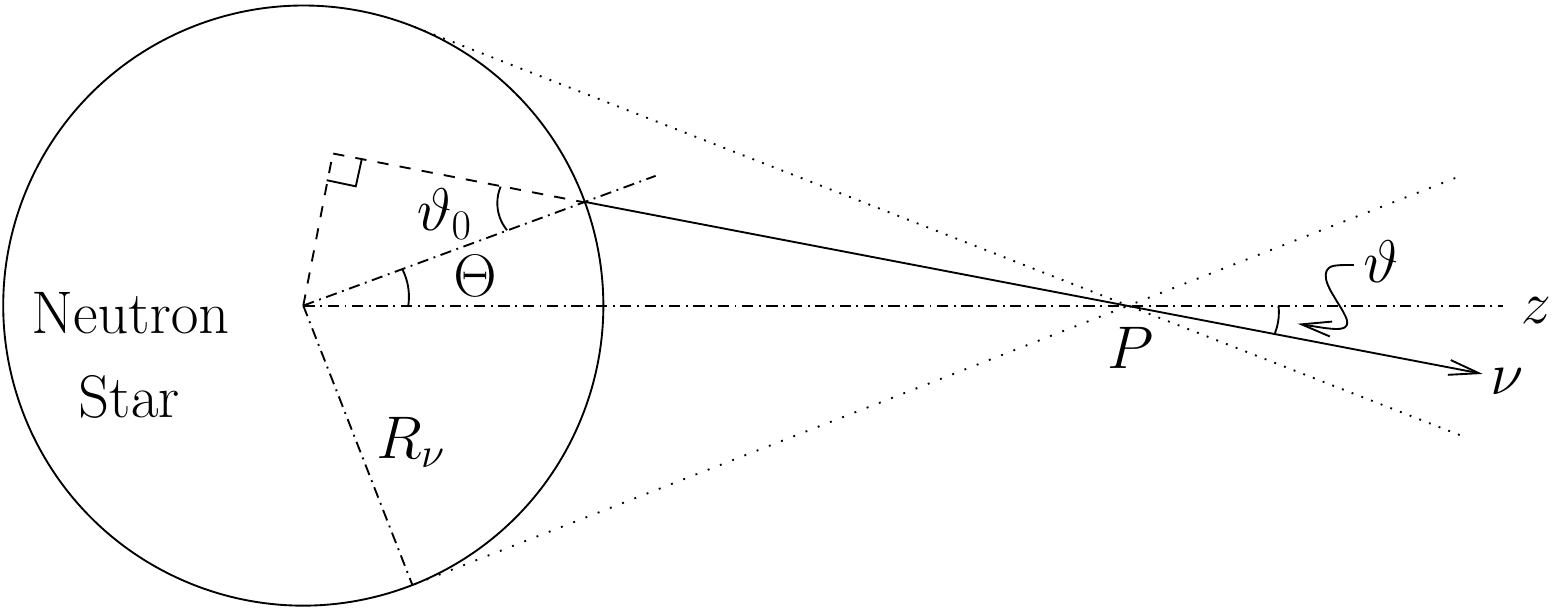}
\caption{Geometric depiction of the bulb model. In a supernova explosion the newly formed protoneutron star, depicted as a sphere, emits neutrinos uniformly and isotropically from its surface at radius R (the neutrinosphere). At emission, neutrinos are in flavor eigenstates. The bulb model assumes both spherical and azimuthal symmetry along the radial direction, with $r$ the distance from any point outside the sphere to the center of the PNS.  Each neutrino state of the neutrino beam is characterized by the momentum and the emission angle $\vartheta_0$. A neutrino emitted  at $\vartheta_0$ (solid line) encounters a radially propagating neutrino (dot-dashed line) with an angle $\vartheta$ at the intersection point. The dotted lines show the cone of neutrinos seen from this point, that contributes to the $\nu\nu$ interaction Hamiltonian. (Figure adapted from \cite{}).}
\label{fig:bulb}
\end{center}
\end{figure}

\textcite{Duan:2005cp} uncovered collective flavor modes in supernovae due to $\nu\nu$ forward scattering using the bulb model; 
see Fig. \ref{fig:bulb}. In this model neutrinos start free streaming at the neutrinosphere, which is taken to be a sharp surface (independent of energy and flavor), and are emitted, assuming both spherical symmetry and azimuthal symmetry along the radial direction $r$. 
  
Using the bulb model \textcite{Duan:2006an} performed "single-angle" and demanding "multiangle" simulations. Indeed computationally, one can treat ${\cal H}_{\nu\nu, {\bf p}}$ Eq.\eqref{eq:nunu} in either of two ways. In the simplified single-angle approximation, the flavor history of a neutrino at a given angle with respect to the radial direction is representative of all angles\footnote{Many took $\pi/4$ and often $0^{\circ}$, which strictly speaking corresponds to noninteracting neutrinos.}. In contrast, multiangle simulations include the full angular dependence of the $\nu\nu$ potential. Note that in binary compact objects, in which calculations are particularly challenging, it is also common to treat ${\cal H}_{\nu\nu, {\bf p}}$ as in single-angle approximation; see \cite{Malkus:2012ts,Malkus:2014iqa,Malkus:2015mda,Frensel:2016fge,Zhu:2016mwa}.

\begin{figure}
\begin{center}
\includegraphics[width=.4\textwidth]{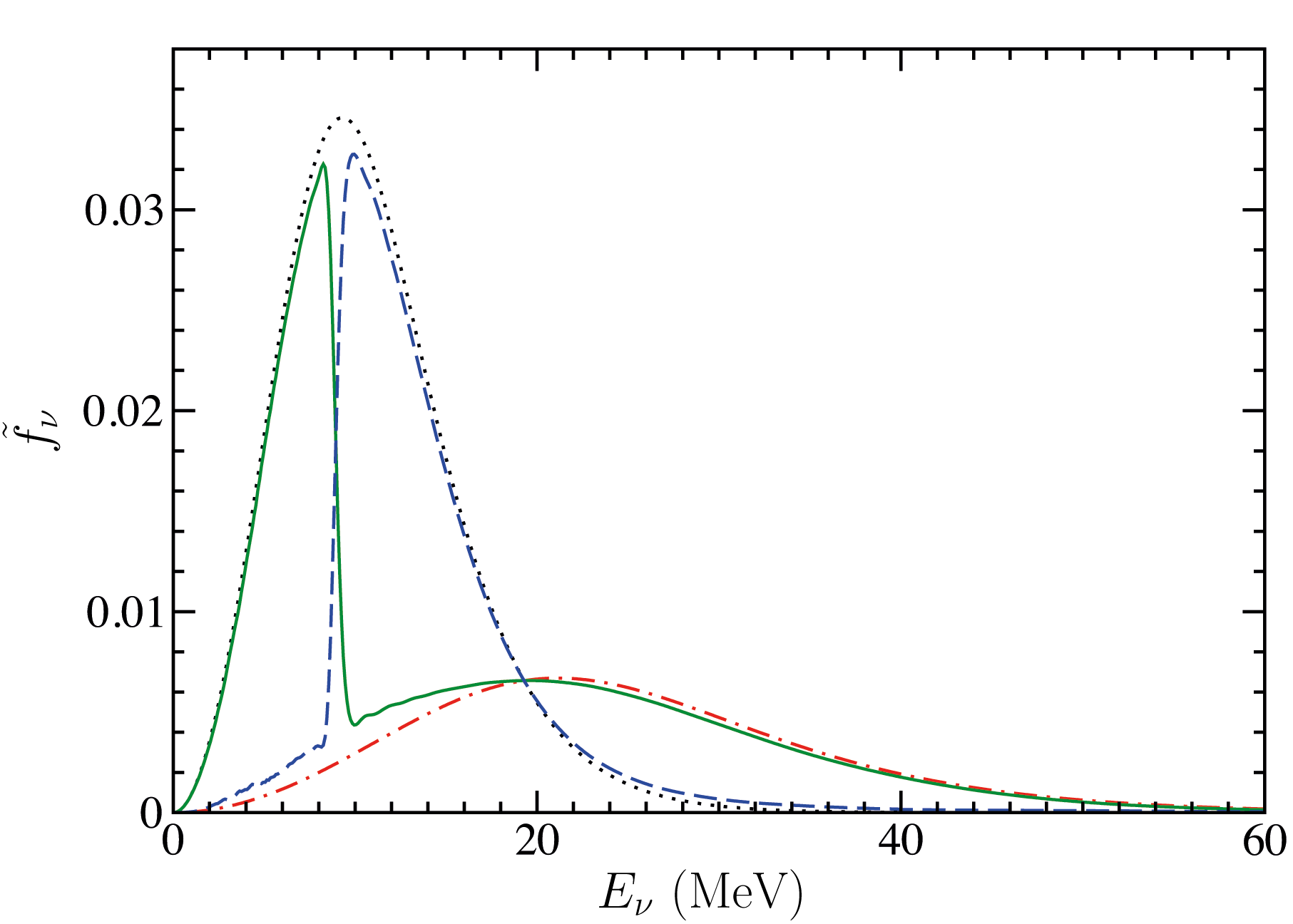}
\caption{Spectral splits in the {\it bulb} model: example of the neutrino spectral swapping in supernovae.  Due to $\nu\nu$ interactions, the quasi-thermal spectra of $\nu_e$ (dotted) and of $\nu_x$ (dot-dashed line) at the neutrinosphere ($r = R_{\nu}$) undergo significant modification. Indeed, there is a critical energy ($E_{\rm crit} = 8$ MeV) for which, if $E < E_{\rm crit}$ the spectra of $\nu_e$ (full line) is unchanged, whereas for $E > E_{\rm crit}$ the $\nu_e$ and $\nu_x$ spectra interchange. The inverse holds for $\nu_x$ (dashed line). The results are obtained at r=250 km. (Figure adapted from \cite{Duan:2010bg}).}
\label{fig:spectralsplits}
\end{center}
\end{figure}

The numerical results of \textcite{Duan:2006an} showed large-scale modes and spectral splits, which are sharp boundary features at the edges of spectral swap intervals  (Fig. \ref{fig:spectralsplits}). 
Note that in these early works, the neutrino and the matter number densities were such that self-induced flavor conversion was intertwined with matter effects.  
Moreover, \textcite{Balantekin:2004ug} found that the neutrino-neutrino refraction impacted the equilibrium electron-fraction $Y_e$, which is important for $r$-process nucleosynthesis in neutrino-driven winds.

For many years the bulb model has received particular attention. Simulations showed puzzling flavor behaviors that triggered intense theoretical work.
Named collective neutrino oscillations, these phenomena occur for sizable $\nu\nu$ interactions and for nonzero mixings (even for extremely small values of $\theta$).
The collective modes were named slow after the identification of fast modes; see Sec. II.G.2.

Three regimes emerge when neutrinos travel from the neutrinospheres, where the neutrino number densities are large, to regions where matter dominates. They are called the synchronization, the bipolar instability \cite{Duan:2005cp,Duan:2006an,Duan:2007mv,Hannestad:2006nj} and  the spectral splits \cite{Duan:2006an,Raffelt:2007cb,Fogli:2007bk,Galais:2011jh,Dasgupta:2009mg}. 

The spin formalism in flavor space provides an image of these three phases. First, as in the early Universe at the epoch of primordial nucleosynthesis, neutrino spins synchronize in a stable collective mode.  
Second, neutrino spins experience an instability where $\nu_e \bar{\nu}_e$ pairs convert into $\nu_x \bar{\nu}_x$ ones due to lepton-number conservation. They perform precession and nutation around ${\rm {\bf B}}$ and behave like a pendulum \cite{Duan:2005cp,Duan:2007mv,Duan:2007fw} or a gyroscopic pendulum \cite{Hannestad:2006nj}. Third,
they undergo either full or no conversion, depending on the neutrino energy, while generating spectral swapping and splits (Fig. \ref{fig:spectralsplits}). 
A depiction of these modes for $3 \nu$ flavors was given in the $e_3$-$e_8$ triangle diagram of
 \textcite{Dasgupta:2009mg} who employed Bloch vectors and the SU(3) algebra.  

Other approaches to these phenomena brought further insight. Using the matter basis,
\textcite{Galais:2011jh} found that with the $\nu\nu$ refraction the adiabaticity parameters depend on the matter angle and phase derivative and bipolar oscillations start when the latter diverges.
Moreover, \textcite{Galais:2011gh} showed that spectral splits arise from a magnetic resonance phenomenon: the swapping emerges 
because the spins satisfy (or do not satisfy) a magnetic resonance condition, depending on the neutrino energy.

\begin{figure}
\begin{center}
\includegraphics[width=.4\textwidth]{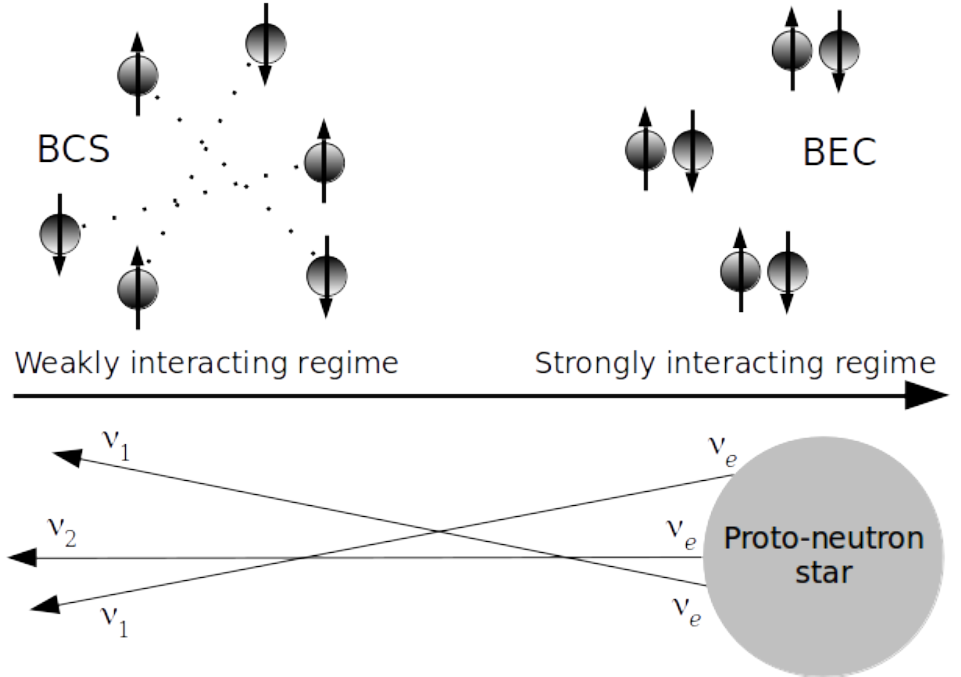}
\caption{Neutrino-neutrino interactions in a supernova: correspondence between neutrinos propagating from dense to dilute regions and the BEC to BCS limits in ultracold atomic gases.  From \cite{Pehlivan:2016lxx}.}
\label{fig:BECBCS}
\end{center}
\end{figure}

\textcite{Pehlivan:2011hp} used a different angle of attack. With an algebraic many-body approach and the Bethe ansatz, they demonstrated that the splits emerged in the transition from a quasiparticle to a particle description. In a subsequent study \textcite{Pehlivan:2016lxx} established that the emergence of spectral splits in the bulb model, when going from the regions where neutrinos strongly interact to those where they weakly interact, is similar to the behavior of Cooper pairs in the BEC-BCS crossover in experiments with ultra cold atomic gases; see Fig. \ref{fig:BECBCS} and Sec. III.G.

While the first studies revealed instabilities only in inverted mass ordering, \textcite{Dasgupta:2009mg} showed that more plausible ratios
of the $\nu_e, \bar{\nu}_e$ and $\nu_x$ fluxes (i.e. other than 1) produced single and multiple spectral splits in both hierarchies. 
One of the key hypotheses of the bulb model is that neutrinos are emitted as pure flavor states at the neutrinosphere assumed as a sharp surface.
\textcite{Hansen:2019iop} pointed out that if one takes into account that neutrino emission is over a region of finite width (a few kilometers), the integration over 
the extended emission region either modifies the growth of slow modes (started in the emission region) or delays the start of flavor transitions. 

Moreover, \textcite{Esteban-Pretel:2008ovd} argued that 
large matter densities introduced decoherence in multiangle calculations because of the angular factor $(1 - \hat{\bf p} \cdot \hat{\bf p}')$ in Eq.\eqref{eq:nunu}. Following these findings, \textcite{Chakraborty:2011nf,Chakraborty:2011gd} showed, using one-dimensional supernova simulations, that matter suppressed collective effects  when
that matter exceeded the $\nu\nu$ number density. This finding should be confirmed by multiangle calculations using multidimensional supernova simulations.

The studies assumed stationarity and homogeneity of the medium where neutrinos propagate. Moreover, they used the mean-field approximation and neglected collisions. 
Thanks to these approximations, the full seven-dimensional problem of neutrino flavor evolution reduces to a more tractable one, typically in two- or three-dimensions, i.e. ($E, r$) or ($E, r, \theta$). But even schematic models with a reduced number of degrees of freedom are often challenging to solve. 

A step toward higher spatial dimensionality was provided by the so-called neutrino line model with two spatial dimensions, either with only two neutrino beams from each initial condition \cite{Duan:2014gfa}, or with multiangles at each point source \cite{Abbar:2015mca}, showing inhomogeneous modes located at larger neutrino densities than homogenous ones.   
\textcite{Abbar:2015fwa} and \textcite{Dasgupta:2015iia} identified temporal instabilities arising in nonstationary models since time can cancel a constant matter term, producing instabilities deep in the supernova.

Moreover, \textcite{Cherry:2012zw,Cherry:2013mv} discovered that a neutrino halo (a small amount of backscattered neutrinos due to collisions) could completely reshape the flavor patterns produced by forward scattering. This finding casts doubts on the treatment of $\nu\nu$ interactions, therby showing the limitations of the mean-field equations used
(see Section \ref{sec:MF}) in which neutrinos forward scatter only from the neutrinosphere (see Sec. \ref{eq:MFder} and F). The halo effect was further studied by \textcite{Sarikas:2012vb} and \textcite{Cherry:2019vkv}.

Furthermore, based on a linearized study of the bulb model \textcite{Raffelt:2013rqa} found azymuthal-angle instabilities. This finding showed that solutions do not necessarily inherit the symmetries of the initial or boundary conditions (unless enforced). This fact was named spontaneous symmetry breaking. Several works have confirmed symmetry breaking solutions with linearized analysis in supernovae \cite{Mirizzi:2013rla,Chakraborty:2014nma}. 

Flavor evolution can even reveal chaotic behaviors. This was mentioned by \textcite{Raffelt:2013isa} 
in a stationary model with two opposing neutrino momenta. In fact, using the same model \textcite{Hansen:2014paa} clearly identified the exponential divergence of Liapunov exponents when considering infinitely close initial trajectories of the neutrino spin vectors.

Thus, after the discovery of novel phenomena in the bulb model, models increased in complexity and included nonstationarity, inhomogeneities, unconstrained symmetries, etc \ldots. These in-depth investigations kept uncovering new features and the richness of flavor evolution in dense media, due to the $\nu\nu$ refraction.  
While theorists thought they were developing a solid understanding, fast modes arrived triggering another proliferation of studies\dots

\subsubsection{Fast modes}
Using a $3 \nu$  two-beam model \textcite{Sawyer:2005jk,Sawyer:2008zs} found that $\nu\nu$ interactions could "speed up" flavor transformation and produce counterintuitive modifications on a short timescale of $t = (2 \sqrt{2} G_{\rm F}  n_{\nu})^{-1}$  of the order of a few nanoseconds. 
Much later \textcite{Sawyer:2015dsa} considered nontrivial angular distributions at the neutrinospheres, with $\bar{\nu}_e$ emitted deeper than $\nu_e$. He identified modes on a tens of meters scale. 

Sawyer's findings triggered enthusiasm again: fast modes 
took place close to the neutrinosphere. They could influence the supernova dynamics and nucleosynthesis. Theorists had finally found the short scale modes they were looking for.

Unlike slow modes, fast modes have the peculiarity that they are triggered neither by the mixings Eqs.\eqref{eq:PP} and \eqref{eq:Hvac}, nor by the matter contribution Eq.\eqref{eq:Hmswfull}. Since only the neutrino emission matters, \textcite{Izaguirre:2016gsx} introduced the angle distribution of the ELN which for $2 \nu$ flavors reads
\beq\label{eq:ELN}
G_{\bf v} = \sqrt{2} G_{\rm F}  \int_0^{\infty} dE ~ {E^2 \over {2 \pi^2 }}~ [\phi_{\nu_e}(E, {\bf v}) -  \phi_{\bar{\nu}_e}(E, {\bf v})] \ .
\eeq

As indicated by the results of \textcite{Sawyer:2015dsa} and pointed out by \textcite{Dasgupta:2016dbv,Izaguirre:2016gsx},
fast modes take place when the angular distributions of $\nu_e$ and $\bar{\nu}_e$ cross each other along a given direction (i.e. $G_{\bf v}$ changes sign): this is an ELN {\it crossing}.
The sufficiency of this condition and/or its necessity have been controversial. \textcite{Morinaga:2021vmc} gave the mathematical proof that in inhomogeneous media the occurrence of an ELN crossing is equivalent to the existence of a fast instability, i.e. the condition is both necessary and sufficient. In addition \textcite{Fiorillo:2023hlk} proved that it is a sufficient condition when the medium is homogeneous.  

In recent years many studies of fast modes have been realized based on the linearized approach; see \cite{Capozzi:2017gqd,Abbar:2019zoq,Abbar:2018shq,DelfanAzari:2019epo,Tamborra:2014aua,Abbar:2017pkh,Chakraborty:2019wxe,Chakraborty:2016yeg,Chakraborty:2016lct,Shalgar:2019qwg,Padilla-Gay:2020uxa,Wu:2017qpc,George:2020veu,Xiong:2020ntn} 
and \textcite{Tamborra:2020cul}.

 \textcite{Sawyer:2005jk} had already advocated that fast modes could bring flavor equilibration of the different neutrino species. The ansatz has been admitted in the literature for some time. In fact it has the advantage that it simplifies the problem since the neutrino spectra emerge from a supernova identical. However, using a two-beam model, \textcite{Abbar:2018beu} evolved for the first time fast modes in the full nonlinear regime showing they do not necessarily lead to flavor equilibration. 

Fast modes behave differently in $3 \nu$ flavors than in $2 \nu$ flavors. For the former the concept of ELN crossing needs to be generalized to $\mu$LN and $\tau$LN crossings. \textcite{Chakraborty:2019wxe} first investigated such effects with the dispersion relation treating 
both time and space. They pointed out the importance of three-flavor effects. The further analysis of \textcite{Capozzi:2020kge}, which 
went up to the nonlinear regime, showed instabilities over a tens of nanosecond scale 
that either were absent (in $2 \nu$ flavors) or got damped. 
These findings emphasize the need for three-flavor analysis of fast modes.   

Since the scale for fast modes is so short, there can be regions requiring the treatment of flavor mechanisms and collisions 
when the medium is dense. Indeed the fast rate exceeds the collision rate even within a supernova core. Using a one-dimensional model, with two momentum modes, \textcite{Capozzi:2018clo} analyzed the interplay between collisions and fast modes and showed that collisions can trigger the conditions for fast conversions. Significant efforts are made to study flavor conversion modes in the presence of collisions; see \textcite{Capozzi:2018clo,Richers:2019grc,Hansen:2022xza,Xiong:2022vsy,Johns:2021qby,Padilla-Gay:2022wck}. Keep in mind, though, that a consistent treatment in multidimensional simulations is numerically challenging and therefore far off.

Fast modes are present in schematic models, but do they occur in detailed supernova simulations?
The first investigation by \textcite{Tamborra:2017ubu}, which was based on one-dimensional simulations, concluded that there was an absence of fast modes. Searches in multidimensional simulations revealed the presence of fast modes, in contrast with this early finding. 

\begin{figure}
\begin{center}
\includegraphics[width=.4\textwidth]{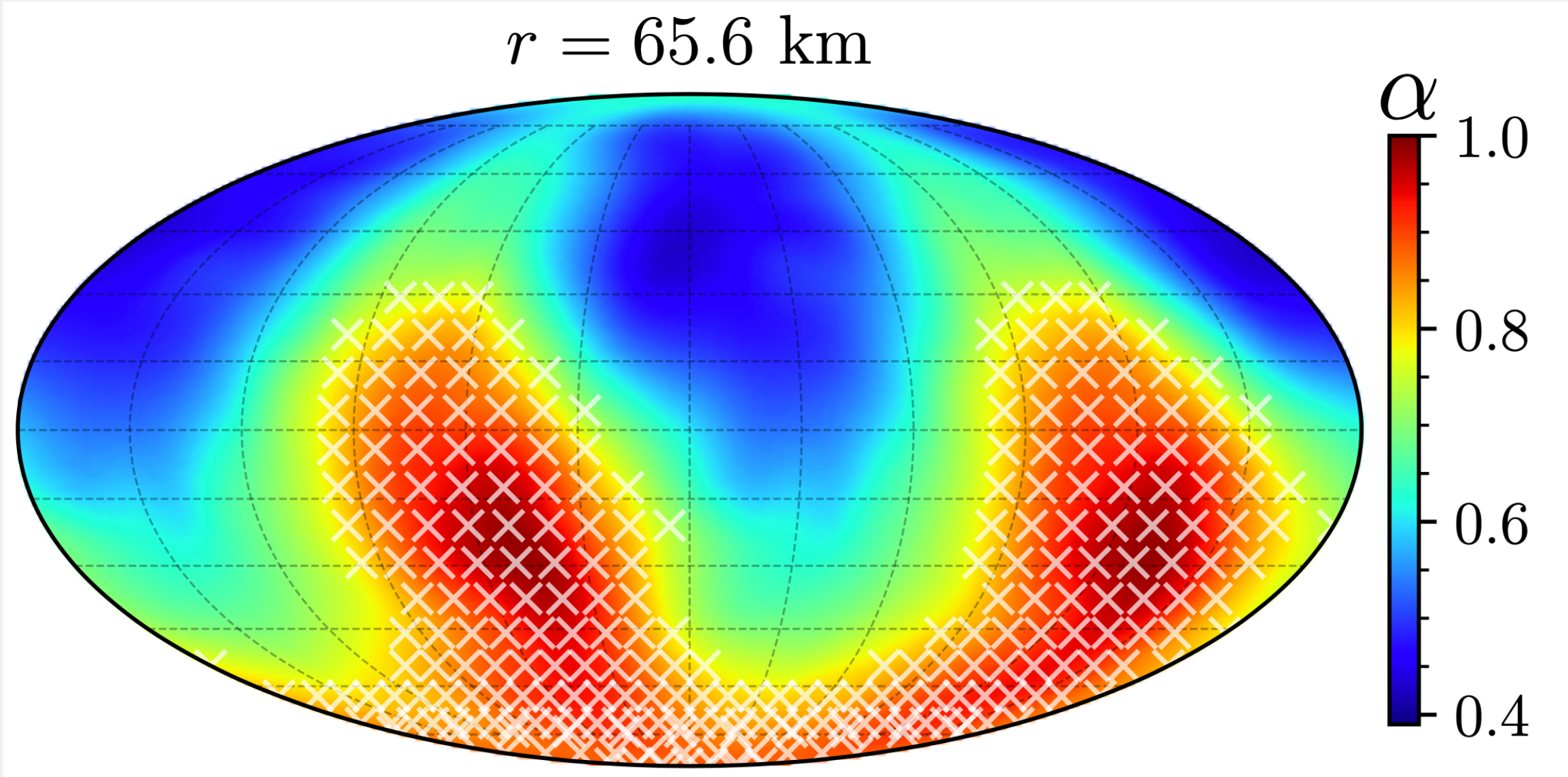}
\caption{Fast modes in multidimensional supernova simulations: Mollweide projection of the $\bar{\nu}_e$-over-$\nu_e$ ratio at a distance $r= 65.6$ km in a snapshot at postbounce time of 200 ms of a 3D supernova model. The white crosses indicate the location where fast modes occur, based on a linearized analysis. From \cite{Abbar:2019zoq}.}
\label{fig:fastmultiD}
\end{center}
\end{figure}

\textcite{Abbar:2019zoq} identified fast-growing modes in two- and three-dimensional simulations when $\alpha = n_{\nu_e}/n_{\bar{\nu}_e}$ is of the order of 1. A linear stability analysis confirmed the presence of fast modes in correspondence with the angular crossings (Fig. \ref{fig:fastmultiD}), even deep in the supernova core. Their influence on the neutrino spectra was found to be small, since the neutrino spectra  are already similar at the location of the crossings.

 \textcite{Glas:2019ijo} and \textcite{DelfanAzari:2019tez} also found ELN crossings nearby the neutrinosphere and confirmed the presence of fast modes in detailed supernova simulations (with full Boltzmann transport) in three- and two-dimensions respectively. Note that multidimensional supernova simulations do not provide full information on the neutrino angular distributions as a function of time. Therefore, methods have 
 been developed, as for instance by \textcite{Dasgupta:2018ulw}, to identify fast modes using the moments of the angular distributions.

Moment methods are commonly used in supernova codes with classical Boltzmann neutrino transport. They require a truncation
of the tower of moment equations and the use of a closure approximation to estimate the unknown from the evolved moments. 
\textcite{Myers:2021hnp} suggested a method to study flavor evolution based on quantum moments, i.e. angular moments of the quantum neutrino distributions.
In the bulb model setting, they showed that calculations with two moments were able to capture features of multiangle simulations, such as the
onset of collective oscillations, sometimes overestimating the coherence. 
The method was applied to fast modes by \textcite{Grohs:2022fyq} in a three-dimensional simulation of a neutron star merger. The results showed
that the method can qualitatively capture some of the aspects of fast modes such as their growth rate.

\textcite{Nagakura:2022kic} proposed a method to investigate fast modes on large scales ($>$ 10 km) in a supernova, using an {\it ad hoc} rescaling of the neutrino number density and studying the convergence when its value tends to the one found using physical conditions. First studies have 
appeared that implement fast modes in detailed one-dimensional \cite{Nagakura:2023mhr} or multidimensional core-collapse supernova simulations \cite{Ehring:2023abs}. The goal is to investigate their impact on the heating rates in the gain region and on the explosion dynamics; see Appendix B. While exploring this role is crucial, at this stage and in light of the approximations done it is too early to draw conclusions. 

Beyond the studies in the supernova context, \textcite{Wu:2017qpc} performed the first analysis of fast modes in accretion disks resulting from binary 
compact object mergers. They found the conditions for fast modes to be generically met because of the excess of $\bar{\nu}_e$ over $\nu_e$ and of the 
geometry of such environments. 

As for nucleosynthesis, \textcite{Xiong:2020ntn} studied the influence of fast oscillations in neutrino-driven winds in a low and a high mass core-collapse supernova.
They showed that partial (or total) flavor equilibration creates more proton-rich conditions ($Y_e > 0.5 $) enhancing the $\nu$p process and mass ejection. 
\textcite{Wu:2017drk}  considered the impact of fast modes on the $r$-process in a neutrino-driven wind nearby a black hole remnant from compact binary mergers. Under
the approximate assumption of flavor equilibration, fast modes produced an increase of lanthanides (more generally nuclei with A $>$ 130) up to a factor of $10^3$ due to the
decrease of $Y_e$, showing a potentially high impact on kilonova light curves. 

On the contrary, using an analysis for a hypermassive BNS merger remnant, 
\textcite{George:2020veu} found lanthanides to be only slightly affected by fast pairwise conversion (under the same flavor equilibration hypothesis).
In the disk outflows of a hypermassive neutron star \textcite{Fernandez:2022yyv} studied the production of heavy $r$-process elements in the disk outflows of an hyper-massive neutron star and found that the abundance of lanthanides and actinides, which are important for kilonova opacities, can be modified up to a factor of 2 if fast modes are included (assuming 
partial or total flavor equilibration). 

In conclusion, while studies have shown that the impact of flavor mechanisms on $r$-process nucleosynthesis is important, further work is needed, 
based also on more consistent treatments, in order to draw definite conclusions about their role. As for fast modes and their impact, their investigation undergoes a fast development where interesting aspects are uncovered. These short scale modes will continue to attract attention in the coming years.        

\section{Flavor evolution: Theoretical frameworks}
Neutrinos propagating in a dense environment constitute a unique weakly interacting, many-body system. Its description benefits from 
good fortune in some respects since one does not have to deal with phenomenological interactions as with atomic nuclei, as it has been since long before the advent of ab initio calculations. This is a specific case, because theoretical approaches developed for many-body systems need to be extended to relativistic particles with mixings. 

The literature is rich with theoretical approaches for the neutrino many-body system (see the review by \textcite{Volpe:2015rla}). These range from the mean-field and the extended mean-field equations, to the linearized equations and a dispersion relation approach, to the neutrino quantum kinetic equations, as we now see.

\subsection{The mean-field approximation}
The mean-field constitutes the simplest approximation to describe neutrino propagation in astrophysical environments. 
Neutrino mean-field equations were derived by 
\cite{Dolgov:1980cq,Samuel:1993uw,Sawyer:2005jk,Serreau:2014cfa,Balantekin:2006tg,Volpe:2013uxl}.  

It is common to determine neutrino propagation using the formalisms of Green's functions, density matrices, neutrino (iso)spins, or neutrino amplitudes. 
The density matrix formalism is widely used.
The $\nu$ and $\bar{\nu}$ one-body densities are defined as 
\begin{eqnarray}\label{eq:rho1}
\varrho_{1, ij}({\bf p}, h, {\bf p}',h') & = & \langle a^{\dagger}_j({\bf p}',h')
a_i({\bf p}, h) \rangle  \ ,   \nonumber \\
\bar{\varrho}_{1, ij}({\bf p}, h, {\bf p}',h') & = & \langle b^{\dagger}_i({\bf p}, h) b_j ({\bf p}',h') \rangle \ ,
\end{eqnarray}
where the quantum expectation value is over the astrophysical or cosmological background; $i,j \in[1,N]$ with $N$ the number of neutrino families. The diagonal elements of the one-body density matrix correspond to the expectation value of the number operator and are the only contributions for particles without mixings. The off-diagonal elements ($i \neq j$) implement coherence due to the mixings. \textcite{Sigl:1992fn} stated that neutrino evolution requires a "matrix of densities". 

The mean-field approximation consists in replacing a two-body interaction term, which has the form of a quadratic operator ${\cal I}_1 {\cal I}_2$, with one-body terms as
\begin{equation}
{\cal I}_1 {\cal I}_2 \rightarrow {\cal I}_1 \langle {\cal I}_2 \rangle + \langle {\cal I}_1 \rangle {\cal I}_2 + \langle {\cal I}_1 \rangle \langle {\cal I}_2 \rangle \ .
\end{equation} 

If one takes the charged- or neutral-current interaction terms of the GWS model, the Hamiltonian is quartic in the creation and annihilation operators. 
In the mean-field approximation, the Hamiltonian becomes quadratic in the creation and annihilation operators and has the general bilinear form \cite{Serreau:2014cfa}
\beq\label{e:Heff}
H_{\rm MF}(t) = \int d {\bf x}\, \bar{\psi}_{i}(t,{\bf x})\Gamma_{ij}(t,{\bf x}) \psi_{j}(t,{\bf x}),
\eeq
where $\Gamma_{ij}(t,{\bf x})$ is the interaction kernel that depends on the specific interaction terms that one considers. The quantities
 $\psi_i$ denote the neutrino field in the mass basis
(for the $i$th mass eigenstate)\footnote{Here we write expressions that take neutrino Dirac fields into consideration. The generalization to Majorana fields is straightforward. Detailed expressions of both the Dirac and Majorana cases were given by \cite{Serreau:2014cfa}.}
\beq\label{eq:psi}
\psi_{j}(t,{\bf x}) = \sum_h \int_{p} [u_j({\bf p}, h)a_j({\bf p}, h)e^{-i p\cdot x} + v_j({\bf p}, h) b_j^{\dagger}  e^{i p\cdot x}  ] \ , 
\eeq
with $p \cdot x = p^{\mu} x_{\mu}$ and $u_j({\bf p}, h)$ and ${v_j(\bf p}, h)$ the four-components complex spinors, solution of the Dirac equation.
As an example, inserting Eq.\eqref{eq:psi} into Eq.\eqref{e:Heff} and taking the case of $\nu_e$-electron scattering, the interaction kernel reads  
\beq\label{e:Gee}
\Gamma_{ee}(t,{\bf x}) = {G_F \over \sqrt{2}} \gamma^{\mu} (1-\gamma^5) \langle   \bar{\psi}_e (t,{\bf x}) \gamma_{\mu} (1-\gamma^5) \psi_e (t,{\bf x})\rangle \ .
\eeq

A theoretical framework to treat the evolution of many-body systems was given by BBGKY hierarchy
the \textcite{Born:1946}, \textcite{Bogoliubov:1946}, \textcite{Kirkwood:1935}, \textcite{Yvon:1935}. 
Introduced for a nonrelativistic many-body system the hierarchy replaces the exact evolution of the quantum many-body system by a hierarchy of integro-differential equations for $n$-body density matrices\footnote{Note that  \textcite{Wang:1984cg} reformulated the hierarchy as a set of equations for $n$-body correlation functions.},
\beq\label{eq:reddes}
\rho_{1\ldots n} = \langle a^{\dagger}_n \ldots a^{\dagger}_1 a_1  \ldots a_n  \rangle \ , 
\eeq 
If one truncates the hierarchy at different levels, one can derive approximate equations of motion, including the mean-field, extended mean-field 
and Boltzmann equations. 

The mean-field approximation consists in neglecting the correlated part of the two-body density\footnote{Note that the reduced density matrices can be referred to as one-particle, two-particle, \ldots, $n$-particle density matrices as well as one-body, two-body, \ldots $n$-particle density matrices.} 
\begin{equation}\label{eq:rho12}
\rho_{12} = \rho_1 \rho_2 - c_{12}  \ , 
\end{equation}
i.e. setting $c_{12}=0$, where $\rho_{12}$ is the two-body density of a generic many-body system, and $\rho_1$ and $\rho_2$ are one-body densities for particles 1 and 2.
Since only the uncorrelated part is retained, the two particles propagate independently. Thus the full many-body system evolves as made up of independent particles. 

In the BBGKY hierarchy the mean-field approximation corresponds to a truncation at the lowest level, neglecting collisions
and more generally two-body (and higher order) correlations. Implementing the correlated part of $\rho_{12}$ (i.e. $c_{12} \neq 0$), collisions are also included. Thus, one gets the Boltzmann equation from
the evolution of the full two-body density $\rho_{12}$.

Note that a simple way to derive mean-field evolution equations is through the Ehrenfest theorem
\begin{equation}
i \dot{\varrho}_{1,ij} = \langle [a^{\dagger}_j a_i, H] \rangle \ ,
\end{equation}
where the neutrino Hamiltonian operator here is in second-quantized form. 

While BBGKY was originally for a nonrelativistic many-body system, \textcite{Calzetta:1986cq} generalized it for relativistic many-body systems which involves, in particular, an infinite hierarchy of equations. \textcite{Volpe:2013uxl} applied the BBGKY hierarchy to a many-body neutrino system, considering neutrino and antineutrino one-body densities  
\eqref{eq:rho1} in an astrophysical environment. They obtained the commonly used mean-field equations, as well as extended mean-field equations, as we discuss; see Sec. III.C. Following these developments, \textcite{Froustey:2020mcq} also included the contribution of $c_{12}$ and derived the Boltzmann equation for particles with mixings (neutrino quantum kinetic equations) for the case of the early Universe.

\subsection{The mean-field Hamiltonian: a derivation}\label{eq:MFder}
We now consider a neutrino propagating in a medium and interacting with the particles composing it. 
One way of deriving the mean-field Hamiltonian for a given scattering process is to coherently add 
the scattering amplitudes between the propagating particle and a particle in the background. More explicitly, the potential is given by 
\beq\label{eq:MFpot}
V_{kr}(\rho) = \sum_{s,p} v_{(kp,rs)} \rho_{sp} \ ,
\eeq
where the sum (or integration) is over the product of the matrix elements $v_{(kp,rs)}$ between the incoming and outgoing single-particle states, and the quantity $\rho_{sp}$ is
\beq\label{eq:rhosp}
 \rho_{sp} =  \langle a^{\dagger}_p  a_s  \rangle \ . 
 \eeq
The quantum numbers $k, p, r, s$ (for example the particle momentum and helicity) characterize the incoming and outgoing
single-particle states in the scattering process.

We now apply Eqs.\eqref{eq:MFpot} and \eqref{eq:rhosp} to the neutrino case and, as an example,  
derive the mean-field term of the matter Hamiltonian associated with $\nu_e$-$e$ scattering; see the upper right tadpole diagram in Fig. 3.

The charged-current interaction term associated with $\nu_e$-$e$ scattering reads 
\begin{align}\label{eq:CC}
H_{\rm CC}  =&  {G_{\rm F} \over{\sqrt{2}}} [ \bar{\psi}_{\nu_e}(t,{\bf x})\gamma^{\mu}(1-\gamma_5) \psi_{e}(t,{\bf x})] \nonumber \\ 
& \times  [ \bar{\psi}_{e}(t,{\bf x})\gamma_{\mu}(1-\gamma_5) \psi_{\nu_e}(t,{\bf x})] \ . 
\end{align}

The first step in determining Eq. \eqref{eq:MFpot} is to evaluate the matrix elements 
\beq\label{eq:ME}
v_{(kp,rs)} \equiv \langle k, p \vert H_{\rm CC} \vert r, s \rangle \ . 
\eeq
More explicitly one needs to calculate
\begin{align}\label{eq:CCF}
v_{(kp,rs)} = & {G_{\rm F}  \over{\sqrt{2}}} \langle \nu_e, e \vert  \int d {\bf x}~ [ \bar{\psi}_{e}(t,{\bf x})\gamma^{\mu}(1-\gamma_5) \psi_{e}(t,{\bf x})] \nonumber \\
& \times  [ \bar{\psi}_{\nu_e}(t,{\bf x})\gamma_{\mu}(1-\gamma_5) \psi_{\nu_e}(t,{\bf x})] \vert \nu_e, e \rangle \ , 
\end{align}
where the Fierz transformation has been applied to Eq.\eqref{eq:CC}.  

By introducing the Fourier expansions of the electron and neutrino quantum fields Eq.\eqref{eq:psi}, one can calculate 
the matrix element using the following general relation:
\begin{align}\label{eq:ME1gen}
\langle e(1) \vert a^{\dagger}(2) a(3)     \vert e(4) \rangle = & (2 \pi)^3 \delta^3({\bf p}_1- {\bf p}_2)2 E_{p_1}  \delta_{h_1,h_2} \nonumber \\
&  (2 \pi)^3 \delta^3({\bf p}_3- {\bf p}_4)  2 E_{p_3} \delta_{h_3,h_4}\ , 
\end{align}
which can easily be obtained from the equal-time anticommutation rules; see Footnote 20.
In Eq.\eqref{eq:ME1gen} the labels 1-4 stand for a set of single-particle quantum numbers (${\bf p}, h$). The momentum of the incoming particle is 
${\bf p}_4$, and that of the outgoing particle is ${\bf p}_1$. 

Using Eq. \eqref{eq:ME1gen} one gets
\begin{align}\label{eq:ME1}
v_{({\bf k}' {\bf p}' , {\bf k} {\bf p})}  = & {G_{\rm F} \over{\sqrt{2}}} \int d{\bf x} ~  [\bar{u}_{\nu_e} ({\bf k}')\gamma^{\mu}(1 - \gamma_5) u_{\nu_e} ({\bf k}) ]  \nonumber \\
& \times  [\bar{u}_{e} ({\bf p}')\gamma_{\mu}(1 - \gamma_5) u_{e} ({\bf p}) ]  e^{i ({\bf p} + {\bf k} - {\bf p}' - {\bf k}')  \cdot {\bf x}} \ ,
\end{align}
where $({\bf p}, {\bf k})  $ and  $({\bf p}', {\bf k}')$ are the momenta of the incoming and outgoing particles respectively.
The first two factors in the integral in Eq.\eqref{eq:ME1gen} depend on spinorial products, whereas the last one ensures momentum conservation. 

The second step to determine the mean-field term is to perform the integration over the degrees of freedom of the electron background and
evaluate
\begin{align}\label{eq:ME2}
V_{\rm CC}(n_e) = &    {G_{\rm F}  \over{\sqrt{2}}}  \sum_{h_e, h_e'} \int_{p_e,p'_e} \int d{\bf x} ~  [\bar{u}_{\nu_e} ({\bf k}')\gamma^{\mu}(1 - \gamma_5) u_{\nu_e} ({\bf k}) ]  \nonumber \\
& \times  [\bar{u}_{e} ({\bf p}')\gamma_{\mu}(1 - \gamma_5) u_{e} ({\bf p}) ]  e^{i ({\bf p} + {\bf k} - {\bf p}' - {\bf k}')  \cdot {\bf x}}  \nonumber \\
& \times \rho_{({\bf p}_e, h_e, {\bf p}_e', h_e')}  \ ,  
\end{align}
with
\begin{align}\label{eq:rhoe}
\rho_{({\bf p}_e, h_e, {\bf p}_e', h_e')} & = \langle  \psi \vert a^{\dagger}_e( {\bf p}_e', h_e')  a_e({\bf p}_e, h_e) \vert \psi  \rangle \, . 
\end{align}
In Eq.\eqref{eq:rhoe} the expectation value is over the wave function describing the background (at a finite temperature $T$ and a chemical potential $\mu$).

If one assumes that the background particles are uncorrelated (independent), then the expectation value can be performed on the single-particle states
describing the electrons. Besides neglecting correlations, one often makes the following ansatz for the background:
\begin{align}\label{eq:rhoehom}
\rho_{({\bf p}_e, h_e, {\bf p}_e', h_e')} &  =  (2 \pi)^3 \delta^3({\bf p}_e - {\bf p}_e') \delta_{h_e,h_e'} 2 E_{p} \rho_{e, p} \ , 
\end{align}
that is, that the medium is homogeneous and only forward-scattering is considered, the electron (and therefore neutrino) momenta are unchanged. 
The quantity $\rho_{e, p}$ is the  distribution of the electron occupation numbers which follows a Fermi-Dirac function.  
We often say that the mean-field approximation corresponds to coherent forward scattering.  

By plugging Eq.\eqref{eq:rhoehom} into Eq. \eqref{eq:ME2}, the spinorial products in Eq. \eqref{eq:ME2} can be evaluated thus giving 
\beq\label{eq:prod}
8 p^{\mu} (k_{\mu} - m s_{\mu}) = 8 E_{p} E_{k} (1 - \hat{\bf p} \cdot \hat{\bf k}) (1 - h_{\nu}) \ ,
\eeq
where we have introduced the four-vector
\begin{align}\label{eq:Snu}
s_{\mu} = h_{\nu} \Big({\vert {\bf k} \vert \over{m}}, {E_{k} \over{m \vert {\bf k} \vert}} \Big) \ ,
\end{align}
and have already imposed momentum conservation. 
Assuming that electrons are relativistic, one gets
\begin{align}\label{eq:Vrhoe}
V_{\rm CC}(n_{e}) = & {G_{F} \over{\sqrt{2}}} (2 \pi)^3 \delta^3({\bf k} - {\bf k}')  8E_{k} (1 - h_{\nu}) \int_{p} (1 - \hat{\bf p} \cdot \hat{\bf k}) \rho_{e, p}  \ , \nonumber \\
= & 2a\sqrt{2} G_{\rm F}  \int { {d{\bf p}} \over{(2 \pi)^3}}(1 - \hat{\bf p} \cdot \hat{\bf k}) \rho_{e, p} \ , 
\end{align}
with\footnote{Note that the factor $a$ goes away when one calculates the neutrino mean-field evolution equations. It ensures that the neutrino momenta are unchanged and that neutrinos have the right helicity. } $a = (2 \pi)^3 \delta^3({\bf k} - {\bf k}')  E_{k} (1 - h_{\nu}) $.
If the medium is isotropic the angular dependence in Eq.\eqref{eq:Vrhoe} averages out. 

Since the total number of electrons in the medium is given by
\begin{align}\label{eq:Ne}
N_e \equiv & \sum_{h_e} \int_{p} \langle a^{\dagger}_e ({\bf p}, h_e)  a_e({\bf p}, h_e)  \rangle = 2 V \int {d{\bf p} \over{(2 \pi)}^3} \rho_{e, p}  \ ,
\end{align}
Eq. \eqref{eq:Vrhoe} becomes
\begin{align}\label{eq:MFee}
V_{\rm CC}(n_e) = \sqrt{2} G_{\rm F}  n_e \ , 
\end{align}
where ${n}_e = N_e/V$ is the electron number density. This is the well-known mean-field Hamiltonian responsible for the MSW effect in matter Eq. \eqref{eq:Hmsw}. 

We now consider a $\bar{\nu}$ propagating in an electron background. When the same procedure is used, the derivation of the mean-field contribution requires
the calculation of the matrix element \eqref{eq:ME1}, this time involving the $b$ and $b^{\dagger}$ operators. 
This introduces a negative sign arising from anticommuting the antiparticle operators.
One gets then the well-known potential
\begin{align}\label{eq:MFeean}
V_{\rm CC}(n_e) = - \sqrt{2} G_{\rm F}  n_e \ ,
\end{align}
with an opposite sign with respect to the neutrino case.
Similarly, one can determine that the potential for $\bar{\nu}_e$-positron ($\nu_e$-positron) scattering has a positive (negative) sign.

Following the procedure just outlined, one can derive any mean-field contribution to the neutrino Hamiltonian, such as those coming from $\nu\nu$ interactions or from NSIs. 

\subsection{Beyond the usual mean-field}
The mean-field equations \eqref{eq:LVN} and \eqref{eq:MF} with vacuum mixings, the standard charged- and neutral-current $\nu$-matter and $\nu\nu$ interactions respectively have been widely used in studies of flavor evolution in dense astrophysical environments. More than a decade ago, 
some explored extensions of such equations to establish the robustness of the mean-field approximation and the possible necessity to go beyond.

We first mention that, as pointed out by several researchers, the Hamiltonian with $\nu\nu$ interactions is analogous to an interacting system of spins that have a spin-exchange interaction
and feel an external magnetic field.
Following \textcite{Balantekin:2006tg}, one can introduce the operators\footnote{The indices here indicate $\nu_e$ or $\nu_x$. The flavor (iso)spin operators in the Cartesian basis $(x, y, z)$ are related to the ones in cylindrical basis $(+, -, z)$ as usual, i.e. $J_{\pm} = J_x \pm i J_y$. }  
\begin{align} \label{eq:J}
J^+_{\bf p}  & = a_x^\dagger({\bf p}) a_e({\bf p}) ~~
J^-_{\bf p}   =  a_e^\dagger({\bf p}) a_x({\bf p}) \\ \nonumber
J^0_{\bf p}  & =  \frac{1}{2}\left[a_x^\dagger({\bf p})a_x({\bf p})-a_e^\dagger({\bf p})a_e({\bf p}) \right] \ , 
\end{align}
which satisfy the commutation relations
\begin{align} \label{eq:SU2}
[J^0_{\bf p},J^\pm_{\bf q}]  & =   \pm \delta^3({\bf p}-{\bf q})J^\pm_{\bf p} \\ \nonumber
[J^+_{\bf p},J^-_{\bf q}]  & =   2 \delta^3({\bf p}-{\bf q})J^0_{\bf p} \ , 
\end{align} 
and are the generators of as many SU(2) algebras as the number of distinct neutrino momenta.

The neutrino Hamiltonian for 2$\nu$ can be written as \cite{Balantekin:2006tg,Pehlivan:2011hp}
\begin{align}\label{eq:HJs}
H & =  \sum_{\omega} \omega { \rm {\bf B}}_{\rm vac} \cdot {\bf J}_{\omega} + \mu \sum_{{\bf p}, {\bf q}} (1 - \cos\theta_{{\bf p}{\bf q}})  {\bf J}_{\bf p} \cdot {\bf J}_{\bf q}  \ . 
\end{align} 
The first term in Eq.\eqref{eq:HJs} is the vacuum contribution that depends on the effective vacuum magnetic field ${\bf B}_{\rm vac}$ (see Appendix A) and ${\bf J}_{\omega}$, the flavor (iso)spin operator of the neutrino with the vacuum oscillation frequency $\omega$. The second term in Eq.\eqref{eq:HJs} is the $\nu\nu$ interaction term in the mean-field approximation, where neutrinos either keep or exchange their momenta. The $\nu\nu$ coupling is $\mu = \sqrt{2} G_{\rm F} /V$, while $\theta_{{\bf p}{\bf q}}$ is the angle between ${\bf p}$ and ${\bf q}$.
The matter term is neglected in Eq.\eqref{eq:HJs}. 

\textcite{Balantekin:2006tg} provided a derivation of the mean-field equations using the algebraic formulation of the neutrino Hamiltonian \eqref{eq:HJs}. 
They used a coherent-state path-integral approach and showed that the mean-field equations correspond to the saddle-point approximation of the path-integral for the full many-body system. Moreover, they identified contributions beyond the mean-field as corrections to the saddle-point solution.  

Later \textcite{Volpe:2013uxl} used the BBGKY hierarchy to derive mean-field equations\footnote{In the context of atomic nuclei, the neutrino mean-field equations correspond to the so-called time-dependent Hartree-Fock approximation. If the initial state for the many-body system is a Slater determinant, it remains a Slater determinant at all times \cite{Ring:2004}.} for the $\nu$ and $\bar{\nu}$ one-body density matrices \eqref{eq:rho1} in a dense astrophysical environment. 
Moreover, thanks to the hierarchy, \textcite{Volpe:2013uxl} pointed out that the neutrino evolution equations had further terms at the mean-field level, namely, two-point correlators from wrong-helicity contributions due to neutrino masses and from pairing (or abnormal) densities\footnote{\textcite{Sigl:1992fn} mentioned such correlations but discarded them. Note that neutrino-antineutrino correlations were included in the neutrino evolution equations in the context of baryogenesis via leptogenesis by \textcite{Fidler:2011yq}.}. For Dirac neutrinos the latter read
\begin{eqnarray} \label{e:kappa}
\kappa_{ij}(t,{\bf q},h,{\bf q}',h') &= \langle b_{j}(t,{\bf q'},h')   a_{i} (t,{\bf q},h)  \rangle \  ,  
\end{eqnarray}
and their Hermitian conjugates. 
\textcite{Volpe:2013uxl} showed that in the presence of pairing correlators one can cast the extended mean-field evolution equations, as in Eq. \eqref{eq:LVN}, as
\beq\label{eq:evgen}
i \dot{\cal R} = [{\cal H},{\cal R}]  \ , 
\eeq
by introducing the quantities ${\cal H}$ and ${\cal R}$ which are the generalized Hamiltonian and density. These includes both the $\nu$ and $\bar{\nu}$ contributions and have a $2 N \times 2 N$ matrix form; see \textcite{Volpe:2013uxl} for their explicit expressions. Such a system of $\nu$ and $\bar{\nu}$, with pairing correlators,
can be described in terms of independent quasiparticles, by introducing a Bogoliubov transformation \textcite{Vaananen:2013qja}.

Besides pairing correlators, wrong-helicity contributions have also been discussed.
By the early 1990s, \textcite{Rudzsky:1990} had already derived quantum kinetic equations 
for Wigner distribution functions and included both neutrino flavor and spin. 
In their derivation of neutrino quantum kinetic equations, \textcite{Vlasenko:2013fja} pointed out contributions from the correlators   
\beq\label{eq:coherence}
\zeta_{ij}(t,{\bf q})=\langle a^\dagger_j(t,{\bf q},+)a_i(t,{\bf q},-)\rangle \ ,
\eeq
and named them spin coherence. They are due to the neutrino mass and suppressed by the factor $m/E$. 

 \textcite{Serreau:2014cfa}  derived the most general mean-field equations for inhomogeneous and 
 anisotropic media while considering Dirac as well as Majorana neutrinos.
Such equations included contributions either from pairing or from wrong-helicity correlators referred to as helicity (instead of spin) coherence. Using the approach of \textcite{Serreau:2014cfa}, \textcite{Kartavtsev:2015eva} also included contributions from neutrino electromagnetic properties. 

Spin or helicity coherence requires anisotropy of the medium to be nonzero. 
The corresponding generalized equation \eqref{eq:evgen} can again be cast in the form Eq.\eqref{eq:LVN} but with the generalized Hamiltonian having both flavor and helicity structure \cite{Rudzsky:1990,Vlasenko:2013fja,Serreau:2014cfa}.
Helicity coherence couples $\nu$ with $\bar{\nu}$, i.e., active and sterile neutrinos if $\nu$ are Dirac particles, or neutrinos and antineutrinos if $\nu$ are Majorana particles. 

The impact on flavor evolution of the supplementary terms from the correlators \eqref{e:kappa} and \eqref{eq:coherence} has been investigated as well.
\textcite{Kartavtsev:2015eva} pointed out that the pairing correlators do not influence flavor because the large kinetic contributions cannot be removed\footnote{Usually, the diagonal contributions, proportional to the identity matrix are subtracted and do not impact neutrino flavor.}. This fact makes the influence of the off-diagonal contributions from pairing correlators small.  

Concerning helicity coherence, simulations in $3 \nu$ flavors with detailed astrophysical inputs from binary neutron star merger remnants \cite{Chatelain:2016xva} or supernovae \cite{Tian:2016hec} showed that nonlinear feedback does not operate in detailed settings; see Sec. II.F.2. As a consequence, according to current understanding helicity coherence does not seem to influence the neutrino flavor in media.    

\subsection{Linearization}
Linearization is a widespread approach. It is used in many domains of physics, such as in nuclear physics, condensed matter 
and hydrodynamics. 
In the linearization procedure, which is a small amplitude approximation, one solves eigenvalue equations that are numerically more tractable than the full nonlinear problem.

The first application of linearization to the neutrino mean-field equations in a supernova was by \textcite{Sawyer:2008zs}. Afterward,
\textcite{Banerjee:2011fj} derived a linearized version of the equations of motion in the bulb model. The procedure has since 
been widely employed in the study of both slow and fast modes. \textcite{Vaananen:2013qja} provided an alternative derivation of the linearized equations, by generalizing the random-phase-approximation (RPA) commonly used in studies of atomic nuclei and metallic clusters. Subsequently \textcite{Izaguirre:2016gsx} recast the linearized equations in a dispersion-relation approach. 

\subsubsection{The linearized equations}
We now have a closer look at the linearized version of the equations of motion for supernova neutrinos.
Here we follow \textcite{Banerjee:2011fj} and consider the bulb model, which includes neutral-current $\nu\nu$ interactions; see Sec. II.G.1.
 For 2$\nu$ flavors, the neutrino flux matrices can be rewritten as  
\beq\label{eq:S}
\rho_{\ell, r} = {\phi_{\nu_e, \ell, r} + \phi_{\nu_x, \ell, r} \over 2} + g_{\ell}
 \Big(
\begin{tabular}{cc}
$s_{\ell, r}$  & $ S_{\ell, r} $   \\
$S^*_{\ell, r} $   & - $ s_{\ell, r}  $    \\ 
\end{tabular}
\Big) \ ,
\eeq
where $\ell = ({\omega, u}) $. The last matrix on the rhs of Eq.\eqref{eq:S}, which is referred to as ${\bf S}_{\ell, r}$ is an Hermitian matrix and its 
prefactor 
\beq\label{eq:gfac}
g_{\ell} = {(\phi^0_{\nu_e, \ell} - \phi^0_{\nu_x, \ell}) \over 2 } \ .
\eeq
depends on the fluxes at the neutrinosphere.
The $s_{\ell, r}$ factor represents, for a given mode, how much the flavor content has changed with respect to the initial condition.
The elements of  ${\bf S}_{\ell, r}$ satisfy 
the normalization condition $s^2_{\ell, r} + \vert S_{\ell, r} \vert^2 = 1$. The quantities
$(1+s)/2$ are the survival probabilities. 
\begin{figure}
\begin{center}
\includegraphics[width=.4\textwidth]{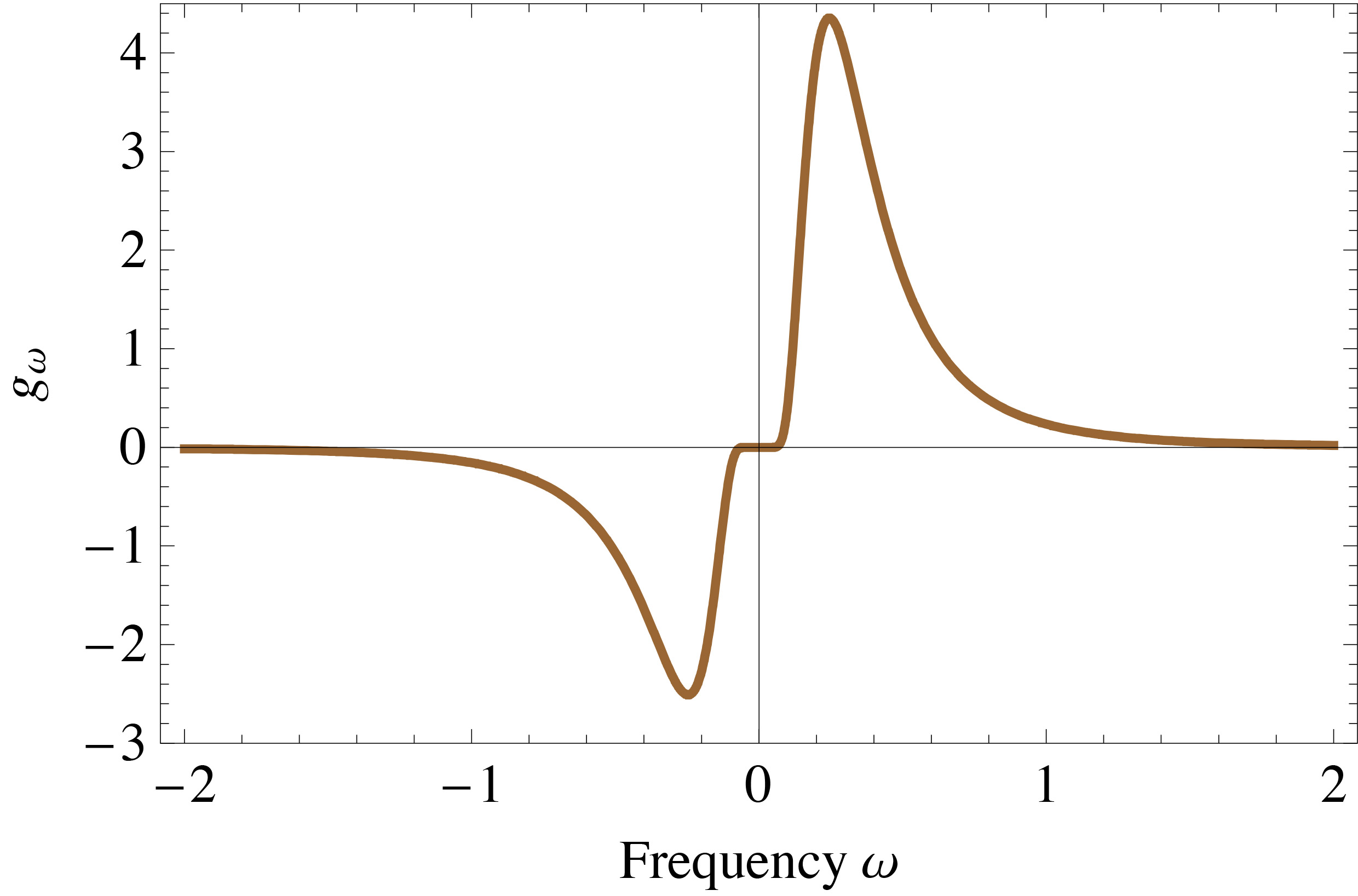}
\includegraphics[width=.42\textwidth]{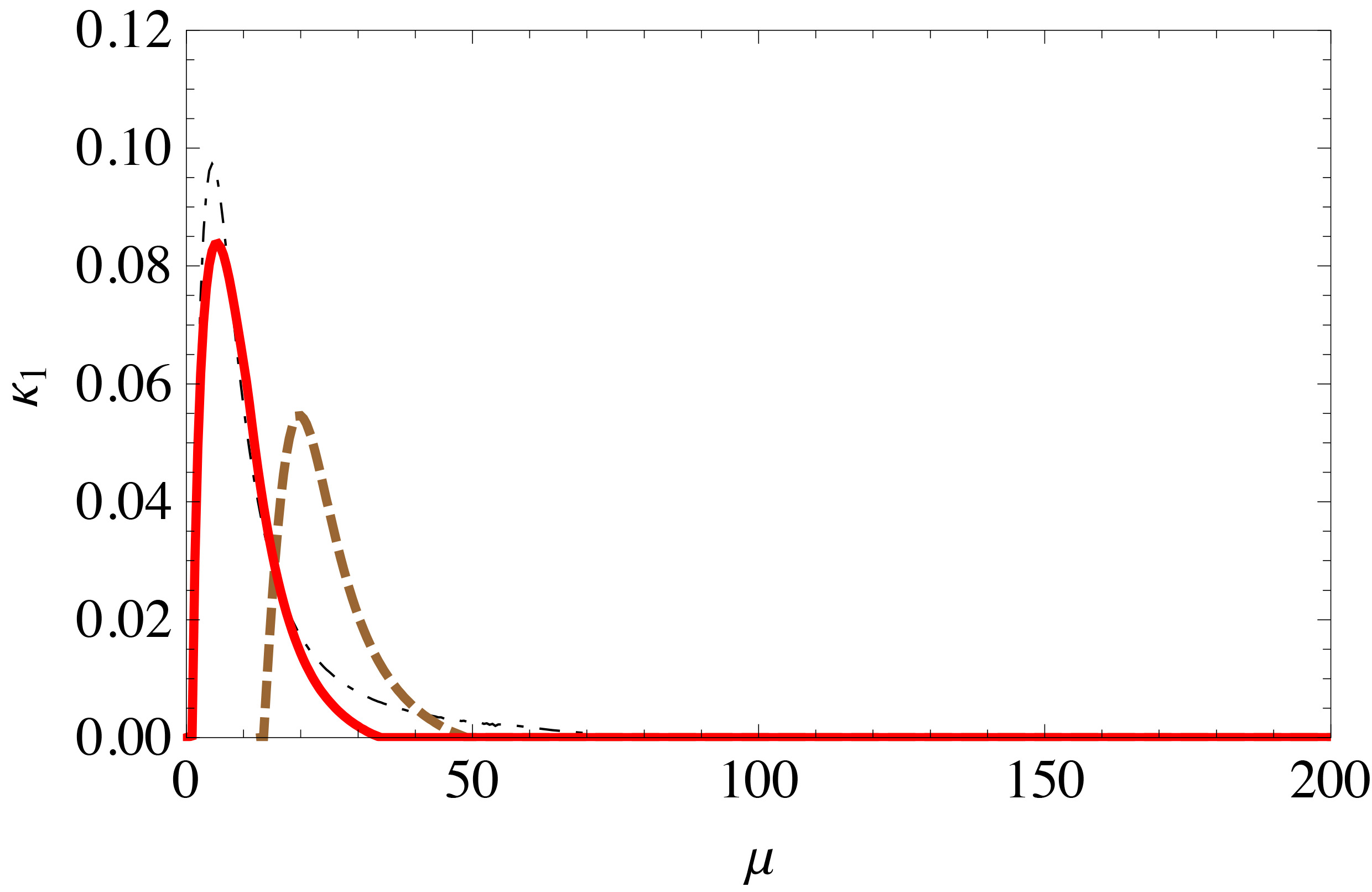}
\caption{Linearization. Upper panel: example of the solution of the eigenvalue equations for neutrino Fermi-Dirac distributions at the neutrinosphere. The positive frequencies correspond to neutrinos, whereas the negative ones correspond to antineutrinos. Lower panel: the quantity $\kappa_1$, as a function of the $\nu\nu$ interaction coupling constant $\mu$, is the imaginary part of one of the two unstable solutions.  Adapted from \textcite{Banerjee:2011fj}.}
\label{fig:linearised}
\end{center}
\end{figure}
The indices $(\omega, u,$ and $r$) are the vacuum oscillation frequency,
$u = \sin^2 \vartheta_R$ with $u \in [0,1]$ characterizes the neutrino emission at the neutrinosphere (at $r = R$), and
$r$ is the distance defining the $\nu\nu$ intersection point along the symmetry direction; see Fig. \ref{fig:bulb}.

In the linearization procedure one considers the initial state and performs small variations around it. 
The initial states at the neutrinosphere are in our case
${\bf S}_{\ell, R} = {\rm diag} (1, -1)$ and $\rho_{\ell, R} = {\rm diag} (\phi^0_{\nu_e}, \phi^0_{\nu_x})$ Eq. \eqref{eq:S}. 
The small amplitude approximation consists in the following assumption
\beq\label{eq:RPA}
s_{\ell, r} \simeq 1 ~~~~ \vert  S_{\ell, r}  \vert \ll 1 .   
\eeq
Under this hypothesis, the neutrino mean-field equations \eqref{eq:LVN} with the contributions from vacuum equation \eqref{eq:Hvac}, matter equation \eqref{eq:Hmswfull}, and $\nu\nu$ interaction equation \eqref{eq:nunu} become, when $r \gg R_{\nu}$,
\begin{align}\label{eq:lin}
i \partial_r S_{\ell, r} = ~& [\omega~ + ~u (\lambda_r ~ + ~\epsilon \mu_r)] ~ S_{\ell, r} \nonumber \\
~~& - \mu_r \int_0^1 d u' \int_{- \infty}^{+ \infty}  d \omega' (u + u')  g_{\ell'} S_{\ell', r} \ ,
\end{align}
with the positive (negative) frequencies corresponding to neutrinos (antineutrinos); see Fig. \ref{fig:linearised}. 
In Eq.\eqref{eq:lin} the third term on the rhs is the total lepton number 
$\epsilon = \int~ du ~ d\omega ~g_{\ell}$ (normalized to $\phi^0_{\bar{\nu}_e }$). The second and last terms on  the rhs of Eq.\eqref{eq:lin} are the matter and $\nu \nu$ terms with the following coefficients:
\beq\label{eq:lambdamat}
\lambda_r = \sqrt{2} G_{\rm F} (n_e - n_{\bar{e}}) {R^2\over {2 r^2}} ~~~ \mu_r = { \sqrt{2} G_{\rm F}  \phi^0_{\bar{\nu}_e }R^2\over {8 \pi r^4}} \ .
\eeq

One seeks for solutions of Eq. \eqref{eq:lin} of the type
\begin{align}\label{eq:exp}
S_{\ell, r} = Q_{\ell, r} e^{- i \Omega r} \ , 
\end{align}
which leads to the eigenvalue equations 
\begin{align}\label{eq:eigen}
[\omega ~ + ~ u (\lambda_r ~ + ~\epsilon \mu_r) -  \Omega] Q_{\ell, r}  & =  \nonumber \\ 
- \mu_r \int_0^1 d u' \int_{- \infty}^{+ \infty}   d \omega' (u + u') & g_{\ell'} Q_{\ell',r}  \ .
\end{align}
If the eigenvalue $\Omega \in {\Re }$, the initial condition is stable and the system performs small oscillations around it. 
If $\Omega \in {\cal C}$, one faces an instability in flavor space\footnote{Note that the linearized equations admit the pair ($\Omega$, $\Omega^*$) as solutions.}: the system deviates exponentially from the initial state. This is often called a runaway solution. 
Figure \ref{fig:linearised} gives an example of the application of linearized equations in a supernova context.

Therefore, a complex eigenvalue indicates the start of flavor modification when neutrinos depart from the neutrinosphere.
However, we emphasize that linearization does not provide any information on the full nonlinear regime. Indeed the linearized equations are inherently based on the small amplitude approximation \eqref{eq:RPA}.
Only the full numerical solution of the equations of motion tells us how significant flavor conversion is at large scales.

In the study of atomic nuclei or metallic clusters, linearized equations are obtained with the RPA.
With this approach one determines small variations of the matter density around the initial state. If the eigenvalues 
are real, this indicates that the initial state is a true ground state; 
if they are complex, then the initial state is not a ground state of the system. The latter situation is, in fact, what
one looks for, in the neutrino case. In addition, in the RPA, one can face "spurious"
solutions that are numerical artifacts; see \cite{Ring:2004}. These were also found in the neutrino
context when for example multiangle calculations of the $\nu\nu $ interaction do not include a sufficiently
large number of angle bins \cite{Sarikas:2012ad}. 
 
Here is how \textcite{Vaananen:2013qja} generalized RPA to neutrinos. 
At initial time, the density matrices $\varrho $ and $\bar{\varrho}$ correspond to a stationary state
\beq\label{eq:rhoini}
[h^{0},\varrho^{0}]= 0 ~~~~~~[\bar{h}^{0},\bar{\varrho}^{0}]= 0 \ .
\eeq 
Since the RPA is a small amplitude approximation, one performs small variations $\delta \varrho(t)$ of the density\footnote{A similar approach is taken for $\delta \bar{\varrho}$.} around $\varrho^{0}$
\beq\label{e:rho1}
\delta \varrho = \varrho^0 + \delta \varrho(t) = \varrho^{0} +  \varrho'e^{-i\Omega t} +  \varrho'^{\dagger} e^{i\Omega^* t} .   
\eeq
where $\varrho'$ stands for the off-diagonal terms of the density matrix.
The mean-field Hamiltonian around this solution changes accordingly
\beq\label{e:linh}
h(\varrho) =  h^{0} + {\delta h \over {\delta \varrho}}\Big|_{\varrho^{0}}\delta \varrho + \ldots ~~~~ \bar{h}(\bar{\varrho}) =  \bar{h}^{0} + {\delta \bar{h} \over {\delta \bar{\varrho}}}\Big|_{\bar{\varrho}^{0}}\delta \bar{\varrho} + \ldots \ .
\eeq
By implementing Eq. \eqref{e:rho1} and retaining terms at the lowest order, one obtains linearized equations of motion\footnote{A detailed derivation and explicit expressions for the $A, B, \bar{A}, \bar{B}$ matrices were given by \textcite{Vaananen:2013qja}.}
that can be cast in the following matrix form (known in the RPA)\footnote{In the application of the RPA approach to atomic nuclei, the initial state is the nucleus ground state, while the variations around it determine the excited states. The quantities $\rho'  $ and $\bar{\rho}'$ are called the forward and backward amplitudes and correspond to particle-hole and hole-particle excitations. The RPA and its numerous variants (for example 
quasiparticle RPA, continuum RPA, second RPA) are used to study the excited states of atomic nuclei, for example. the giant resonances, or to calculate the transition matrix elements of single $\beta$, 2$\beta (2 \nu)$ and  2$\beta (0 \nu)$ decay.}:
\begin{align}\label{eq:linmatrix}
\left(
\begin{array}{ll}
A & B \\
\bar{B} & \bar{A} \\
\end{array} 
\right) \left(
\begin{array}{l}
\varrho'  \\
\bar{\varrho}' \\  
\end{array}
\right) 
 & = \Omega
\left(
\begin{array}{l}
\varrho'  \\
\bar{\varrho}' \\  
\end{array}
\right). 
\end{align}
Note that linearized equations were also cast in a similar matrix form by \textcite{Sarikas:2012ad}. 
The condition for the applicability of linearized equations by \textcite{Vaananen:2013qja} is that the system satisfies Eqs.\eqref{eq:rhoini} at initial time. This can hold not only in the flavor basis but also in the matter basis. In the latter case, in principle, one could apply the linearized equations
at a time later than the initial time at the neutrinosphere (as is usually done).

\subsubsection{A dispersion-relation approach}
A third formulation of the linearized equations was suggested by \textcite{Izaguirre:2016gsx} for fast modes. They cast them in a dispersion-relation approach where the neutrino modes are neutrino flavor (iso)spin waves, described by a four-vector $c$ and a "polarization" vector, in the matter and the neutrino backgrounds. Instead of Eq.\eqref{eq:eigen} one seeks plane waves (in an homogeneous and stationary background) 
\beq\label{eq:pw}
S_{{\bf v}}(t, {\bf r}) = Q_{\bf v} (\Omega, {\bf K}) e^{- i (\Omega t - {\bf K} \cdot {\bf r})} \ .
\eeq
After linearizing the mean-field equations \eqref{eq:LVN}, one gets the following equation
\beq\label{eq:lin2}
v^{\mu}k_{\mu} Q_{\bf v} = - \int {d{\bf v} \over {4 \pi}} v^{\mu} v'_{\mu} G_{\bf v}  Q_{\bf v'} 
\eeq
where  $v^{\mu} = (1, {\bf v})$ and $(\Omega, {\bf K})$ is replaced\footnote{Going to a comoving frame to get rid of the background contribution.} by $(\omega, {\bf k})$. By considering
\beq\label{eq:Qn}
Q_{\bf v} = a^{\mu}k_{\mu}/ v^{\mu}k_{\mu} \ ,
\eeq
\begin{figure}
\begin{center}
\includegraphics[scale=0.4]{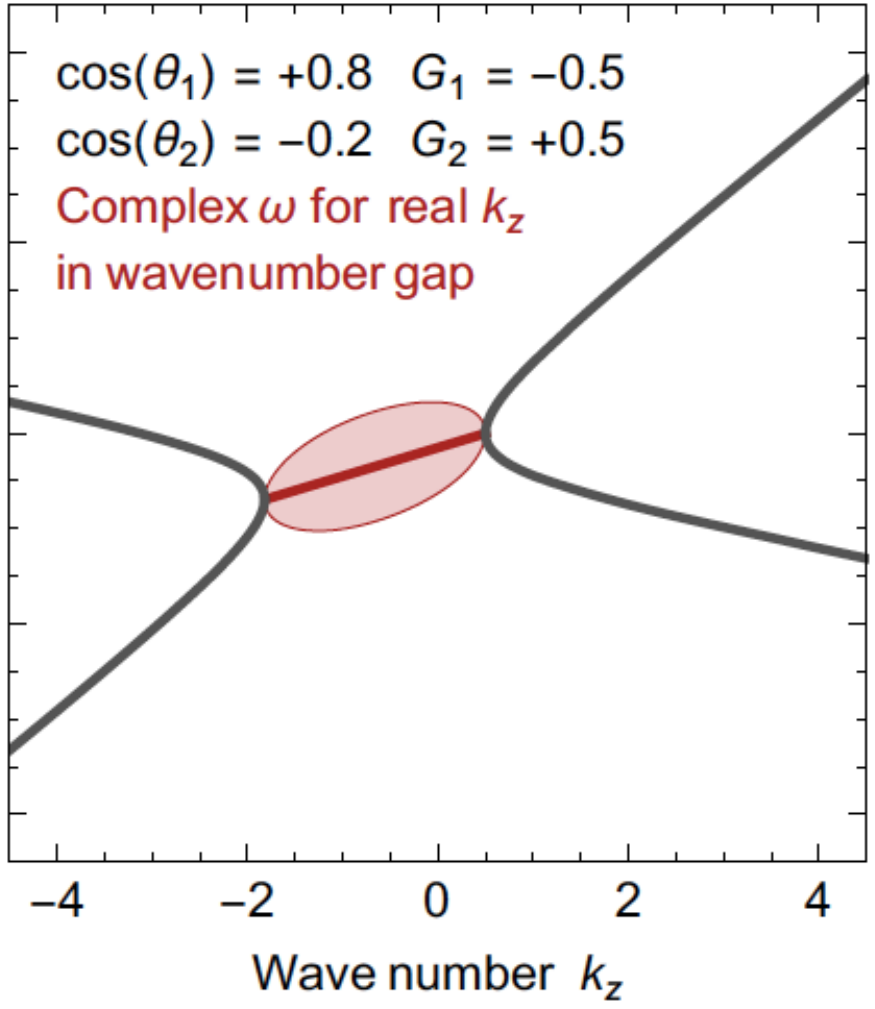}
\caption{Dispersion-relation approach: results for a two-beam model with angular modes $G_1$ and $G_2$. The red blob corresponds to complex values of $\omega$ for real ${\bf k}$ with ${\bf k} = (0,0,k_z)$ for which fast modes grow in time (temporal instability). Adapted from \textcite{Izaguirre:2016gsx}.}
\label{fig:disrel}
\end{center}
\end{figure}

\noindent
one can recast Eq. \eqref{eq:lin2} as a dispersion relation $\omega = \omega (k)$ as
\beq\label{eq:Pi}
\Pi^{\mu\nu} a_{\nu} = 0
\eeq
while seeking nontrivial solutions such that ${\rm det}[\Pi^{\mu\nu}] = 0$.
The "polarisation tensor" in Eq. \eqref{eq:Pi} reads
\beq\label{eq:PiTensor}
\Pi^{\mu\nu} = \eta^{\mu\nu} + \int {d {\bf v} \over {4 \pi}} G_{\bf v} {v^{\mu}v^{\nu} \over {\omega - {\bf v} \cdot {\bf k}}} \ , 
\eeq
with $\eta^{\mu\nu} = {\rm diag} (+,-,-,-)$.  
Equation \eqref{eq:Pi} is a quartic equation in $\omega$ whose roots give four possible dispersion-relation branches: (a) $(\omega, {\bf k}) \in {\Re }$ for stable solutions, (b) $\omega $ or ${\bf k}$ $ \in {\cal C}$ for an unstable solution that grows in time, or space (Fig. \ref{fig:disrel}), and (c) $\omega, {\bf k} \in {\cal C}$ for a mode growing in space and time.

A classification of instabilities was suggested by \textcite{Capozzi:2017gqd} following the one for plasmas by \textcite{Sturrock:1958zz} and \textcite{Briggs:1964}. According to this classification, the linear instabilities can be any of four categories: the {\it completely stable} or {\it stable with damping} instabilities correspond to spatially stable modes,  whereas the {\it absolute} and  {\it convective} instabilities grow either 
temporally (while still embracing the point of its origin) or spatially, moving away from its origin while growing in amplitude and extent. Note that many flavor evolution studies (for slow or fast modes) have been performed by evolving either space or time. 

Among the applications was the study by \textcite{Martin:2019gxb} in a dynamic one-dimensional model. They 
showed that fast modes evolve coherently in the nonlinear regime in time and space, when the corresponding ELN crossings undergo absolute or convective instabilities in the linear regime.
The work by \textcite{Capozzi:2017gqd} was deepened by \textcite{Yi:2019hrp} who pointed out that the complex dispersion-relation branches are bounded by critical points and emphasized their usefulness in identifying fast modes. 

\subsection{Neutrino quantum kinetic equations}
The study of the interplay between collisions and flavor modes is numerically challenging. Thus its investigation is still in the early stages.
First it requires a consistent theoretical framework where one goes from the dense collision-dominated regime to dilute regions in which mean-field equations are sufficient.
Second, the dimensionality remains high in astrophysical environments, whereas in the early Universe isotropy and homogeneity reduces the dimensionality of the problem, making it numerically tractable. 

Numerous researchers have derived neutrino quantum kinetic equations (QKEs) for the early Universe (\cite{Stodolsky:1986dx,Rudzsky:1990,Sigl:1992fn,McKellar:1992ja,Vlasenko:2013fja,Blaschke:2016xxt,Froustey:2020mcq}; 
see also the review by \textcite{Volpe:2015rla}). Such QKEs are being used for 
the study of neutrino flavor evolution in dense astrophysical environments. 

\textcite{Rudzsky:1990} gave an early derivation of the Boltzmann equation for neutrino Wigner distribution functions including neutrino mixings and the MSW potential.
He considered distribution functions as a matrix in the neutrino spin and isospin space, allowing neutrino oscillations in both flavor and spin. The latter become zero when the neutrino rest mass is zero. 

\textcite{Sigl:1992fn} derived QKEs for a matrix of densities for the early Universe while implementing antineutrinos for the first time.
They included (anti)neutrino scattering on neutrons, protons, electrons,
positrons, $\nu$ and pair annihilation in a perturbative approach using the assumption of {\it molecular chaos}. 
In this approximation, the collision timescale is short enough that correlations do not develop between collisions: 
the incoming and outgoing particles in the collision integrals are free single-particle states.

\textcite{Vlasenko:2013fja} gave an alternative derivation of the flavored quantum Boltzmann equations for Majorana 
neutrinos using the closed time path (or "in-in") formalism and the two-particle-irreducible 
effective action. \textcite{Blaschke:2016xxt} extended their results and obtained the full collision term for neutrinos for anisotropic media.
Extending the work  by \textcite{Volpe:2013uxl} done in the context of supernova neutrinos, \textcite{Froustey:2020mcq} rederived 
the neutrino quantum kinetic equations with the BBGKY hierarchy and found that the full collision term agreed with that 
of \textcite{Blaschke:2016xxt}.

While the aforementioned works use different theoretical frameworks and approximation schemes, the neutrino QKEs
obtained by different researchers agree when common assumptions are made. 
Using density matrices, the general form of QKEs,  including collision terms (proportional to $G_{\rm F} ^2$) and mean-field contributions (linear in $G_{\rm F}) $, reads 
\begin{align}\label{e:bolts}
i (\partial_t + \mathbf{v} \cdot \mathbf{\nabla}_{\bf x}) \varrho_{\bf x, p} = [{\cal H}_{\bf x, p}, \varrho_{\bf x, p}] + i C[\varrho, \bar{\varrho}] \ , 
\end{align}
and likewise for $\bar{\varrho}_{\bf x, p}$.  In Eq.\eqref{e:bolts} ${\cal H}_{\bf p}$ includes the mixings and the mean-field Hamiltonian describing neutrino propagation in either an astrophysical environment or the early Universe\footnote{Note that for the latter the mean-field Hamiltonian includes extra contributions from the gauge boson propagators; see \cite{Sigl:1992fn,Froustey:2020mcq}.}. The last term in Eq.\eqref{e:bolts} is the collision term $C$.
While the first term on the rhs of Eq.\eqref{e:bolts} usually introduces coherence, the second term is responsible for the production and absorption of neutrinos and of kinematical decoherence among neutrinos with different momenta.   \\

\noindent
{\it A separation of scales?}\\
 In dense matter, an important length scale is the neutrino mean free path, $\lambda = (\sigma \rho)^{-1}$, with $\rho$ the matter number density and $\sigma$ the interaction cross section of a neutrino with a particle of the medium. Close to matter saturation density at $\rho =  3 \times 10^{14}$ g/cm$^{3}$ and for a typical cross section $\sigma = 6 \times 10^{-41}$ cm$^{2}$, a 10 MeV neutrino has a mean free path of about a meter.  
Depending on the neutrino energy (and flavor), the idealized location where neutrinos start free streaming in a supernova core is at densities of about $\rho = 10^{10}$-$10^{12}$   g/cm$^{3}$. At $\rho = 10^{10}$ g/cm$^{3}$, $\lambda$ is of tens of kilometers.

The flavor length scale is another important quantity. For a long time, the MSW resonance has provided the only flavor length scale in flavor studies. Considering for example typical supernova matter profiles, the $H$-resonance is located at approximately $\rho_{\rm res} \approx 10^{3}$ g/cm$^{3}$, whereas the $L$-resonance is at about $\rho _{\rm res} \approx= 1$ g/cm$^{3}$ (for a 10 MeV neutrino). If the MSW effect were the only flavor phenomenon, the mean free path and the flavor length scales would be well separated. 

For many years, this argument has supported the use of two distinct theoretical treatments, one for the dense region, where the particles of the medium act as a random scatterer, and one for the dilute region, where the neutrinos free stream. This translates, formally, in the use of relativistic Boltzmann transport equations \cite{Lindquist:1966,Bruenn:1985en,Cardall:2013kwa} for the former and of mean-field equations for the latter.  

It is common practice numerically to simplify the seven-dimensional transport neglecting, in particular, the mixing and mean-field terms. However, flavor studies usually separate the two regions of applicability and treat the neutrinosphere as an idealized sharp surface, although the neutrino decoupling region\footnote{This is conventionally defined as the surface where the optical depth is 2/3.} is  built up by collisions and is energy and flavor dependent. 

Since the work of \textcite{Duan:2006an}, more than 15 years of investigation has introduced new scales in the problem showing that
the separation of the collision and the flavor scales does not necessarily hold in the presence of neutrino-neutrino interactions. This is particularly true in presence of fast modes, which take place close to the neutrinosphere and have wavelengths shorter than the collision one. 

\subsection{Flavor instabilities in the presence of collisions}
In the supernova context, studies  that include collisions as well as contributions from the neutrino mixings and the mean field have started. 
\textcite{Richers:2019grc} studied the influence of collisions on slow modes. But it is the influence of collisions on fast modes, in particular, that is being investigated because of their short scale; see \textcite{Capozzi:2018clo,Hansen:2022xza,Shalgar:2020wcx,Martin:2021xyl,Johns:2021qby,Padilla-Gay:2022wck}.

Collisions usually introduce decoherence and are thought to damp flavor conversion. However things are more subtle since collisions also
contribute to determine the neutrino angular distributions that are the key for fast modes. 
There are now two examples indicating that collisions might enhance flavor conversion: scattering-enhanced fast modes and collisional instabilities.

\textcite{Capozzi:2018clo} highlighted the notion that collisions can trigger fast modes. 
\textcite{Martin:2021xyl} studied, in the linear and nonlinear regimes, an inhomogeneous gas model with monoenergetic neutrino undergoing direction changing $\nu$-nucleon elastic scattering. They showed that collisions can make ELN distributions tend asymptotically isotropic as a function of time, thus suppressing or killing fast modes. \textcite{Hansen:2022xza} found that collisions can enhance or suppress fast flavor conversions, depending on the hypothesis that neutrino emission is close to isotropic, or forward peaked. \textcite{Johns:2022bmu} argued that  inhomogeneous models are better suited than homogeneous ones when one investigates the role of collisions on fast modes. 
The importance of inhomogeneities was also stressed by \textcite{Sigl:2021tmj}.
Indeed neutrino angular distributions, which are crucial in triggering fast modes, are determined by advection and inhomogeneity on one hand, and collisions on the other. 

Moreover, collisional instabilities have emerged. As first pointed out by \textcite{Johns:2021qby}, collisional instabilities are a new kind of flavor instability.
In the transition region where neutrinos are still partially coupled to the medium,  
a difference between the collision rates of neutrino and antineutrinos can cause neutrino coherence to grow 
and trigger an instability on a microsecond timescale. If neutrino angular distributions are made anisotropic, the interplay of
fast modes and collisions produces a collisional instability with even shorter timescales. 

In an homogeneous, axisymmetric model, \textcite{Padilla-Gay:2021haz} showed that fast modes evolve as a gyroscopic pendulum in flavor
space analogously to slow modes. In a subsequent work \textcite{Padilla-Gay:2022wck} added collisions in the damping approximation. 
They showed that the system tends to an asymptotic state, identified by a universal relation that depends only on the largest departure 
of the flavor pendulum from its initial orientation. Moreover the results confirm the presence of collisional instabilities when the neutrino and
antineutrino collision rates are unequal. Progress toward the implementation of collisions in models of increased complexity is ongoing. 
 
Note that steps toward a consistent solution of full QKEs were also made in the early Universe.   
Indeed, for the off-diagonal contributions of the collision term, the damping approximation was extensively used; see \textcite{Dolgov:2002ab,Mangano:2005cc,Gava:2010kz}. 
\textcite{Froustey:2020mcq} and then \textcite{Bennett:2020zkv} recently performed the first calculations with the full collision term, the $\nu$ mixings and the mean-field terms. Also including radiative corrections to the plasma equation of state, this challenging numerical calculation has 
yielded a precise value of the effective number of degrees of freedom $N_{\rm eff} = 3.0440$ at the epoch of primordial nucleosynthesis. 

Moreover \textcite{Cirigliano:2017hmk} investigated the role of anisotropies in a dense neutrino gas with two spatial dimensions, in presence of $\nu\nu$ interactions and collisions, and showed that instabilities are not necessarily suppressed by kinematical decoherence. 
\textcite{Hansen:2020vgm} showed that $\nu$-$\bar{\nu}$ asymmetry can significantly grow due to the nonlinearity of the evolution and can influence $N_{\rm eff}$ in the presence of small anisotropies. 

We emphasize that the computational cost of implementing quantum neutrino transport in detailed core-collapse supernova simulations is high.
As long as the MSW effect or slow modes were the main focus of flavor studies, the argument of separation of scale held as 
previously mentioned, since slow modes 
took place far out in the star's layers to impact the supernova engine; see \textcite{Dasgupta:2011jf}). 
However, unlike slow modes, fast modes can influence supernova dynamics. 
In a full quantum neutrino transport, the treatment of fast modes would require a range of dynamical length scales of many 
orders of magnitude, in addition to a large number of angle and energy bins which are usually necessary to fully track the evolution of neutrino flavor mechanisms. 

More work is needed to achieve a definitive understanding of the impact of collisions on flavor evolution in dense astrophysical environments. In this respect, a unified and consistent solution of the full neutrino QKEs represents a long-term goal. 

\subsection{Neutrinos in the presence of strong gravitational fields}
The theoretical description of neutrino propagation discussed thus far is in flat spacetime. In core-collapse supernovae, accretion disks around black holes or compact binary mergers, there is a compact central object producing a strong gravitational field. Gravity modifies neutrino propagation and impacts flavor evolution. 
To investigate its role, the neutrino equations of motion need to be extended to curved spacetime, as done by \textcite{Cardall:1996cd,Piriz:1996mu,Grigoriev:2004bm,Dvornikov:2013pta,Deaton:2018ser,Chatelain:2019nkf}. 

To date the influence of gravity has received limited attention, although the first works exploring its role date back to the 1980s.
Stodolsky \cite{Stodolsky:1978ks} first considered the problem of finding the quantum mechanical phase acquired by a particle propagating along a classical trajectory in the presence of a gravitational field.  To discuss and compare matter and light interferometry, he introduced the quantum mechanical phase along a path, from spacetime point A to spacetime point B,
\begin{equation}\label{e:phi}
\Phi = \int_A^B m ds,
\end{equation}   
where $m$ is the particle mass. In Eq.\eqref{e:phi} the quantity $ds$ is the infinitesimal line element along the particle worldline
\begin{equation}\label{e:ds}
ds^2 = g_{\mu\nu} dx^{\mu}dx^{\nu},
\end{equation}   
with $g_{\mu\nu}$ the metric tensor and $x^{\mu}$ a coordinate system. The covariant phase equation \eqref{e:phi} can be rewritten as 
\begin{equation}\label{e:phi2}
\Phi = \int_A^B p^{\mu}dx_{\mu},
\end{equation}  
with $p^{\mu}= m  g^{\mu\nu} {dx_{\nu} \over{ds}}$ the particle's canonical momentum.

Afterward, the impact of gravitational fields on the vacuum oscillation phase was widely investigated. In particular 
\textcite{Ahluwalia:1996ev,Ahluwalia:1998jx,Bhattacharya:1999na}, \textcite{Cardall:1996cd,Fornengo:1996ef,Godunov:2009ce,Chatelain:2019nkf} considered
the case of a Schwarzschild metric of a static and spherically symmetric gravitational field, and \textcite{Wudka:2000rf,Lambiase:2005gt,MosqueraCuesta:2017iln,Visinelli:2015uva} 
focused on the Kerr-Newman metric. Note that early works discussed the possibility to separate the contribution to the oscillation phase from the gravitational field, from the one due to the mixings. This does not appear to be observationally possible. Note also that \textcite{Dvornikov:2004jxk} pointed out a new mechanism called spin light that neutrinos emit in the presence of gravitational fields.
  
\begin{figure}
\begin{center}
\includegraphics[scale=0.7]{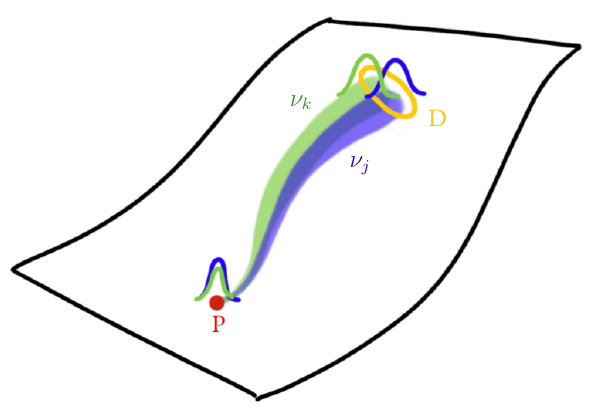}
\caption{Decoherence in curved spacetime: drawing of neutrino wave-packet propagation from a production point P to a "detection" 
point D in the presence of strong gravitational fields. The widths correspond to the trajectory distributions due to the finite wave-packet width.  Each wave-packet associated with a mass eigenstate follows a trajectory close to null geodesics.  From \textcite{Chatelain:2019nkf}.}
\label{fig:decoherence}
\end{center}
\end{figure}

In core-collapse supernovae the role of trajectory bending and energy redshift have been studied by few authors. For example, \textcite{Fuller:1996kt} argued that the difference in the 
gravitational redshift between $\bar{\nu}_e$ and $\nu_e$ can increase the electron fraction and impact $r$-process nucleosysnthesis above the nascent protoneutron star. 
\textcite{Yang:2017asl} found that trajectory bending nearby a compact source in a supernovae produced a neutrino halo similar to the one identified by \textcite{Cherry:2012zw}
due to $\nu\nu$ interactions. \textcite{Caballero:2011dw} provided the full nucleosynthetic outcomes of $r$-process elements in 
black-hole-accretion disks models. Their results showed the importance of the inclusion of trajectory bending and neutrino energy redshift when determining element abundances.

Decoherence is another important aspect of neutrino flavor evolution. Indeed, in a wave packet description of neutrino propagation the wave-packets associated with neutrino mass eigenstates
can decohere thereby suppressing flavor oscillations; see \cite{Giunti:2007ry}. In vacuum one quantifies decoherence by wave packet separation through the "coherence length" $L_{\rm coh} \simeq (4 \sqrt{2} E^2 / \vert \Delta m ^2 \vert)  \sigma_x $ for Gaussian wave packets, with $\sigma_x$ the intrinsic wave-packet dispersion. 
Note that the Daya-Bay experiment investigated the effects of a wave-packet description of vacuum oscillations and set the first limit on its width, finding it not to be significant \cite{DayaBay:2016ouy}. 

For supernova neutrinos, since neutrinos travel over large distances decoherence effects by wave-packet separation can be sizable as discussed e.g. in \textcite{Kersten:2015kio}. \textcite{Akhmedov:2017mcc} investigated such effects in the density matrix formalism and showed that, in vacuum, they induce a damping of the off-diagonal terms.
Moreover they studied decoherence effects in presence of dense matter and neutrino backgrounds for the cases of adiabatic and 
nonadiabatic evolution. 

Extending the formalism to curved-spacetime, \textcite{Chatelain:2016xva} investigated the impact of wave-packet decoherence in a Schwarzschild metric (Fig. \ref{fig:decoherence}). They pointed out that, in curved spacetime, instead of the coherence length a coherence proper time $\tau_{coh} $ should be used to quantify decoherence. This is defined as the time at which the difference between the proper times at a "detection" point D satisfies $\tau = \sigma_t  \sqrt{B(r_D)} $\footnote{Here $B(r) = 1 - r_s/r$ and $r_s = 2 M$ are the Schwarzschild radius and $M$ the mass of the central object.}. With matter and $\nu\nu$ interactions neglected, decoherence was found to produce modifications of the proper time by several tens of a percent.

\subsection{Connections: from atomic nuclei to quantum devices}  
Several researchers 
have unraveled interesting connections between a weakly interacting neutrino gas and other many-body systems \cite{Pehlivan:2011hp,Volpe:2013uxl,Vaananen:2013qja,Mirizzi:2015fva}, or investigated the role of many-body correlations and of entanglement \cite{Bell:2001wk,Bell:2001uw,Bell:2003mg,Friedland:2003dv,Friedland:2003eh,Pehlivan:2014zua,Birol:2018qhx,Lacroix:2022krq,Cervia:2019res,Patwardhan:2021rej,Roggero:2021fyo,Roggero:2021asb,Roggero:2022hpy,Martin:2021bri,Amitrano:2022yyn,Martin:2023ljq}.
Besides being interesting from the theoretical point of view, these studies brought novel ways to approach the problem of 
neutrino propagation in dense environments and opened new numerical treatments, in particular using quantum devices.

We first emphasize that the algebraic formulation of the neutrino Hamiltonian in dense environments allowed an interesting connection 
to be established between dense neutrino gases and condensed matter.
Indeed, following \textcite{Balantekin:2006tg}, \textcite{Pehlivan:2011hp} considered the Hamiltonian in Eq.\eqref{eq:HJs} with the single-angle approximation
\begin{align}\label{eq:HJ}
H & =  \sum_{\omega} \omega { \rm {\bf B}}_{\rm vac} \cdot {\bf J}_{\omega} + \mu \sum_{{\bf p}, {\bf q}}  {\bf J}_{\bf p} \cdot {\bf J}_{\bf q}  \ . 
\end{align}

\textcite{Pehlivan:2011hp} showed that the neutrino Hamiltonian \eqref{eq:HJ} (without matter and for constant $\mu$) is formally equivalent to the reduced BCS Hamiltonian in the theory of superconductivity \cite{Bardeen:1957mv}  
\beq\label{eq:BCS}
H_{\rm BCS} = \sum_k \epsilon_k t_k^z + G T^+ T^- \ ,
\eeq  
which describe Cooper pairs of  of valence electrons in a lattice.
The first term depends on the single-particle energies  $\epsilon_k $ of the Cooper pairs that electrons with spin-up ($\uparrow$) or spin-down ($\downarrow$) can occupy. The second term depends on
the constant $G$, representing the pairing interaction strength and on the quasispin operators 
\beq\label{eq:Tz}
t_k^z  = (c^{\dagger}_{k,\downarrow} c_{k,\downarrow} - c^{\dagger}_{k,\uparrow} c_{k,\uparrow} - 1) \ ,  
\eeq  
and
\begin{align}\label{eq:Tpm}
T^+ = \sum_{k} t_k^+ & = \sum_k c^{\dagger}_{k,\downarrow} c^{\dagger}_{k,\uparrow} ~~~~T^-= \sum_k t_k^- = \sum_k c_{k,\uparrow} c_{k,\downarrow} \ , 
\end{align}
that obey SU(2) commutation relations.
This result highlighted that the neutrino Hamiltonian is exactly solvable since, as pointed out by \textcite{Richardson:1966zza}, the BCS Hamiltonian has analytical solutions
thanks to the algebraic Bethe ansatz method, whereas \textcite{Gaudin:1976sv} showed the exact solvability of the model because of the number of quantum invariants. 
The work of  \textcite{Pehlivan:2011hp} was further elaborated by \textcite{Pehlivan:2014zua} who included neutrino magnetic moments coupled to magnetic fields. 
Moreover, \textcite{Fiorillo:2023mze} discussed the notion that pendulumlike solutions, 
identified in the context of both slow and fast modes, are analogous to soliton solutions in the BCS case. 

More connections between neutrino gases in dense systems and other domains have been uncovered. \textcite{Volpe:2013uxl} established a formal connection between neutrinos propagating in dense media and atomic nuclei, or metallic clusters, through the BBGKY hierarchy. 
Moreover they pointed out that extended mean-field equations include the pairing correlators \eqref{e:kappa} that are formally analogous to pairing correlations in the BCS theory for superconductivity in condensed matter, or to pairing in atomic nuclei. 

\begin{figure}
\begin{center}
\includegraphics[scale=0.5]{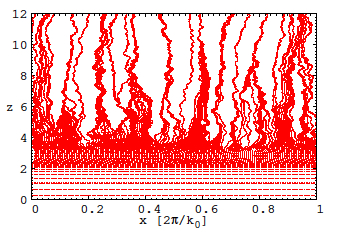}
\caption{Analogy between nonlinear fluids and supernova neutrinos with $\nu\nu$ interactions. 
As the translation symmetry is broken in a two-dimensional model, the streamlines of the $\nu_e$ flux along the vertical direction become irregular, showing large variations and converging toward preferred directions. This behavior is analogous to the transition from the 
laminar to the turbulent regime. Adapted from \cite{Mirizzi:2015fva}.}
\label{fig:hyd}
\end{center}
\end{figure}

An analogy with fluids was pointed out by \textcite{Mirizzi:2015fva} in a two-dimensional model of supernova neutrinos. They discussed that the instability produced by $\nu\nu$ interactions, breaking the spatial symmetry, has an analogy with nonlinear fluid instability. In particular, the transition from a coherent to an incoherent regime in flavor behaves like a streaming flow that changes from the laminar to the  turbulent regime;
see Fig. \ref{fig:hyd}. 
 
One of the crucial open questions is how many-body correlations evolve as a function of  the number of particles $N$ since the Hilbert space of the corresponding wave function increases exponentially with $N$. Developments based on the Bethe ansatz have brought the first comparisons of mean-field and exact results (for a small number of particles), showing in some cases significant differences.  
\textcite{Cervia:2019res,Patwardhan:2021rej} 
employed concepts from quantum information theory, in particular entanglement entropy, to quantify the entanglement between neutrino states\footnote{Note that numerical solutions with the Bethe ansatz turned out to be unstable when the number of particles $N$ is  larger than 10.}.

The entanglement entropy of a neutrino with frequency $\omega$ with the rest of the neutrinos is defined as
\beq\label{eq:Entropy}
S ({\omega}) = - {\rm Tr}[{\varrho^{(red)}_{\omega}} {\rm log}\varrho^{(red)}_{\omega}] = - \sum_{s= {\pm}} \lambda_{s, \omega} {\rm log}(\lambda_{s, \omega} ) \ , 
\eeq     
with the reduced density matrix obtained by tracing over all other neutrinos
\beq\label{eq:rhored}
\varrho^{(red)}_{\omega} = {\rm Tr}_{(\omega' \neq \omega)}\varrho \ , 
\eeq
and the eigenvalues given by\footnote{Remember that $\varrho^{(red)}_{\omega} = {1 \over 2} ({\bf 1} + {\bf \sigma} \cdot   {\bf P}_{\omega}$) Eq.\eqref{eq:poldes}; see Appendix A.}
\beq\label{eq:lambda}
\lambda_{\pm, \omega} = {1 \over 2} (1 \pm \vert  {\bf P}_{\omega} \vert) \ ,
\eeq
with ${\bf P}_{\omega}$ the polarization vector; see Appendix A.

\begin{figure}
\begin{center}
\includegraphics[scale=0.7]{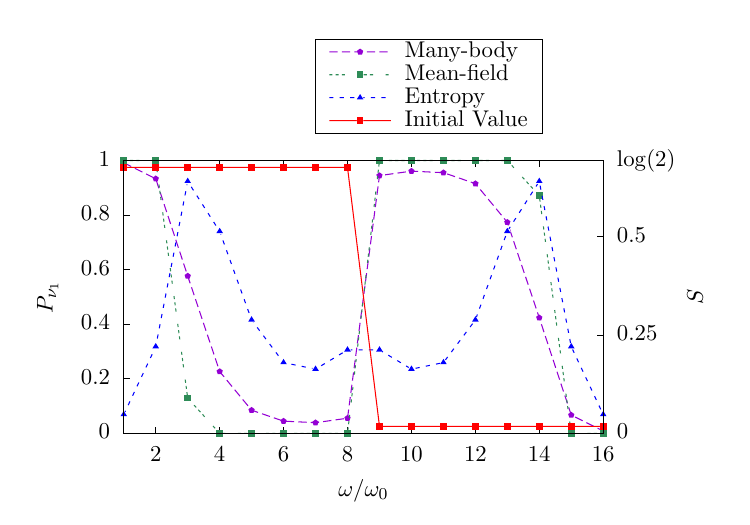}
\caption{Probability that each neutrino is found in the $\nu_1$ state, $P_{\nu_1}$ when $\mu \le  \omega_{0}$ (the latter being the vacuum oscillation frequency) for a number of neutrinos $N = 16$. Shown are the initial configurations (a $\nu_e$ at $\omega_i$ with $i=1,8$ and $\nu_x$ at $\omega_i$ with $i = 8, 16$). The many-body results (purple line) are compared to the mean-field ones (green line). The entanglement entropy $S$, that encodes information on the deviation from the mean-field results of the full many-body ones, peaks at the spectral-split frequencies $\omega/\omega_{0} = 3$  and 14, showing that correlations are important around the spectral-split frequencies. Adapted from \cite{Patwardhan:2021rej}.}
\label{fig:manybody}
\end{center}
\end{figure}

If the neutrino mode is maximally entangled with its environment, then $\vert {\bf P}_{\omega} \vert = 0$, and the entanglement entropy   
$S ({\omega}) = \log(2) $. In the mean-field approximation, the many-body wave-function is the factorized product of single-particle wave-functions,
giving $\vert {\bf P}_{\omega} \vert = 1$ and $S ({\omega}) = 0$. Thus the entanglement entropy gives information on the 
deviations from the mean-field approximation due to many-body correlations. Figure \ref{fig:manybody} shows the entanglement entropy for a system of the order of ten particles. 
The results show that the entropy is the highest for the neutrinos whose energies are the closest to the spectral split; 
see Figure \ref{fig:spectralsplits} and Sec. II.G.1. 

After extending the Hamiltonian \eqref{eq:HJ} to $3\nu$ flavors using the SU(3) algebra, \textcite{Siwach:2022xhx} realized 
the first study of the entanglement of a many-body system with $N = 5$. 
The results show that the role of correlations increases with $N$ and can
reach the maximum entanglement entropy $S ({\omega}) = log(3)$ (instead of $log(2)$ for 2$\nu$ flavors) as found by \textcite{Patwardhan:2021rej}.  

Furthermore advanced numerical techniques such as tensor network methods allow the time evolution of the neutrino 
many-body problem to be treated for a larger number of particles than with conventional methods. \textcite{Cervia:2022pro} 
showed that, depending on the initial conditions and the number of spectral splits, some speedup can be obtained 
(for $N$ up to 20-25) compared to direct numerical integration methods if one uses a time-dependent variational principle on a tensor network. 

\textcite{Roggero:2021fyo,Roggero:2021asb} and \textcite{Roggero:2022hpy,Martin:2021bri}
performed further investigations of entanglement entropy for neutrino systems following the real-time dynamics of systems of a 
larger size, up to $10^2$ and $10^3$ particles.

New developments also concern numerical methods that are at variance with forward integration techniques. An example is 
the recent effort to employ an inference procedure, as in the statistical data assimilation explored by \textcite{Armstrong:2020gxk,Rrapaj:2020lmx,Armstrong:2021bxq}, which searches for the optimization of a cost function. The method does not require knowledge of the initial conditions, but rather constraints at some locations of the coordinate axis (not necessarily at the bounds) that parametrize the model equations of motion. 

Finally, recent studies have created the exciting possibility of investigating correlations and entanglement of strongly correlated neutrinos on quantum computers. 
\textcite{Hall:2021rbv} studied the evolution, the entanglement entropy of a pair and concurrence of a four-particle neutrino 
system using a quantum device for the first time.
\textcite{Amitrano:2022yyn} considered trapped-ion qubits.

The results obtained thus far for small $N$ are exploratory and indicate the potential impact of many-body correlations 
in this context. However, keep in mind that, to definitely assess the role of many-body correlations and the 
reliability of the mean-field approximation, we need to evaluate to a large number of
neutrinos. This is computationally challenging.  Having highlighted aspects of our understanding of neutrino flavor evolution in dense environments, we now turn to observations.

\section{Past and future observations}
\subsection{SN1987A}
On February 23, 1987, Sk-69$^{\circ}$202 exploded in the Large Magellanic Cloud, a satellite galaxy of the Milky Way, producing SN1987A \cite{Arnett:1990au,Bethe:1990mw,Raffelt:1996wa}. It was the first naked-eye sighting since Kepler's supernova (Ia) in 1604. Measurements based on the expanding photosphere method agreed within 10 $\%$ \cite{Schmidt:1992yr} that the supernova was located at 50 $\pm$ 5 kpc from the Earth. Nowadays the distance to the LMC is currently known within 1$\%$, i.e., $49.59 \pm 0.09 ~(stat) \pm 0.54 ~(sys)$ kpc, based on eclipsing binary systems \cite{Pietrzynski:2019xx}.

SN1987A was unique in many respects. Its progenitor, the first known, was a blue supergiant, whereas supernovae progenitors were thought to be red supergiants. This became the {\it blue problem} \cite{Arnett:1990au,Bethe:1990mw}; see \cite{Xiang:2023xnb}) for a recent discussion. 
The earlier SN1054 seen by Chinese astronomers left a pulsar in the Crab Nebula. It has taken slightly over 30 years to finally have indications for a compact object, likely a neutron star, at its location \cite{Alp:2018oek,Cigan:2019shp,Page:2020gsx}. 
The inner ring, large mixing and asymmetrical ejecta of SN1987A indicated strong asphericity in the explosion; see \textcite{Arnett:1990au,Podsiadlowski:1992yr,Janka:2012wk} and Fig.\ref{fig:SN1987A}.  

\begin{figure}
\begin{center}
\includegraphics[scale=0.5]{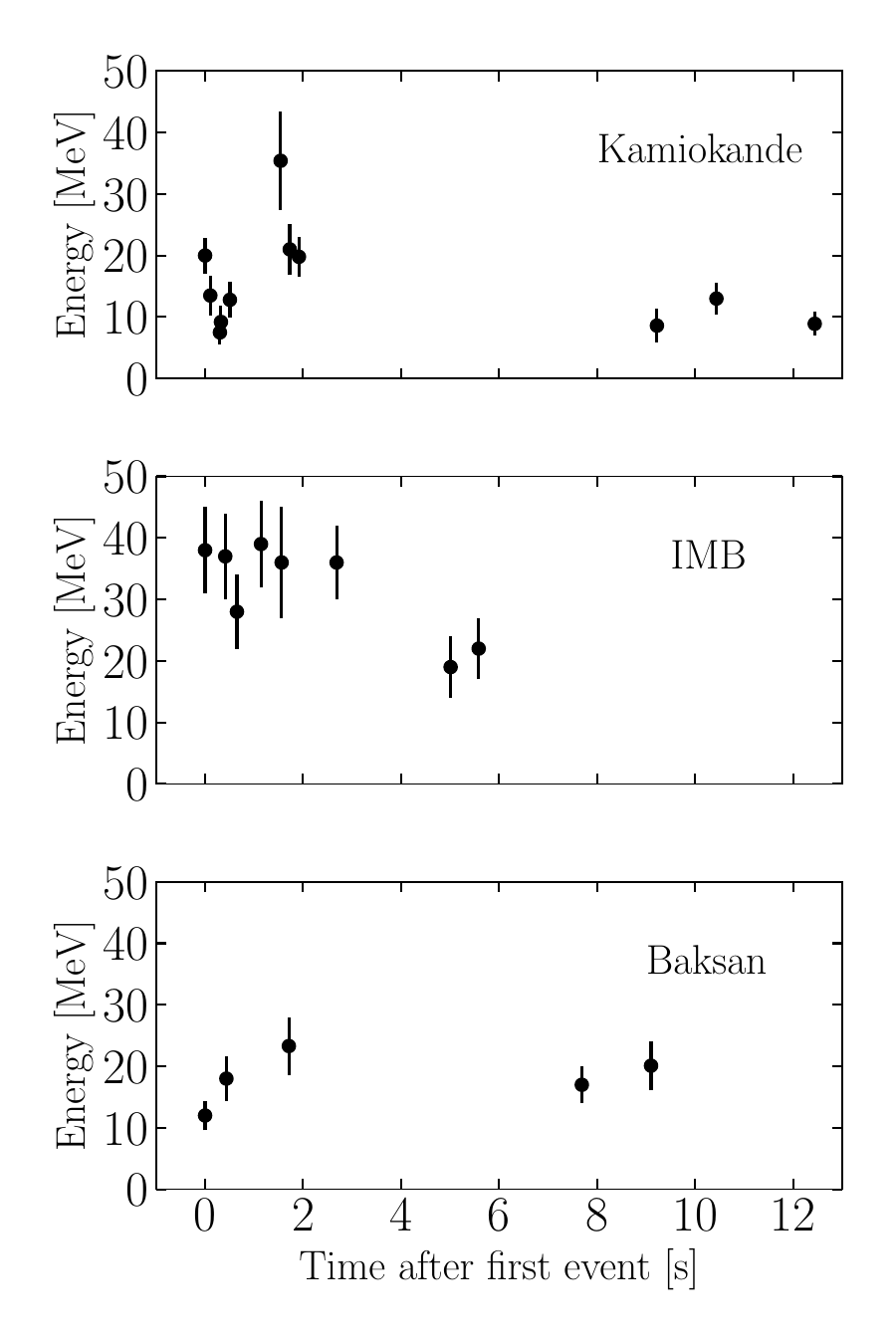}
\caption{Neutrino events from SN1987A at Kamiokande, IMB and Baksan. The energies correspond to secondary positrons produced in 
inverse $\beta$-decay. Events have been shifted at the same $t=0$ (clock relative offsets are unknown). From \textcite{Fiorillo:2022cdq}.}
\label{fig:SN1987Aev}
\end{center}
\end{figure}

The closest supernova in the past several hundred years, SN1987A was observed in all wavelengths from gamma rays to radio 
waves and neutrinos from the collapse of the stellar core were detected for the first time. \textcite{Suzuki:2008zzf} 
provided a description of this pioneering observation on its 20th anniversary.  
The water Cherenkov detector KII \cite{Kamiokande-II:1987idp} observed a neutrino burst\footnote{The probability that the observed burst is a random fluctuation over a constant background is of about $6 \times 10^{-7}$ \cite{Raffelt:1996wa}.} of 11 events of energy ranging from 
7.5 to 36 MeV in 13 s, while IMB \cite{Bionta:1987qt} measured eight events in 6 s with 20-40 MeV. BST \cite{Alekseev:1988gp} 
detected a burst of five events in 9 s and, about 5 h before, the Mont Blanc LSD \cite{Aglietta:1987it} recorded five events during 7 s with energy $\ge$ 7 MeV (Fig. \ref{fig:SN1987Aev}). Since no corresponding signals have been found in the other detectors, 
the LSD events remain controversial. 

Assuming energy equipartition among the neutrino species, early analyses \cite{Sato:1987yi,Schramm:1990pf,Arnett:1990au} and subsequent studies based on two-dimensional likelihoods \cite{Loredo:2001rx,Pagliaroli:2008ur,Vissani:2014doa,Ivanez-Ballesteros:2023lqa,Fiorillo:2023frv} showed the total energy associated with the neutrino events from SN987A to be $5 \times 10^{-52}~$erg  at best-fit value, confirming Colgate and White's hypothesis, and that the spectra agree reasonably well with thermal expectations with $\bar{\nu}_e$ temperatures $T = 4$~MeV (best fit). Analysis of the neutrino time signal agreed with the expected supernova pulse duration of about 10 s and exponential neutron star cooling. Therefore, SN1987A events confirmed the global 
picture\footnote{Detailed information on the time, energy and angular distribution of SN1987A events, and the related backgrounds were given by \textcite{Vissani:2014doa}. Note that the KII and IMB neutrino events showed a forward-peaked angular distribution, instead of isotropic (likely a statistical fluctuation). The analysis of Baksan data points to a higher neutrino luminosity than KII and IMB. As for the time signal, the data in KII and Baksan show a double peak structure separated by 7s, whereas the one seen by IMB did not \cite{Bionta:1987qt}. This is likely a statistical fluctuation.} of the neutrino emission during a gravitational core-collapse supernova explosion. 

Moreover, the Bayesian analysis of the time signal by \textcite{Loredo:2001rx} and by \textcite{Pagliaroli:2008ur} 
supported the {\it delayed-shock model} over the favored {\it prompt-shock model} by unambiguously showing the  presence of an accretion phase. From nuclear matter equations of state, the formation of a neutron star (instead of a black hole was favored, with the mass \cite{Sato:1987yi} and radius \cite{Loredo:2001rx,Pagliaroli:2008ur} compatible with expectations. The radius of the neutrino emitting surface was found to be 18 km \cite{Raffelt:1996wa}. 

The observation of neutrinos from SN1987A was confirming suggestions and results obtained by the supernova modelers. 
\textcite{Baade:1934} who first suggested that in supernovae the considerable energy comes from the gravitational collapse of the inner core into a neutron star. \textcite{Hoyle:1960zz} proposed that stars could die due to thermonuclear runaway (SN Ia) of degenerate material or implosion of the stellar core (SN II and Ib/c).
\textcite{Colgate:1960} pointed out that the collapse could be followed by core bounce and shock formation. The shock would expel most of the star's mass by propagating into the mantle. 

To support the prompt-shock model, \textcite{Colgate:1966ax} hypothesized that most of the gravitational binding energy of the imploding core, $E_b \sim G M_{\rm NS}^2/R_{\rm NS}$, (with $M_{\rm NS}$ and $R_{\rm NS}$ the mass and radius of the newly formed neutron star) namely 
$[1.5, 4.5] \times 10^{-53}~$erg, would be emitted with neutrinos. A few percent of the energy deposited by neutrinos back into matter energy, could drive the supernova explosion. 
In the early 1970s, \textcite{Freedman:1973yd} suggested that neutrinos could coherently scatter on nuclei with a large cross section at neutrino energies below 100 MeV. He pointed out that this coherent effect might impact $\nu$ emission during stellar collapse and neutron star cooling. This gave a new insight on the role of $\nu$ in dense astrophysical ennvironments. Later, \textcite{Bethe:1984ux} shaped the model for supernova explosions into the delayed neutrino-heating mechanism.   

The information on the approximately 24 neutrinos from SN1987A has been a noteworthy laboratory for 
particle physics and astrophysics. 
The neutrino signal gave the start of the explosion, the energy released in the gravitational collapse, 
the temperature at the neutrinosphere, and information on the explosion mechanism,  on the one hand, for example 
rejecting the favored prompt bounce-shock model and supporting the delayed neutrino-heating mechanism, 
and provided limits on neutrino properties and nonstandard physics on the other hand. 

But, is the delayed neutrino-heating mechanism that drives the explosion of most core-collapse supernovae? 
Early one-dimensional calculations faced shock stagnation. 
SN1987A observations of asymmetries and the presence of strong hydrodynamic mixing processes, during the explosion, 
supplied momentum to the development of multidimensional simulations. 
With the advent of two-dimensional simulations, investigations showed the region behind the shock to be convectively 
unstable, the shock itself to be unstable to radial perturbations (the SASI) and finally since 
neutrino-driven convection is turbulent, turbulence could aid the shock propagation at larger radii;
 see \textcite{Mezzacappa:2022hmk} and references therein. 

Thanks to this impressive progress in ascertaining the explosion mechanism, in recent years a consensus has 
been emerging across the supernova community. Indeed simulations show that the majority of supernovae explode because of the delayed neutrino-heating mechanism, where neutrinos efficiently reheat the shock aided by convection, turbulence, 
and the SASI  \cite{Mezzacappa:2022hmk}. However, there is not yet a consensus on the role 
of the SASI for all progenitors; see \textcite{Burrows:2012yk,Burrows:2019zce}. 
More generally different groups might not obtain similar outcomes for the 
same progenitor because of differences in numerical treatments, for example of the neutrino transport or on the progenitor. 
Rotation and magnetic fields can also help explosions, and their effects are being studied; see 
\textcite{Kotake:2005zn,Kuroda:2020bdq}. Notwithstanding, the answer to this long-standing unresolved question 
appears to be forthcoming.
 
Analyses of SN1987A events have produced a wealth of information on nonstandard neutrino properties, 
interactions, and particles \cite{Raffelt:1996wa,ParticleDataGroup:2020ssz}. For example, neutrinos flew through space for 
1.6 $ \times 10^5$ yr, yielding a bound on the neutrino lifetime of $\tau/m > 5 \times 10^{5} $ s/eV \cite{Kamiokande-II:1987idp,Bethe:1990mw}. The nonobservation of a $\gamma$-ray signal over background, corresponding to the neutrino time signal gave stringent bounds on the neutrino lifetime from radiative decays \cite{Raffelt:1996wa,Giunti:2014ixa,ParticleDataGroup:2020ssz}. Optical brightening followed 
neutrino emission by a few hours. Neutrino propagation through space, at nearly the same speed as that of 
photons, gave a tight constraint on the neutrino speed  $c_{\nu}$, i.e. $ \vert ( c - c_{\nu})/c \vert < 2 \times 10^{-9}$ \cite{Longo:1987ub}. Moreover, the absence of a dispersion of the neutrino pulse gave upper limits on the neutrino charge and 
the $\nu_e$ mass (about 20 eV now excluded). 

\begin{figure}
\includegraphics[scale=0.5]{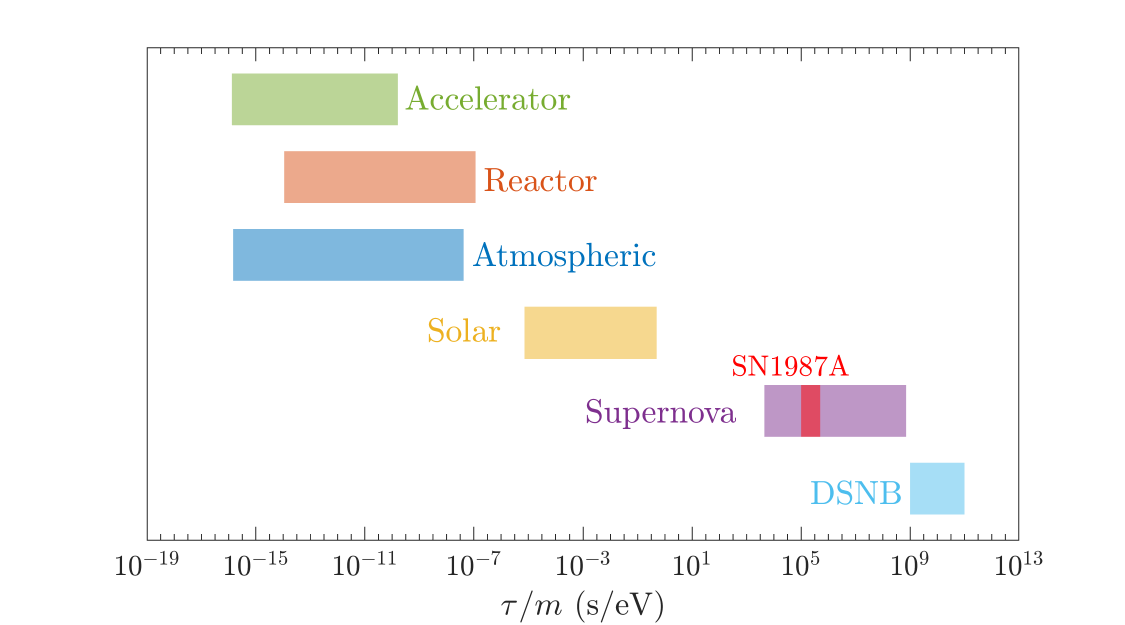}
\caption{Neutrino two-body nonradiative decay. The sensitivity to the lifetime-to-mass ratio for experiments exploiting 
different neutrino sources is shown. The bands are obtained by assuming an exponential depletion of the initial neutrino flux, with only $1 \%$ or $99 \%$ of it reaching the detectors. The red band for SN1987A assumes a distance of 50 kpc. Typical neutrino energies and distances are assumed for the other neutrino sources. From \textcite{Ivanez-Ballesteros:2023lqa}.}
\label{fig:nudecay}
\end{figure}

As for neutrino nonradiative two-body decay, Fig. \ref{fig:nudecay} shows a comparison of the sensitivity of experiments 
using different neutrino sources to the lifetime-to-mass ratio $\tau / m$. Neutrinos from core-collapse supernovae and the DSNB have a unique sensitivity window. As for SN1987A, \textcite{Ivanez-Ballesteros:2023lqa} performed a full 3$\nu$ flavor study of 
neutrino nonradiative two-body decay. Using spectral information of the neutrino events, they derived the bound $\tau/m \ge 2.4~ (1.2) \times 10^5$ s/eV at 68 C.L. (90) $\%$ C.L. for $\nu_2$ and $\nu_1$ (inverted mass ordering), competitive to cosmological bounds. 

Furthermore, energy-loss arguments on the shortening of the SN1987A $\nu$ time signal associated with the neutron star 
cooling gave limits on the neutrino magnetic moment \cite{Barbieri:1988nh}, right-handed neutrinos or currents, axions 
and axionlike particles (see \cite{Chang:2018rso,Payez:2014xsa}), and dark photons (see \cite{Mahoney:2017jqk} . Considering Majoron-like bosons decaying to neutrinos, the absence of a SN1987A flux with $E > $100 MeV provides a strong bound on neutrino Majoron-coupling for Majorons with 
100 eV $< m_{\phi} < 100 $ MeV \cite{Fiorillo:2022cdq}. 

The impact of flavor mechanisms on SN1987A $\nu$ data was also investigated. 
\textcite{Jegerlehner:1996kx} studied the sensitivity of SN1987A events to the MSW effect. Note that in this early analysis the priors for $\langle E_{\bar{\nu}_e} \rangle $ and $\langle E_{\bar{\nu}_{\mu,\tau}} \rangle $ were based on supernova simulations with incomplete microphysics, which was yielding too large 
$\langle E_{\bar{\nu}_{\mu,\tau}} \rangle $, i.e. in the range 24-27 MeV. In their analysis including MSW, \textcite{Jegerlehner:1996kx}  found that hot $\bar{\nu}_e$ spectra were marginally compatible with observations and correctly pointed to an inconsistency.

\textcite{Lunardini:2004bj} performed a model-dependent analysis of the spectral modifications of SN1987A events 
due to the MSW effect and found the effect to be significant.
However, the likelihood analysis by \textcite{Vissani:2014doa} based on modern supernova fluxes 
showed the impact of the MSW phenomenon to be small and comparable to effects from other inputs 
such as backgrounds and detector thresholds. 
Furthermore, with a similar analysis, \textcite{Ivanez-Ballesteros:2023lqa} showed that the best fit and confidence 
contours of the $\bar{\nu}_e$ average energy and luminosity also depended on the pinching parameter, 
producing variations that were at the same level as the other inputs. Finally an analysis of the time and energy of the SN1987A neutrino events by \textcite{DedinNeto:2023hhp} agreed on a preference for the accretion model, as in previous findings, and showed a slight preference for the MSW mechanism (normal mass ordering). 

Besides the studies on the impact of flavor phenomena, detailed comparisons are being realized between SN1987A data 
and supernova simulations. 
\textcite{Olsen:2021uvt} compared three one-dimensional supernova models to KII data based on a Bayesian analysis
and found that the sparse data prevented discrimination among them.   
However, \textcite{Li:2023ulf} confronted outcomes of multidimensional supernova models with SN1987A data during the first second and 
found disagreement even when considering different flavor scenarios (MSW effect, flavor equilibration due to the $\nu\nu$ interaction, etc.).
On the contrary \textcite{Fiorillo:2023frv} set up a detailed comparison of one-dimensional supernova 
simulations with the SN1987A observations and found an agreement with the early neutrino signal. The late neutrino time signal appears to be too short, which might point to the role of convection in the protoneutron star. 
Further progress toward obtaining consistent successful three-dimensional supernova models will shed more light on 
the consistency between modern simulations and the neutrino events from SN1987A. 
The ensemble of the results extracted thus far from SN1987A observations is emblematic of how patience 
for such rare events can be rewarded with scientific knowledge and progress.   

\subsection{From the next supernova}
If a supernova explodes in the future, 
neutrinos emitted in the first instants of the core-collapse will be detected several hours before optical emission and will guide optical instruments \cite{Super-Kamiokande:2016kji}. If the supernova is nearby, presupernova neutrinos from thermal \cite{Odrzywolek:2003vn} and weak processes \cite{Patton:2017neq} in the late stages of the stellar evolution might be observed preceding core collapse and could give advanced warning \cite{Yoshida:2016imf} as well as information on the supernova progenitor; see the review by \textcite{Kato:2020hlc}).  SK could detect about 
200 presupernova neutrinos 12 h before the collapse of a 15-25 $ {\rm ~M}_{\odot}$ star at 0.2 kpc (such as Betelgeuse); whereas SK-Gd and KamLAND results could reach 0.6 kpc \cite{Super-Kamiokande:2016kji} and 0.69 kpc \cite{KamLAND:2015dbn} respectively. 
Moreover, if it is close enough, one could measure the neutrino light curve up to 100 s, as first pointed out by \textcite{Li:2020ujl}, which would yield interesting information on the late cooling phases of the newly formed protoneutron star. 

The observation of the next supernova will benefit the SNEWS network \cite{SNEWS:2020tbu} and dark matter detectors (see e.g. \cite{DarkSide20k:2020ymr,Lang:2016zhv}. From the complementarity of the technologies available, we 
measure the time and energy of the neutrino flavors through inverse $\beta$-decay, neutrino-nucleus scattering, neutral current scattering on electrons as well as protons \cite{Beacom:2002hs}, and coherent neutrino-nucleus scattering \cite{COHERENT:2017ipa}.   

For inverse $\beta$-decay the cross sections are precisely known \cite{Strumia:2003zx,Ricciardi:2022pru}. On the contrary, the cross sections 
associated with charged-current $\nu$-nucleus interactions, particularly on $^{16}$O, $^{40}$Ar, $^{56}$Fe, $^{208}$Pb are still affected by theoretical uncertainties, that can reach several tens of a percent; see Fig. 5 of \textcite{Vaananen:2011bf} for the lead case. The two exceptions include deuterium whose cross sections are known within a few percent precision \cite{Balantekin:2003ep}, and $^{12}$C \cite{Volpe:2000zn,Hayes:1999ew}. 

\textcite{Volpe:2003fi,Volpe:2006in} suggested using a novel technique, i.e. the low 
energy beta-beam, also to perform $\nu$-nucleus cross section measurements
at energies of interest for supernovae. More than 20 years after the proposal at the Spallation Neutron Source \cite{Avignone:2001ti,Barbeau:2021exu}, neutrinos from decay-at-rest muons are being used for such experiments. The results of the first two measurements on the inclusive $\nu_e$ charged-current cross section on $^{127}$I \cite{COHERENT:2023ffx} and $\nu_e$ on $^{nat}$Pb \cite{COHERENT:2022eoh} have been released. More nuclear targets will be studied, including $^{40}$Ar, which is important for the DUNE experiment. 

Besides providing a better knowledge of the spin and spin-isospin weak nuclear response to neutrinos, in the energy range of interest for the detection of supernova neutrinos, such measurements could shed further light on the issue of the quenching of the axial-vector coupling constant, and 
also on forbidden states, as pointed out by \textcite{Volpe:2005iy}. 
Using effective field theories and powerful quantum many-body techniques, \textcite{Gysbers:2019uyb} 
showed that the quenching of the allowed Gamow-Teller transitions arose 
from strong correlations in the nucleus and weak interaction coupling to two nucleons. 
This result represents a major step forward, since this quenching, observed in beta decay and in charge-exchange reactions, 
has been a controversial issue for 50 years.

\begin{figure}
\begin{center}
\includegraphics[scale=0.7]{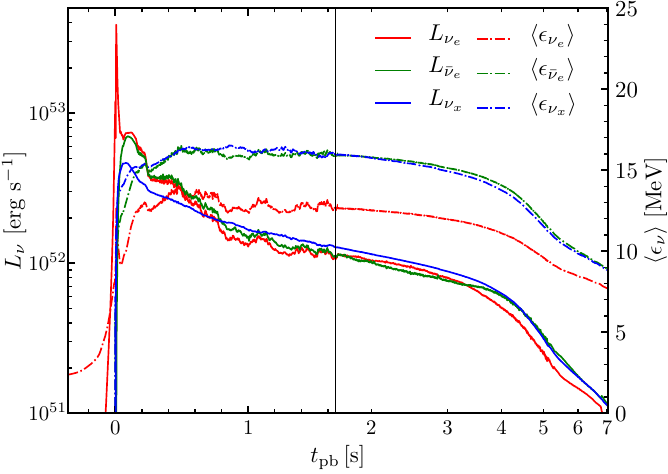}
\caption{Neutrino supernova time signal. The luminosities $L_{\nu}$ and the average energies $\langle \epsilon_{\nu} \rangle$ of $\nu_e$, $\bar{\nu}_e$ and $\nu_x$ (the last of which stands for $\nu_{\mu}, \bar{\nu}_{\mu}, \nu_{\tau}, \bar{\nu}_{\tau}$) are presented. 
The results correspond to the three-dimensional simulation of a nearly 19 ${\rm ~M}_{\odot}$ progenitor followed from about 7 min before the onset of explosion up to 7 s. There are three characteristic phases: the $\nu_e$ neutronization peak (about 20 ms), the $\nu_e$, $\bar{\nu}_e$ and $\nu_x$ ($\bar{\nu}_x$) energies emitted during the accretion phase (about 500 ms) and the cooling of the newborn neutron star. From \textcite{Bollig:2020phc}.}
\label{fig:3DSN}
\end{center}
\end{figure}

The future observation of a supernova time signal will result in many fundamental lessons.
 As supernova simulations show, according to the current paradigm, the neutrino time signal involves three main characteristic phases: the neutronization burst, the accretion phase and the cooling of the newly born proton-neutron star.
An example is shown in Figure \ref{fig:3DSN} which provides the outcome of a self-consistent three-dimensional simulation by \textcite{Bollig:2020phc} that reaches an explosion energy at infinity of 1 bethe (10$^{51}$ erg) and a $^{56}$Ni amount of $\le 0.087 {\rm ~M}_{\odot}$ 
which is in agreement with SN1987A observations (within 10$\%$).  

During the first tens of milliseconds, $\nu_e$ are emitted due to the capture of the electrons by the nuclei, that, which with iron photodissociation contribute to the core collapse. These neutrinos form the so-called neutronization peak. The neutrino emission is powered by the accretion flow (accretion phase) during which all neutrino flavors are produced. The signal presents
fast variations and directional dependence due to large convection and hydrodynamic instabilities and the LESA. 
Finally, when the explosion is ongoing and the shock stagnation is overcome, 
neutrinos are emitted from the cooling of the proton-neutron star and the luminosity drops. 
In Fig. \ref{fig:3DSN} one can see how much the hierarchy of the average energies  
$\langle E_{\nu_e} \rangle < \langle E_{\bar{\nu}_e} \rangle < \langle E_{\nu_x} \rangle $ is satisfied.

The $\nu_e$ flux from the 50 ms neutronization burst represents less than 1$\%$ of the total neutrino luminosity. 
The accretion phase that lasts from a few tens to a few hundreds milliseconds 
and the neutron star cooling of about 10 s take away most of the gravitational binding energy.  
Each of these phases has its own specificities from the point of view of the supernova evolution 
and the neutrino flavor conversion, as we discuss later in the review. 

If the supernova is close enough, the precise measurement of the time signal will be crucial to definitively 
assess the explosion mechanism through the identification 
of oscillations with high frequencies, correlated with SASI, whose measurement requires a 
precise time resolution such as the one for the IceCube detector; 
see for example \cite{Walk:2019miz,Tamborra:2014hga,Muller:2014rpb}. In this respect, modifications from flavor evolution should not swamp the signature.   

The measurement of early ($ < 20 $ ms) stages of neutrino emission would give information on the bounce time \cite{Halzen:2009sm}. Note that this is key to estimating the burst time of the gravitational waves \cite{Pagliaroli:2009qy} which are produced mainly 
by the oscillations of the newly formed protoneutron star  \cite{Abdikamalov:2020jzn}. The concomitant multimessenger event of neutrino and gravitational waves from a core-collapse supernova was also discussed by \textcite{Halim:2021yqa}.

From the point of view of flavor evolution, the neutronization-burst represents a unique phase. Only the MSW effect appears to influence the neutrino spectra. Neither fast nor slow modes, as we understand them now, are at work. The former requires crossings in the neutrino and antineutrino 
angular distributions, whereas the latter  requires $\nu \bar{\nu}$ pairs (in the bulb model\footnote{In more complex models as well, the $\nu\nu$ interaction did not appear to influence this early phase.}). Moreover there are effects neither 
from shock waves, since shock waves reach the MSW region after 1 to 2 s, nor from turbulence. Therefore, the 
neutronization burst appears to be a good laboratory to explore nonstandard properties. 
These include nonstandard $\nu\nu$ interactions \cite{Das:2017iuj} and neutrino 
nonradiative decay \cite{Ando:2004qe,deGouvea:2019goq}. 

Since flavor mechanisms produce neutrino spectral modifications (see Sec. II.B), 
an important question to ask is: With what precision will we be able to reconstruct the supernova neutrino fluxes when the next 
supernova explodes?; see \cite{Vaananen:2011bf,Lu:2016ipr,Lujan-Peschard:2014lta,GalloRosso:2020qqa}. 
The answer depends on the observatories that are operating at the time and the supernova distance. 

Note that a precise determination and reconstruction of the supernova neutrino spectra might not be trivial in the likelihood analysis 
where the ensemble of the parameters are left free to vary even in the simplest case with the MSW effect. 
Indeed, \textcite{Minakata:2008nc} pointed out the presence of parameter degeneracies that can in principle be broken by combining detection channels \cite{GalloRosso:2017hbp}. However, while most of the neutrino parameters appear to be precisely measurable (for a supernova at 10 kpc) identifying the neutrino pinching for some of the flavors might be more tricky \cite{GalloRosso:2017mdz}. 

As for unknown neutrino properties, the neutrino signal from the next supernova could be a good laboratory to determine the neutrino mass ordering for which there are currently hints of a low statistical significance. The passage of the shock wave can be pictured as it goes 
through the MSW region; see Sec. II.D. Shock wave effects can be important  and can produce 
distortions of the positron or electron time signals, depending on the neutrino energy and mass ordering; see \cite{Takahashi:2002yj,Fogli:2003dw,Lunardini:2003eh,Fogli:2004ff,Kneller:2007kg} and the 
reviews by \textcite{Duan:2009cd,Horiuchi:2018ofe}. 
In addition, the rise time of the neutronization burst 
can be used to determine the neutrino mass ordering in a detector like IceCube \cite{Serpico:2011ir}.
Although these signatures are interesting, it is likely that an Earth-based experiment like JUNO, DUNE or Hyper-K will measure the neutrino mass ordering before the next supernova explodes. In particular, 
it should be measured at about 3 $\sigma$ after six \cite{JUNO:2015zny} or ten years \cite{Abe:2011ts}. 

As for CP violation in the lepton sector, hints for $\sin \delta < 0$ (90 $\%$ CL) indicates that the CP violating phase should soon 
be discovered through the DUNE and Hyper-K experiments. The effects of the Dirac CP violating phase was studied in the context of core-collapse supernovae.  \textcite{Akhmedov:2002zj} concluded that there should be no impact of the Dirac phase on the $\nu_e$ fluxes in a supernova, even if the $\nu_{\mu}$ and $\nu_{\tau}$ fluxes are unequal. In contrast to 
such findings, \textcite{Balantekin:2007es} demonstrated that the Dirac phase can impact the electron neutrino fluxes if the muon and tau neutrino fluxes differ because of radiative corrections or of nonstandard interactions such as flavor-changing neutral currents. 
The result relies on a factorization condition of the neutrino Hamiltonian,  
${\cal H}(\delta) = S^{\dagger}{\cal H}(\delta = 0) S$ with $S^{\dagger}(\delta) = {\rm diag}(1, 1, e^{i \delta})$. 
Physics that breaks such a condition can engender CP violation effects on the $\nu_e$ fluxes.

These findings were generalized in the presence of $\nu\nu$ interactions by \textcite{Gava:2008rp}, 
beyond the mean field to the full many-body problem \cite{Pehlivan:2014zua} and on the neutrino degeneracy parameter in the early Universe \cite{Gava:2010kz}. In the supernova numerical calculations showed the impact of the phase to be small \cite{Balantekin:2007es}. However, the combined effect of the Majorana CP violating phase(s) and the neutrino magnetic moment could trigger sizable effects, creating 
the possibility of new resonances, as pointed out by \textcite{Popov:2021icg}.

Even when the mass ordering and CP violation are precisely measured, supernova neutrinos will remain interesting probes for nonstandard physics. 
Indeed there are numerous flavor mechanisms related to other key unknown neutrino properties\footnote{These are not the main focus of this review.} that have been extensively discussed in the literature, such as sterile neutrinos, nonstandard interactions, or the neutrino magnetic 
moment (these properties can make sizable modifications to the neutrino spectra in the 
presence of strong magnetic fields in core-collapse supernovae or nearby compact objects); see \textcite{Nunokawa:1996gp,Pehlivan:2014zua,Giunti:2014ixa,Sasaki:2021bvu}.  

In recent years studies have explored the prospects for discriminating among supernova models from future observations using Bayesian inference techniques. Such investigations have included supernova simulations and/or neutrino flavor evolution, with different levels of sophistication.
\textcite{Hyper-Kamiokande:2021frf} examined the possibility of distinguishing among five (one-dimensional or multidimensional) supernova models from different groups, including the MSW effect. By considering likelihood ratios of a pair of models of the accretion phase (500 ms), they showed 
a strong capability to discriminate among them with the Hyper-K detector (300 events), with a supernova at about 100 kpc (60 kpc) and normal (inverted) mass ordering. 

In a subsequent work \textcite{Olsen:2022pkn} considered seven one-dimensional models (differing regarding progenitor mass or the equation of state) and either included the MSW effect or did not. The results of their Bayesian analysis using a neutrino emission up to 9 s showed that the models could be distinguished from each other using  a detector like SK for a core-collapse supernova at a known distance of 25 kpc. The analysis was pushed further by \textcite{Saez:2024ayk} who employed the 300 ms neutrino signal (supernova at 10 kpc), including the MSW effect only, from 18 two- and three-dimensional supernova models (progenitor masses between 9 ${\rm M_\odot}$ and 60 ${\rm M_\odot}$) and the events in DUNE
$\nu$-$^{40}$Ar, Hyper-K, and JUNO (inverse beta decay and also $\nu$-proton scattering for the latter). Their Bayesian analysis showed that Hyper-K stands out as being mostly powerful in distinguishing models (progenitor mass and equation of state). Finally, by considering the accretion phase signal \textcite{Abbar:2024nhz} analyzed the possibility of discriminating among various flavor scenarios beyond the MSW effect, including, in particular, flavor equipartition, spectral swapping, and fast flavor conversion due to the $\nu\nu$ interaction. With several hundred events from a galactic supernova in a water Cherenkov detector (inverse beta-decay and elastic scattering), distinguishing flavor mechanisms appears to be promising when one assumes that the neutrino spectral parameters are unknown. Observations of the next supernova through its neutrinos will yield crucial information for particle physics and astrophysics  
particularly on the explosion mechanism, the supernova location, the equation of state, the progenitor mass,
nonstandard neutrino properties, flavor mechanisms in dense environments, and new physics.

\subsection{Toward the discovery of the diffuse supernova neutrino background}\label{sec:dsnb}
Complementing the a single supernova is the DSNB, which is made of neutrinos emitted by past core-collapse supernovae. The DSNB is nearly isotropic and constant in time. 
It depends on cosmology, astrophysics and particle physics; see the reviews by \textcite{Ando:2004hc,Beacom:2010kk,Lunardini:2010ab,Mathews:2019klh,Suliga:2022ica,Ando:2023fcc}. 

The DSNB flux, including a progenitor dependence, reads 
\beq\label{eq:dsnb}
\phi_{\nu_{\alpha}}(E_{\nu}) = {c} \int \int dM~dz~ \Big | {dt_{\rm c} \over {dz}} \Big |~R_{\rm SN}(z,~M) ~ 
{\phi_{\nu_{\alpha}}(E'_{\nu}, M)} \ ,
\eeq
where $z \in [0, z_{max}]$ is the cosmological redshift, $c$ is the speed of light,
$E_{\nu}'$ is the neutrino energy at the star location at redshift $z$, related to the energy $E_{\nu}$ on 
Earth through $E_{\nu}' = E_{\nu}(1 + z)$ and $\phi_{\nu_{\alpha}}(E_{\nu}', M)$ is the time-integrated neutrino flux for a progenitor of mass $M$. 

A progenitor mass dependence of the DSNB flux was first included by \textcite{Lunardini:2012ne}.
Note that the most general expression for the DSNB flux should also have an explicit dependence on the galactic chemical evolution, 
as considered by \textcite{Nakazato:2015rya,Ashida:2023heb}. 
In Eq. \eqref{eq:dsnb} usually $z \in [0,5]$. However only the lowest redshifts, i.e. $z \in [0, 2]$ give the most 
important contribution to the DSNB flux.
Moreover note that the integral on the progenitor mass covers the range $M \in [8, 125] M_{\odot}$ (taking $100 M_{\odot}$ 
does not introduce any significant difference). 

The first factor in Eq.\eqref{eq:dsnb} is the cosmological time that depends on the cosmological model. The lambda cold dark matter 
($\Lambda$ CDM) model is assumed\footnote{Note that \textcite{Barranco:2017lug} investigated the influence of cosmological models other than the $\Lambda$CDM model 
on the DSNB, but the star-formation rate used is not consistent with them.}. The expansion history of the Universe is then 
\beq\label{eq:cosmic}
\Big | {dz \over {dt_{\rm c} }} \Big | = H_0 ( 1 + z) \sqrt{ \Omega_{\Lambda} + (1+z)^3 \Omega_m} \ , 
\eeq
where $\Omega_m$ and $\Omega_{\Lambda}$ are the matter and the dark energy cosmic energy densities 
and $H_0 = 70~$km s$^{-1}$ Mpc$^{-1}$  is the Hubble constant. DSNB predictions show that the DSNB is not sensitive to variations compatible with the Hubble tension\footnote{There is currently a tension between the Hubble constant value extracted with the "distance ladder method" 
$H_0 = 74.03 \pm 1.42 ~{\rm km~ s}^{-1} {\rm Mpc}^{-1}$ and the one obtained from the cosmological microwave background (CMB), i.e. $H_0 = 67.4 \pm 0.5 ~{\rm km ~s}^{-1} {\rm Mpc}^{-1}$ \cite{DiValentino:2021izs, Abdalla:2022yfr}.}.

The second important input in Eq.\eqref{eq:dsnb} is the evolving core-collapse supernova rate\footnote{The number per unit time per unit comoving volume} $R_{\rm SN}(z, M) $ which is related to the  star-formation rate history $\dot{\rho}_*(z) $ according to
\beq\label{eq:CCSNrate}
R_{\rm SN}(z, M) = \dot{\rho}_*(z) { \phi(M) dM \over {\int^{125 M_{\odot}}_{0.5 M_{\odot}}  \phi(M) M dM}} \ , 
\eeq
where $\phi (M)$ is the initial mass function. 

In his seminal work \textcite{Salpeter:1955it}  introduced the power law initial mass function
\beq\label{eq:IMF}
\phi(M) \sim {M}^{\chi} \ ,
\eeq
for $M \in [0.5, 1]~ M_{\odot}$. The quantity $\phi(M) d M$ gives the number of stars in the mass interval $[M, M + d M]$. Since, the Salpeter initial mass function (IMF) was employed, $\chi$ has been determined with an uncertainty of about $10 \%$.  \textcite{Baldry:2003xi} introduced a modified  broken power law for the IMF with $\chi =- 1.5$ at $ 0.1 ~M_{\odot} \le M \le 0.5~M_{\odot} $ and $\chi = - 2.12$ for $M > 0.5~ M_{\odot} $.  
Note that such a modified IMF gives a similar result for $R_{\rm SN}(z, M)$ \cite{Horiuchi:2008jz}. 
The universality of the IMF at high masses, i.e. the fact that it is the same in all environments and throughout cosmic time, is currently 
under discussion; see by \textcite{Ziegler:2022ivq}.

As for the cosmic star-formation history, it can be deduced from observations (see e.g. \cite{Hopkins:2006bw,Reddy:2007bs,Rujopakarn:2010gn}) and 
described by a piecewise continuous form of a broken power law by \textcite{Yuksel:2008cu} (see also \cite{Madau:2014bja,Singh:2020tmt})
\beq\label{eq:SFR}
\dot{\rho}_*(z) = \dot{\rho}_{0} \Big[(1+z)^{\alpha \eta} + \Big({1+z \over{B}}\Big)^{\beta \eta} + \Big({1+z \over{C}}\Big) ^{\gamma \eta}  \Big]^{-1/\eta}  \ ,
\eeq
with $\alpha = 3.4, \beta = -0.3$ and $\gamma = -3.5 $ the logarithmic slopes at low, intermediate and high redshift. The quantity $\eta = -10$ is the smoothing function
and the constants defining the redshift breaks are $B = 5000$ and $C = 9$. In the cosmological model considered by \textcite{Yuksel:2008cu},  $\Omega_m = 0.3$, $\Omega_{\Lambda} = 0.7$, and $H_0 = 70~$km s$^{-1}$ Mpc$^{-1}$.

\begin{figure}
\begin{center}
\includegraphics[width=.48\textwidth]{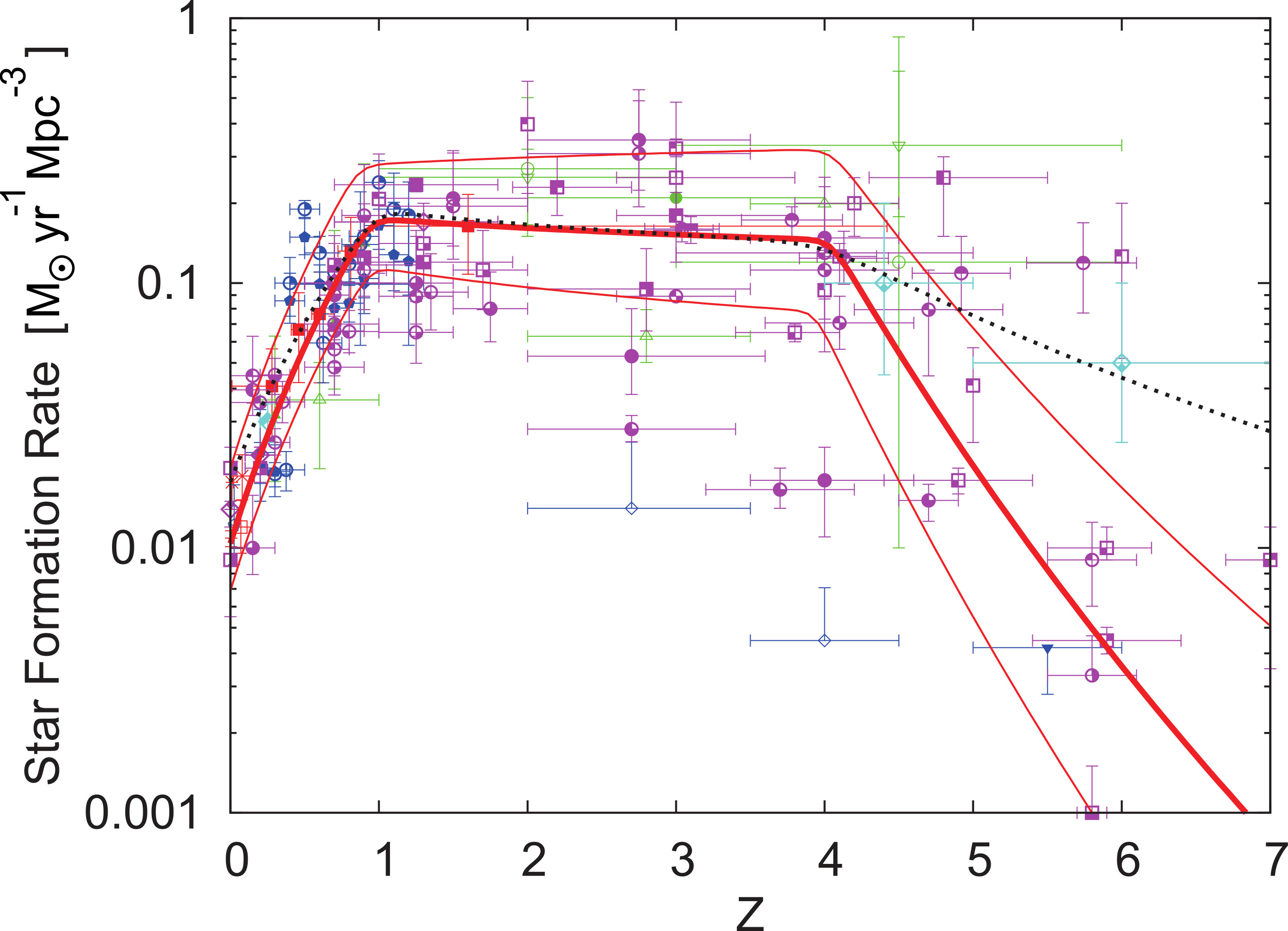}
\caption{A piecewise linear star formation rate fitted to observed dust-corrected data. The symbols show the ensemble of the data in IR, optical, UV, x-ray/$\gamma$-ray and radio bands (red, blue, magenta, light-blue, and green points, respectively. The $\chi^2$ fit to the star-formation rate and the $\pm 1 \sigma$ upper and lower limits as a function of redshift, are shown as the thick and thin solid red lines respectively. The dotted black line shows the star-formation rate obtained with the full data used by \textcite{Yuksel:2008cu}. Note that only the low $z$ part contributes to the DNSB. From \textcite{Mathews:2014qba}.}
\label{fig:CCSNrz}
\end{center}
\end{figure}

Concerning the cosmic star-formation rate history, several parametrizations are available in the literature. The one given by Eq.\eqref{eq:SFR} from \textcite{Yuksel:2008cu,Horiuchi:2008jz} includes gamma-ray burst data  $z > 4$ and is commonly employed. The parametrization of \textcite{Fogli:2004gy} is outdated, whereas that of \textcite{Priya:2017bmm} presents kinks. \textcite{Mathews:2014qba} suggested an alternative parametrization by including only the subset of the star-formation rate data corrected for extinction by dust\footnote{Moreover \textcite{Mathews:2014qba} argued that the "supernova rate problem" could be solved by the inclusion of contributions from binaries, failed supernovae, and electron-capture ONeMg supernovae.}\footnote{Note that in their early work, \cite{Fukugita:2002qw} discussed the constraints on the star-formation rate that could obtained from the expected bounds on the DSNB by SK. }; see Fig. \ref{fig:CCSNrz}.

Currently, the local core-collapse supernova rate is known with the following precision
\begin{align}\label{eq:RSNz}
R_{\rm SN}(0) & = \int^{125 {\rm ~M}_{\odot}}_{8 {\rm ~M}_{\odot}} R_{\rm SN}(0, {\rm M}) {\rm d}{\rm M} \nonumber \\
& = (1.25 \pm 0.5) \times 10^{-4} {\rm yr}^{-1} {\rm Mpc}^{-3} \ .
\end{align}
Moreover there is a disagreement by a factor 2 at $0 \le z \le 1$ between  the core-collapse supernova rate deduced from the star-formation rate history and the one from 
direct core-collapse supernova observations \cite{Horiuchi:2011zz}. This is known as the "supernova rate problem". 
The uncertainty on the evolving core-collapse supernova rate constitutes one of the largest uncertainties in the DSNB predictions; see Fig. \ref{fig:CCSNrz}.

The third and last important factor is the neutrino flux from one single-supernova with the progenitor mass $M$.
\textcite{Lunardini:2009ya} pointed out the relic supernova background can receive a significant contribution from failed supernovae (collapsing into a black hole). 
Indeed, owing to the compression of baryonic matter during black hole formation, the supernova generates large neutrino fluxes with higher average energies 
and larger differences among flavors than optical supernovae, depending on the soft or stiff equation of state, as pointed out by \textcite{Sumiyoshi:2007pp}.

Although the fraction of supernovae that turn into a black hole is subleading, this contribution influence the tail of the DSNB spectrum 
and contributes substantially to the DSNB rates.
The dependence on the progenitor mass was first included by \textcite{Lunardini:2012ne} using detailed long-term supernova simulations.
If one includes the neutrino spectra from core-collapse supernovae that leave a neutron star or a black hole, Eq.\eqref{eq:dsnb} becomes
\begin{equation}
\begin{aligned}
    \phi_{\nu_{\alpha}}(E_{\nu}) = &  {c} \int  dz~ (1+z) \Big | {dt_{\rm c} \over {dz}} \Big | \nonumber \\
    &  \times \left[ 
\int_\Omega~{\rm dM}~R_{\rm SN}(z,{\rm ~M}) ~ {\phi_{\nu_{\alpha}}^{\rm NS}(E'_{\nu}, {\rm~M})} \nonumber \right.  \\
& +  \left. \int_\Sigma~{\rm dM}~R_{\rm SN}(z,{\rm ~M}) ~ {\phi_{\nu_{\alpha}}^{\rm BH}(E'_{\nu}, {\rm~M})} \right] \ , 
\end{aligned}
\end{equation}
where $\Omega$ and $\Sigma$ correspond to the range of masses for which the collapse gives a neutron star (NS) or a black hole (BH).
Thus, the BH fraction is given by
\begin{equation}
    f_{\rm BH} = \frac{\int_\Sigma d \rm M \phi(\rm M)}{\int_{8 \rm M_\odot}^{125 \rm M_\odot} d \rm M \phi(\rm M)}.
\end{equation}
Note that \textcite{Schilbach:2018bsg} investigated only  the DSNB coming from black hole accretion disks.

The fraction of failed supernovae is currently under debate.
\textcite{OConnor:2010moj,Horiuchi:2014ska,Ertl:2016sag,Horiuchi:2017qja} argued that the star compactness could be a good indicator of the fraction of supernovae leaving black holes. This is in contrast to \textcite{Ertl:2015rga} who suggested as indicators two parameters, $M_4$ and $\mu_4$, giving the enclosed mass and its derivative (s = 4, dimensionless entropy per nucleon), to better predict successful explosions in the neutrino-driven explosion mechanism.
On the observational side, a decade long search for failed supernovae found a fraction of failed supernovae of $ f_{\rm BH} = 0.16 ^{+0.23}_{ - 0.12} $ at 90 $\%$ C.L.; see \textcite{Neustadt:2021jjt} and references therein.

\begin{figure}
\begin{center}
\includegraphics[width=.45\textwidth]{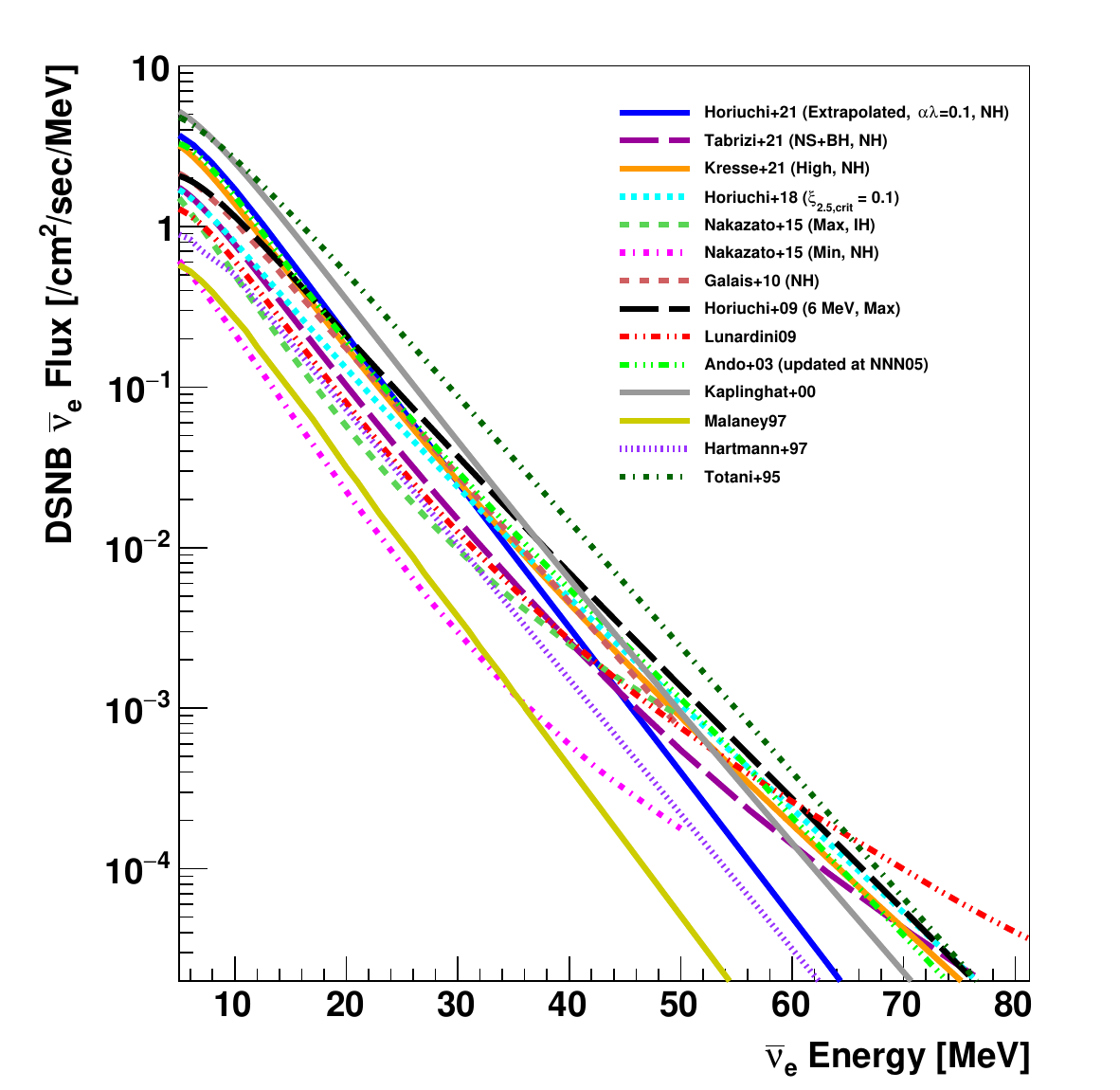}
\caption{Current spread on DSNB $\bar{\nu}_e$ flux predictions as a function of neutrino energy. The results correspond to the predictions by
\textcite{Horiuchi:2020jnc,Tabrizi:2020vmo,Kresse:2020nto,Horiuchi:2018ofe,Nakazato:2015rya,Galais:2009wi,Horiuchi:2008jz,Lunardini:2009ya,Ando:2003ie}.
The labels refer to neutrino normal mass ordering (NH), inverted mass ordering (IH), neutron star (NS) and black hole (BH). 
 From \cite{Super-Kamiokande:2021jaq}.}
\label{fig:DSNBpred-SKpap}
\end{center}
\end{figure}

Predictions of the DSNB flux and rates have different levels of sophistication with respect to astrophysical inputs (including the black-hole fraction, the contribution from the binaries, the evolving core-collapse supernova rate, the shock waves and the shock wave revival time, dense matter equation of state) on the one hand, and the neutrino flavor mechanisms, the neutrino properties and the new physics 
on the other hand. Besides the  previously mentioned astrophysical uncertainties, the neutrino fluxes from a single supernova with progenitor mass $M$ also have important variations. 
Figure \ref{fig:DSNBpred-SKpap} shows the current spread on the predictions of the DSNB fluxes. 

Some twenty years ago, \textcite{Ando:2004hc} pointed out that most of the DSNB flux would come from low redshifts, i.e. $ 0 < z < 1$; whereas the higher redshifts ($z >  1$) would contribute at low energy ($E <$10 MeV). They 
emphasized that the early SK limit by \cite{Malek:2002ns} was just above predictions when realistic
evolving core-collapse supernova rate and neutrino spectra are employed.

Concerning the black hole fraction, its value is either treated in a parametrized form or obtained from extensive supernova simulations. For example in \textcite{Tabrizi:2020vmo,Moller:2018kpn,Ivanez-Ballesteros:2022szu,Priya:2017bmm} 
implemented the progenitor mass dependence and 
the fraction of failed supernovae based on one-dimensional supernova simulations. Note that, according to the detailed simulations by 
\textcite{Kresse:2020nto} the BH fraction\footnote{\textcite{Priya:2017bmm} also considered a more conservative value of $ f_{\rm BH} = 0.09$.} ranges from 17 $\%$ to 41 $\%$. 
\textcite{Kresse:2020nto,Moller:2018kpn,Horiuchi:2017qja,Horiuchi:2020jnc} performed extensive supernova simulations  (100 to 200 models) to include a detailed progenitor mass dependence. 
\textcite{Kresse:2020nto} and \textcite{Horiuchi:2020jnc} also included helium-star progenitors to account for contributions from 
binaries currently unknown, showing it can be significant. 
In their study \textcite{Horiuchi:2017qja} argued the compactness to be a good indicator of the fate of a supernova, as previously mentioned, and found that 
the DSNB detection could test small values, namely $ f_{\rm BH} > 20 \%$. 

Several have investigated the sensitivity of the DSNB to the nuclear equation of state \cite{Moller:2018kpn,Ashida:2022nnv,Nakazato:2015rya}.  
\textcite{Moller:2018kpn} showed it to be statistically poor. Conversely it was found, in a joint analysis of Hyper-K, DUNE and JUNO events, that the BH fraction could be confirmed at $90 \%$ C.L. if it was $ \ge 20 \%$, with a sensitivity to the local supernova rate of 20-33 $\%$ (20 years running). 
\textcite{Ashida:2022nnv} examined the dependence of the DSNB fluxes on the equation-of-state for both its NS and BH contributions. Their results showed that Hyper-K could test the $f_{\rm BH}$-$f_{\rm HNS}$\footnote{The quantity $f_{\rm HNS}$ is the fraction of high mass neutron stars to the total number of supernovae producing a neutron star.} at the $3 \sigma$ level  for 10 years running.
In a subsequent investigation, besides the dependence on the equation of state, \textcite{Ashida:2023heb} proposed a new model for the chemical evolution of galaxies, where the IMF varies with the galaxy type. They found a significant increase at low energy (extra contribution from early-type galaxies) and at higher energy (more frequent BHs) that could be discriminated from the other models considered through a Bayesian analysis of the events in SK-Gd and Hyper-K experiments.

Some of the DSNB predictions are anchored to the SN1987A neutrino signal. 
\textcite{Lunardini:2005jf} first used neutrino spectral information from SN1987A and direct information on the supernova rate, finding the DSNB flux to be a factor of 4 smaller than the early upper limit given by \cite{Malek:2002ns}. \textcite{Yuksel:2007mn} discussed the compatibility of the spectra detected in Kam-II and IMB expeirments and showed that the two were consistent. 
Their DSNB predictions, which were based upon inferential statistical methods,  showed a significant flux in the detectable window ($E \in [10,18]$ MeV).  Using SN1987A analysis to model neutrino emission, \textcite{Vissani:2011kx} obtained DSNB predictions 
finding about one event per year for SK to one to three events per year when Gd is included. 

Predictions for the DSNB neutrino spectra and rates are influenced by flavor conversion mechanisms as well.
The established MSW effect is routinely implemented in predictions 
\cite{Ando:2004hc,Galais:2009wi,Chakraboty:2010sz,Priya:2017bmm,Moller:2018kpn,Horiuchi:2017qja,Kresse:2020nto,Ivanez-Ballesteros:2022szu,Tabrizi:2020vmo,DeGouvea:2020ang,Ekanger:2022neg,Ashida:2022nnv,Ashida:2023heb}. 
To date only a few studies have implemented flavor effects beyond the MSW mechanism. 

\textcite{Lunardini:2012ne} and
\textcite{Chakraboty:2010sz} studied the influence of collective neutrino oscillations due to the $\nu\nu$ interaction that produces single or multiple spectral splits,
depending on the initial neutrino fluxes and the neutrino mass ordering.  They 
found small sensitivity to these effects and the mass ordering. 
 \textcite{Galais:2009wi} investigated shock waves and $\nu\nu$ interaction effects 
 in the bulb model and found variations up to $10 \%$-$20\%$ due to the shock waves.
\textcite{Nakazato:2013maa} found that the DSNB rates also depend on the shock wave revival time. 

Note that the DSNB is also influenced by the way in which one describes the 
late time emission from the cooling of the neutron star (about 50 $\%$ of the entire supernova neutrino emission). Using information from late time 3D simulations, \textcite{Ekanger:2022neg} showed the importance of precise knowledge of the neutron star cooling phase, which can make the DSNB event rate in SK vary by a factor of 2 to 3.
Furthermore, \textcite{Ekanger:2023qzw} included recent data on the evolving star-formation rate as well as long-term axisymmetric models for supernova neutrino emission.
Their findings corroborated the fact that the DSNB detection lies in the near future. 

The DSNB is also a laboratory for nonstandard neutrino properties and particles; see e.g. \cite{Goldberg:2005yw,Farzan:2014gza,Reno:2020cgj,deGouvea:2022dtw}). 
For example, the presence of a background of strongly interacting sterile neutrinos can leave an imprint on the DSNB as dips 
that could be detected by Hyper-K \cite{Balantekin:2023jlg}. 
The DSNB is also sensitive to nonradiative two-body decay in the window $\tau/m \in [10^9, 10^{11}]$ s/eV; see Fig.\ref{fig:nudecay} and \textcite{Ando:2003ie,Fogli:2004gy,Tabrizi:2020vmo,Ivanez-Ballesteros:2022szu,DeGouvea:2020ang}. This observational window for the lifetime-to-mass ratio is unique compared to terrestrial experiments, astrophysical sources like the Sun or a supernova,  
and cosmological probes such as big bang nucleosynthesis or the CMB; for the latter see e.g. \cite{Chen:2022idm}.

In this context, \textcite{Fogli:2004gy} performed a three-flavor study of neutrino nonradiative decay showing that it can have an important impact
on DSNB predictions associated with inverse beta decay. 
In a 2$\nu$ framework of neutrino nonradiative decay, \textcite{DeGouvea:2020ang} found that 
the DSNB has a sensitivity to $\tau/m = 10^9$ s/eV 
even when one implements $40 \%$ uncertainty on the DSNB fluxes.  
In a 2$\nu$ framework \textcite{Tabrizi:2020vmo} 
showed that when combining DSNB detection channels at DUNE, JUNO and Hyper-K one could break degeneracies on DSNB predictions
between the no-decay and the decay case for neutrino normal mass ordering (with a strongly hierarchical mass pattern).
\textcite{Ivanez-Ballesteros:2022szu} performed a full $3 \nu$ investigation including current uncertainties on the evolving core-collapse supernova
rate and different scenarios for the fraction of failed supernovae. They 
revealed that the DSNB flux and rates can be suppressed by several factors
in inverted mass ordering, if neutrinos decay. On the contrary, predictions in normal ordering (for any mass patterns) 
are essentially degenerate in presence or absence of decay in the window $\tau/m \in [10^9, 10^{11}]$ s/eV.
Thus, the possibility that neutrino can nonradiatively decay adds a layer of complexity to the interpretation of the upcoming DSNB measurements, regardless of whether it is detected. 

The SK-Gd experiment has been running since 2020. The JUNO experiment should start by 2025. The Hyper-Kamiokande experiment will start taking data in 2027. 
The DUNE experiments should start in around 2030. Note that the DSNB represents a background in dark matter detectors, referred to
as part of the {\it neutrino floor} or {\it fog}. 

Reduction of the backgrounds is crucial for the detection of the weak DSNB flux signal. 
At low energies, solar $\nu_e$s represent an irreducible background 
in the detection of DSNB $\nu_e$s whereas reactor $\bar{\nu}_e$ constitute a background
that depends on the location of detectors searching for the DSNB $\bar{\nu}_e$s. 
\textcite{Priya:2017bmm} pointed out the necessity of a careful treatment of the neutral-current scattering of atmospheric neutrinos in the SK-Gd experiment. In-depth studies, that also use learning algorithms are ongoing for the suppression of this and of other backgrounds, e.g. from cosmogenic ray muon spallation and production of exotic nuclei such as $^9$Li \cite{Super-Kamiokande:2021jaq}. 
In the case of JUNO, it has been shown that the use of pulse-shape analysis, combined with learning numerical algorithms, 
significantly improves the signal efficiency \cite{Cheng:2023zds}. Moreover, a new detector called THEIA (Titan goddess of light, 
25 or 100 kton) is under study that would combine the specificities of Cherenkov and 
scintillator detectors \cite{Theia:2019non}. THEIA-100 would have an unprecedented capability for background discrimination (for example, atmospheric neutrino neutral-current interactions).

As for the expected number of events, this ranges typically between 10 and 20 events in the SK-Gd experiment, 
about 10-40 in the JUNO experiment, five to ten events in the DUNE experiment, 25 to 145  in the 
Hyper-K experiment (ten years running). Such numbers should be considered as indicative of the most conservative to the most optimistic cases, while some of the 
predictions can yield an even larger number of events. 

At present, there is a fluctuation over background at 1.5 $\sigma$ from a combined analysis of SK-I through SK-IV  \cite{Super-Kamiokande:2021jaq}. 
The first result from the SK-Gd experiment, namely the SK VI-SK VII phases, combined with SK I-SK IV data, shows an excess at 2.3 $\sigma$ in the DSNB signal window (model-dependent analysis) \cite{Harada:2024}. The DSNB discovery could be imminent and, in any case, appears to
lie in the foreseeable future. 

\section{Conclusions and perspectives}
In our journey from the beginnings of neutrino physics and astronomy we went through some of the discoveries that paved our knowledge of these elusive particles and opened new horizons.
After the breakthroughs in the evidence for neutrino oscillations and the solution of the solar neutrino problem, experiments and theory achieved milestones in our understanding of neutrino masses and mixings, of neutrinos from 
stellar and cosmological environments while setting important limits on new physics. 

Neutrino physics and astrophysics remain a very active domain of research. The recent results of the IceCube Collaboration are also 
attracting attention due to their observations on ultrahigh energy neutrinos.

Among the most challenging unsolved issues is the evolution and flavor modification of neutrinos from dense compact objects. 
What makes this problem so interesting and challenging is that, besides shock waves and turbulence inherent to exploding massive stars, one has sizable neutrino-neutrino interactions that render neutrino flavor evolution a complex nonlinear many-body problem. Efforts to solve it are motivated not only by theoretical interest but also by observations.

To date, investigations of shock wave effects have mostly used parametric matter density profiles of one-dimensional supernova simulations. Dips, or bumps, are characteristic features of the neutrino time signals due to the shock wave passage in MSW regions and, in particular, in the $H$-resonance one. The identification of such structures offers a means to identify the neutrino mass ordering, normal if the shock wave passage is seen in the $\nu_e$ time signals, inverted, if seen in the $\bar{\nu}_e$ one. 
Since an experiment like JUNO, DUNE or Hyper-K is likely to unambiguously measure the neutrino mass ordering before the next supernova, the imprint of the shock waves in the time signals will provide a picture of the explosion dynamics. 

Keep in mind that multidimensional supernova simulations present strong anisotropies that 
can produce large angular variations of the front and reverse shocks. Moreover, downflows colliding with hot matter that expands due to convection can induce multiple shocks. As a consequence, 
the exact structures might be direction dependent and may possibly evolve chaotically. 
Therefore, further investigations are necessary to assess whether the generic features of the shock wave passage, identified in one-dimensional studies, remain, when one implements information from multidimensional supernova simulations. 

Turbulence also contributes to the not yet fully understood core-collapse supernova explosion mechanism. It is another important aspect that impacts flavor evolution since it introduces
matter density fluctuations that might produce neutrino depolarization, as mentioned. 
Their characteristics (amplitude, scale and power spectrum) 
should be extracted from multidimensional supernova simulations. Since this is a difficult numerical task, most of the available studies have used parametric matter profiles where fluctuations are superimposed. Thus far only one investigation has exploited information from a two-dimensional supernova simulation, finding weak indications that depolarization takes place, in contrast to all previous findings.
New studies are called for, with input from two- and three-dimensional simulations, to establish whether 
neutrino probabilities have a loss of memory effect, or not, 
due to turbulence, in an exploding supernova.

Studies have uncovered the fact that, besides the established MSW effect present in compact objects, multiple MSW resonances are produced by shock waves or turbulence in supernovae, or more generally because of nonlinear feedback. In particular the MSW-like phenomena that were pointed out include the matter-neutrino resonances, the resonance due to helicity coherence, or the $I$- and synchronized $I$-resonances triggered by nonstandard neutrino-matter interactions. 

Another feature that impacts the neutrino flavor in dense media is neutrino-neutrino interactions, which were first studied in the 1990s in the context of the early Universe.
Their investigation in core-collapse supernovae and compact binary mergers has triggered intense theoretical activity in the 
past 15 years. 
In fact, novel unexpected flavor phenomena, that are located much deeper than the MSW region, have attracted 
interest because of the potential impact on the supernova explosion mechanism and on nucleosynthesis, besides the impact 
on future observations of supernova neutrinos. 

Neutrino evolution in the presence of neutrino-neutrino interactions is still an unsolved problem. From the studies performed thus 
far we learnt that, when
relaxing an approximation, or going beyond approaches, unforeseen aspects emerge that can overturn how we represent the picture of neutrino flavor evolution.  
The first, widely investigated, bulb model revealed collective slow modes that are triggered by mixing.
With frequencies $\sqrt{\mu \omega}$ at typical distances of $\mathcal{O}(10^2$-$10^3)$ km from the neutrinosphere, 
such modes occur in regions where they cannot induce extra heating to help explosions, whereas they can influence the 
r-process, as shown in numerous studies.

Moreover, the interplay of $\nu\nu$ interactions with other contributions, such as standard and nonstandard neutrino-matter interactions, opens the path to new MSW-like phenomena (such as the matter-neutrino and $I$-resonances). With time we have learned that the inclusion of new degrees of freedom, as in nonstationary models, or models with two-dimensional spatial degrees of freedom 
like the line model, opens up new regions for flavor instabilities. There are also situations where small initial perturbations that do not pertain to the same symmetries as the initial neutrino emission provide solutions that spontaneously break symmetries (such as the azimuthal one). And in some cases even chaotic flavor evolution can emerge. 

If instead of only forward-scattering neutrinos, one includes a small amount of backscattered neutrinos or a better description of the $\nu$ angular emission, then the entire picture can be overturned as it came unexpectedly. The first option cast doubts on the treatment of neutrino evolution and the need to go beyond a boundary-value problem.  For the second, 
crossings of the $\nu_e$ and $\bar{\nu}_e$ angular distributions turned out to trigger short scale flavor modes,  i.e. $\mathcal{O}(1)$ m or much less, close to the neutrinosphere. These fast modes are currently under active investigation. 

It is now established that fast modes occur in two- and three-dimensional supernova simulations near the neutrinosphere and
even inside the proton-neutron star.  If the neutrino spectra are similar, at the fast mode location, as it appears, their influence on the spectra is small.
To date only certain studies have evolved fast modes to the full nonlinear regime.  There are indications that fast modes can influence the $r$-process in binary neutron star mergers and the $\nu p$-process in core-collapse supernovae. In addition, three-flavor effects were shown to be important in determining when flavor evolution is modified on large scales. The study of fast modes, the conditions for their occurrence and impact is at present a fast-developing field.  

With a few exceptions, all the findings concerning flavor evolution in dense environments available in the literature use the mean-field approximation. 
Linearized mean-field equations and a dispersion-relation approach for fast modes are commonly used to study when neutrino flavor modification is triggered. This 
has the advantage of solving an eigenvalue equation close to the initial condition but loses the long-term evolution of the full nonlinear problem.

Special effort has been devoted to checking the validity of the mean-field equations. This has led 
to new evolution equations, to the rederivation of quantum kinetic equations and to the first attempts to solve kinetic equations with the inclusion of mixings in schematic models. 
Neutrinos in the early Universe, where neutrino kinetic equations are needed, represent a different case in many respects. The homogeneity and isotropy of the medium made possible the first consistent calculations of neutrino evolution with the full collision term, the mixings and mean-field terms.

In the supernova context, extended mean-field evolution equations were derived using in particular the coherent-state path integral, the closed-time-path integral and the BBGKY hierarchy. Such equations have included, in particular, contributions by supplementary two-point correlators, i.e. helicity coherence and pairing correlators. For the former calculations based on detailed simulations of binary compact mergers and core-collapse supernovae have shown that they do not trigger significant flavor evolution due to nonlinear feedback, as perturbative arguments also show. For the latter, no flavor modification appears since the kinetic terms dominate. 

The impact of collisions on flavor evolution is currently an open problem that is numerically challenging because of its high dimensionality. For a long time, the argument of the separation of scales between flavor mechanisms (the MSW effect) and the collision-dominated region justified the use of mean-field equations.  With the advent of $\nu\nu$ interaction studies for dense astrophysical environments, the identification of slow and then fast modes has deeply changed our vision. 

Currently, the interplay between the collisions and fast modes is receiving particular attention. Studies of models with lower dimensionality and approximate treatment of collisions (for example, direction changing and neutrino-nucleon collisions only) uncovered the possibility that collisions can trigger fast modes, suppress them, or enhance them, depending, for example, on the assumed angular distribution of the neutrino emission at the neutrinosphere. While the models studied thus far necessarily have many approximations and limitations, there are now indications that we need to progress toward further complexity since, even if the collision rate is smaller than the flavor scale, collisions are important. And in fact even the crossings between the neutrino and antineutrino angular distributions, associated with the occurrence of fast modes should emerge from collisions in a fully consistent treatment. 

Furthermore, collisional instabilities have been uncovered that are triggered by a difference between the collision rates of neutrino and antineutrino in dense regions. 
The understanding of fast modes, collisional instabilities and their interplay will require more work. Moreover, a crucial question involves what is their impact on the supernova, as well as on nucleosynthesis and on observations. Studies to assess their role, including in SN simulations, have just started and will be pivotal in the coming years.

Conceptually these studies have brought crucial progress. We were used to the MSW effect 
that occurs to the vacuum and matter contributions to the neutrino Hamiltonian, and is also due to the observational discoveries in the field. We have learnt that slow modes do not require matter to take place, and fast modes do not need mixings either, but both depend on sizable neutral-current neutrino-neutrino interactions. Theoretical developments have widely broadened our conceptual framework, opening up a vast ensemble of possibilities due to the nonlinear nature of flavor evolution in dense environments.

All these developments are based on theoretical approaches in flat spacetime.
However, strong gravitational fields are present near compact objects. Their impact on flavor evolution is still in the exploratory phase. 
An extension of the equations of motion in curved spacetime has been discussed. 

Several studies have investigated the impact on the vacuum oscillation phase for different metrics and recently on the decoherence by wave packet separation, in a wave-packet treatment of neutrino evolution in curved spacetime. A halo effect was found in an exploding supernova, whereas it was shown that the inclusion of gravity effects (trajectory bending, energy redshift, etc...) influences $r$-process nucleosynthesis in accretion disks around black holes. Gravitational effects on neutrino propagation and flavor evolution merit more attention in the future. 

Furthermore interesting connections between a system of weakly interacting neutrinos and other domains have been uncovered, often opening up new unforeseen possibilities. 
Links have been established to BCS theory for superconductivity. On the one hand, it has been shown that the neutrino Hamiltonian in the spin formalism is
equivalent to the reduced BCS Hamiltonian. On the other hand, extended equations obtained through the BBGKY hierarchy were shown to include
correlations of the pairing type. This also allowed the equations of motion for the neutrino many-body system 
to be formally connected to other systems in condensed matter, 
or to atomic nuclei.  

An algebraic formulation and the Bethe ansatz showed the $\nu\nu$ many-body Hamiltonian to be solvable (under some conditions). This and other works have yielded the first comparison between mean-field and many-body results highlighting the role of many-body correlations, in particular, through the entanglement entropy. Moreover the
first calculations based on an inference procedure and on quantum devices are appearing. The latter open interesting new possibilities.  

In this respect, keep in mind that the results obtained thus far on the role of many-body correlations are exploratory, potentially indicating limitations of the widely employed mean-field approximation. Notwithstanding, only when we obtain results for really large number of neutrinos will we be able to assess the necessity of going beyond mean-field approaches and predictions. This constitutes a long-term goal.

On the observational side, after the approximately two dozen $\bar{\nu}_{e}$ events of SN1987A we are eagerly awaiting the next supernova to precisely measure the neutrino light curves: perhaps this time if the supernova is close and we are patient enough. 
This observation is crucial both for astrophysics and for particle physics. To mention a few examples, we will learn much about the supernova explosion, possibly having definitive evidence for the explosion mechanism and the favored delayed neutrino-heating mechanism, and the onset time of the explosion which is important for gravitational wave detection. We will get a picture of the shock wave passage in the MSW region from the time signal and, eventually, signatures of the SASI. 
With the advent of SNEWS 2.0 we will be able to locate the exploding star through its neutrinos. 

Moreover, from the first investigations based on Bayesian analysis, a
promising capability to distinguish among different supernova models, and even
various flavor scenarios, has emerged. Judging by the steady progress that has
been made on supernova simulations, in our understanding of neutrino properties and
neutrino flavor evolution in dense media and with the advent of new and even
larger size detectors, we will be ready to extract the most when the next core-collapse
supernova is detected through its neutrinos.

The discovery of the diffuse supernova neutrino background might be imminent. It will be the second time ever we have observed neutrinos from core-collapse supernovae, with a unique sensitivity to the evolving core-collapse supernova rate, the fraction of failed supernovae and binaries, flavor mechanisms and nonstandard neutrino properties such as neutrino decay. 
The ensemble of studies performed so far shows that the DSNB encodes crucial information for astrophysics and particle physics.

Much of the information that we will extract from the upcoming observations will depend on different aspects. These go from lowering energy thresholds and reducing backgrounds
in experiments to constraining the number of parameters and theoretical scenarios through the identification of the neutrino mass ordering, better knowledge of the equation of state of dense matter, or better knowledge of the late time neutrino emission in long-term 3D simulations. Moreover, the steady progress in our understanding of neutrino flavor evolution in 
dense environments will also contribute to the assessment of their impact on the DSNB.  If the DSNB's much awaited discovery becomes a reality, this novel window in low energy neutrino astrophysics will be a unique laboratory for astrophysics, particle physics and the search for new physics.  

In this journey, we highlighted aspects of our current understanding of flavor evolution in dense media, setting it in the context of the historical developments in neutrino physics. 
With a somewhat historical perspective on the theoretical progress in this field, we discussed many aspects that now appear 
to be clear, as well as the numerous theoretical challenges that lie ahead.  

Serious progress has been made. But one lesson that we have learned is that new possibilities can always surface 
and completely change the way that we look at this complex problem. Novel developments that have 
recently emerged might once more provide a completely new insight into this interesting subject.

\section{Acknowledgments}
Through the years I had interesting discussions with numerous researchers in the field. In particular, I thank Baha Balantekin, George Fuller, Carlo Giunti, Thomas Janka, Manfred Lindner, Gail McLaughlin, Kate Scholberg, Francesco Vissani, and also Evgeny Akhmedov, Alessandro Mirizzi, Georg Raffelt, Alexey Smirnov, Rebecca Surman, Irene Tamborra as well as many others with whom I had 
discussions at workshops and conferences, or with whom I exchanged correspondence. 

Moreover, I would like to thank very warmly to two of my former PhD students, Am\'elie Chatelain and J\'er\^ome Gava for the enthusiasm that we shared while working together in this interesting domain. 

\section{Appendix: The spin formalism}
The formalism of effective spins has been extensively used to describe neutrino evolution in dense environments, particularly 
in theoretical studies that involve $\nu\nu$ interactions. It provides a useful tool to picture neutrino conversion mechanisms in flavor space.

We consider a two-level system described via the Hamiltonian (see \cite{Cohen})
\beq\label{eq:H22}
H =  \left(
\begin{tabular}{cc}
$H_{11}$ & $H_{12}$     \\
$H_{21} $ &  $ H_{22}$  \\
\end{tabular}
\right) \ ,   
\eeq
with $H_{21} = H_{12}^*$, that evolves
according to the Schr\"odinger equation
\beq\label{eq:Schro}
i {d \over{dt}} \vert \psi(t) \rangle = H  \vert \psi(t) \rangle \ .
\eeq
We also consider a spin 1/2, with  
${\bf S} = {\bf \sigma} /2$, with $\sigma $ the Pauli matrices 
\begin{equation}\label{eq:Paulimat}
\sigma_x =
\left(
\begin{matrix}
 0 & 1 \\
 1 & 0 
\end{matrix}
\right)
~~~\sigma_y =
\left(
\begin{matrix}
 0 & - i \\
 i  & 0 
\end{matrix}
\right)
~~~\sigma_z =
\left(
\begin{matrix}
 1 & 0 \\
 0 & -1 
\end{matrix} 
\right) \ . 
\eeq
that are a basis for $2 \times 2 $ matrices.
One can associate any two-level system with an effective spin 1/2 whose Hamiltonian
in the basis of the eigenstates of the third component of ${\bf S}$, i.e. $S_z$ has the same elements in $H$ Eq.\eqref{eq:H22}.

Toward this aim we remember that the Hamiltonian describing the interaction between a spin 1/2 and a magnetic field reads
\beq\label{eq:Bint}
{\tilde H} = - \gamma {\bf B} \cdot {\bf S} \\ , 
\eeq
with $\gamma$ the gyromagnetic factor.
When one uses \eqref{eq:Paulimat}, the operator \eqref{eq:Bint} reads in matrix form
\beq\label{eq:HB}
{\tilde H} = - {\gamma \over 2} 
\left(
\begin{matrix}
B_z & B_x - i B_y \\
B_x + i B_y & B_z 
\end{matrix}
\right) \ ,
\eeq
with $B_x, B_y$, and $B_z$ the components of the magnetic field.

For Eq.\eqref{eq:Bint} to be identified with Eq.\eqref{eq:H22},
one must introduce an effective magnetic field ${\bf B}$
with components 
\beq\label{eq:B}
{\bf B} =  - { 1 \over \gamma}\left(
\begin{tabular}{c}
2 ${\rm Re}H_{12}$   \\
- 2 ${\rm Im}H_{12} $  \\
$H_{11} - H_{22}$
\end{tabular}
\right) \ . 
\eeq

Following the Ehrenfest theorem, 
one can show that the Schr\"odinger equation
can be replaced by a precession equation for
the expectation values of the effective
spin, subject to the effective magnetic field,
\beq\label{eq:prec}
\dot{\bf P}  = {\bf P} \times  {\bf B} \ .
\eeq

Such a spin formalism is routinely employed to describe neutrino evolution in flavor space;
see for example \cite{Kim:1987ss,Kim:1987bv,Pastor:2001iu,Duan:2005cp,Duan:2010bg,Galais:2011gh}). 
We now consider the evolution equation \eqref{eq:2nu} for 2$\nu$ flavors involving the neutrino wavefunction (instead of the neutrino flavor state)
\beq\label{eq:nuampl}
\vert \psi (t) \rangle =
\left(
\begin{tabular}{c}
$a_{\nu_e} $ \\
$a_{\nu_{\mu}}$ 
\end{tabular}
\right) \ ,
\eeq
where $a_{\nu_e}$ and $a_{\nu_{\mu}}$ the neutrino amplitudes for a neutrino to be $\nu_e$ or $\nu_{\mu} $ at time $t$. 

Thus, from Eqs. \eqref{eq:Paulimat} and \eqref{eq:nuampl}, one has for the {\it polarization vector} 
\beq\label{eq:pol}
{\bf P} = \langle \psi (t) \vert \sigma \vert \psi (t) \rangle =
\left(
\begin{tabular}{c}
2 ${\rm Re}(a_{\nu_e}  a_{\nu_{\mu}}^*) $  \\
2 ${\rm Im}(a_{\nu_e}  a_{\nu_{\mu}}^*) $ \\
$\vert a_{\nu_e} \vert^2 -  \vert a_{\nu_{\mu}} \vert^2$
\end{tabular}
\right) \ .
\eeq
Note that the polarization vector is related to the density matrix through
\beq\label{eq:poldes}
\varrho = {1 \over 2} (P_0 {\bf 1}  + {\bf \sigma} \cdot {\bf P}) \ ,
\eeq
where the first term depends on the trace $tr(\varrho) = P_0$ and the second term depends upon a trace-free part.

With the polarization vector in Eq.\eqref{eq:prec} defined for $2 \nu$ flavors,  
the effective magnetic field depends on the medium that the neutrinos are traversing.
If one includes only the vacuum and the matter terms to the neutrino Hamiltonian in Eq.\eqref{eq:Schro},
the effective magnetic field reads
\beq\label{eq:Bnu}
{\bf B} = \omega  {\bf B_{\rm vac}} + {\bf B_{\rm mat}} \ , 
\eeq
with $\omega$ the vacuum oscillation frequency and
\begin{equation}\label{eq:Bnuex}
{\bf B_{\rm vac}} =
\left(
\begin{matrix}
 \sin 2 \theta   \\
0   \\
-  \cos 2 \theta 
\end{matrix}
\right)
~~~
{\bf B_{\rm mat}} =
\left(
\begin{matrix}
 0  \\
0   \\
 \sqrt{2} G_F n_e 
\end{matrix}
\right)
 \ .
\eeq
Figure \eqref{fig:spins} shows an example of the evolution of the
neutrino flavor isospins in presence of neutrino mixing and neutrino-matter interactions. 
Equations \eqref{eq:prec}, \eqref{eq:Bnu}-\eqref{eq:Bnuex} can easily be extended
in the presence of $\nu\nu$ interactions; see \textcite{Sigl:1992fn,Pastor:2001iu,Duan:2005cp,Duan:2010bg,Galais:2011gh}).
Note that, following \textcite{Duan:2005cp}, the vector ${\bf P}$ in  \eqref{eq:pol} is often referred to as the 
neutrino flavor isospin (NFIS) in studies on $\nu\nu$ interactions
in core-collapse supernovae. The extension to $3\nu$ flavors requires SU(3) and was 
discussed for example by \cite{Dasgupta:2007ws}.  

\section{Appendix: Neutrinos and the heating rate in core-collapse supernovae or the $r$-process}
Neutrinos play roles in the core-collapse supernova dynamics and in nucleosynthesis processes, particularly the $r$-process, 
the $\nu$p process and $\nu$ nucleosynthesis in dense environments.
Studies have shown that flavor evolution impacts such nucleosynthesis processes, and recently have also displayed 
the supernova dynamics when fast modes are considered. 
 
Here we discuss further elements in these connections.
Both for the supernova dynamics and for the nucleosynthetic abundances, 
a key role is played by neutrino interactions on neutrons and protons, i.e.,
\begin{align}\label{eq:nuint}
\nu_e + n &  \rightarrow e^- + p  \ ,    \\  
\bar{\nu}_e + p & \rightarrow e^+ + n \ . \nonumber
\end{align}

A key parameter for the r process is the electron fraction $Y_e$ Eq.\eqref{eq:Ye}. Equations 
\eqref{eq:nuint} are the dominant reactions that set the neutron-to-proton ratio (as in primordial nucleosynthesis) and $Y_e$ in neutrino-driven winds; see \cite{Qian:1996xt,Hoffman:1996aj,Martin:2015hxa}). The ratio of the rates for  Eqs.\eqref{eq:nuint} is (see \textcite{Qian:2013fsa})
\beq\label{eq:ratiorates}
{\lambda_{\nu_e n} \over {\lambda_{\bar{\nu}_e p}}}  = {{\phi_{\nu_e} \over {\phi_{\bar{\nu}_e}}}} {{\langle \sigma_{\nu_e n} \rangle} \over {\langle \sigma_{\bar{\nu}_e p}}  \rangle}
\approx {L_{\nu_e} \over {L_{\bar{\nu}_e}}} {{\epsilon_{\nu_e} + 2~ \Delta_{np}} \over{\epsilon_{\bar{\nu}_e} -  2~ \Delta_{np}} } \ , 
\eeq
where $\langle \sigma_{\bar{\nu}_e p}  \rangle$ and $\langle \sigma_{\nu_e n} \rangle$ are the neutrino flux-averaged cross sections, 
$\epsilon_{\bar{\nu}_e} \equiv {\langle E_{\bar{\nu}_e}^2 \rangle / { \langle E_{\bar{\nu}_e}} \rangle}$ (a similar approach is taken with 
$\epsilon_{\nu_e}$)
and $\Delta_{np} = 1.293$ MeV is the neutron-proton mass difference.
The last equality on the rhs of Eq. \ref{eq:nuint} comes from the approximate expressions $\phi_{\nu_e} = {L_{\nu_e} / \langle {E_{\nu_e} \rangle} }$ (a similar approach is taken for $\bar{\nu}_e$) and from the low energy $\nu$-nucleon cross sections $\sigma_{\nu_e n} \approx (E_{\nu_e} + \Delta_{np})^2$ and  $\sigma_{\bar{\nu}_e p} \approx (E_{\nu_e} - \Delta_{np})^2$;
for a discussion of the $\nu$-nucleon cross sections, see \textcite{Strumia:2003zx,Ricciardi:2022pru}).

Concerning core-collapse supernova neutrinos and the supernova dynamics, 
knowledge of the neutrino luminosities, spectra and angular distributions is necessary to calculate the heating rate in 
the {\it gain region}. Introduced by \textcite{Bethe:1992fq}, the gain radius separates the region in the supernova
where energy loss due to electron capture is less than energy gain due to neutrino absorption. Within the gain radius the region cools through neutrino emission,
whereas outside matter heats because of neutrinos.

The neutrino heating rate per gram in the gain region reads (see \textcite{Mezzacappa:2022hmk,Janka:2017vcp})
\begin{equation}\label{eq:heatrate}
q_{\nu} = {X_n \over{\lambda^0_a}} {{L_{\nu_e}  \langle E_{\nu_e}^2 \rangle} \over{4 \pi r^2}   \langle s_{\nu_e} \rangle }
+  {X_p \over{\bar{\lambda}^0_a}} {{L_{\bar{\nu}_e}  \langle E_{\bar{\nu}_e}^2 \rangle }  \over{4 \pi r^2}   \langle s_{\bar{\nu}_e} \rangle } \ , 
\end{equation}
where the so-called flux factor $s_{\nu_e}$ and $s_{\bar{\nu}_e} $, in the denominators, implement the fact that neutrinos close to the neutrinosphere do not necessarily move radially outward.
The quantities $X_n$ and $X_p$ are the neutron and proton mass fractions and $\lambda^0_a$ and $\bar{\lambda}^0_a$ are constants related to the weak interaction coupling constants. 
Eq.\eqref{eq:heatrate} indicates that knowledge about neutrino angular distributions, spectra, and luminosities is necessary to determine the neutrino heating rate. 

In conclusion, from \eqref{eq:ratiorates} and \eqref{eq:heatrate} it is clear that spectral modifications due to flavor mechanisms 
can influence the electron fraction, and therefore nucleosynthetic abundances in a dense environment, as well as 
the heating rate, and consequently, the dynamics of a core-collapse supernova. 

\bibliography{references}

\end{document}